\documentclass[aps,showpacs,pre,amsmath,amsfonts,amssymb,superscriptaddress,nofootinbib,notitlepage,a4paper]{revtex4}

\usepackage{cancel}

\usepackage{graphicx}
\usepackage{psfrag}
\usepackage{hyperref}

\newcommand{\beq}{\begin{equation}} 
\newcommand{\eeq}{\end{equation}}
\newcommand{\bem}{\begin{multline}}
\newcommand{\bes}{\begin{split}}
\newcommand{\ees}{\end{split}} 
\newcommand{\bea}{\begin{eqnarray}}
\newcommand{\eea}{\end{eqnarray}}

\def\s{{\sigma}}
\def\t{{\tau}}
\def\g{{\gamma}}
\def\b{{\beta}}
\def\a{{\alpha}}

\def\us{{\underline{\sigma}}}

\def\ut{{\underline{\tau}}}

\def\di{{\partial i}}
\def\da{{\partial a}}

\def\dima{{\partial i \setminus a}}
\def\dami{{\partial a \setminus i}}

\def\tc{{\theta_{\rm IT}}}
\def\td{{\theta_{\rm sp}}}
\def\tKS{{\theta_{\rm KS}}}

\def\aIT{{\alpha_{\rm IT}}}
\def\aSP{{\alpha_{\rm sp}}}
\def\aKS{{\alpha_{\rm KS}}}
\def\aALG{{\alpha_{\rm alg}}}

\def\ind{{\mathbb{I}}}

\def\zv{{z_{\rm v}}}
\def\ze{{z_{\rm e}}}
\def\zc{{z_{\rm c}}}

\def\hf{{\widehat{f}}}
\def\ha{{\widehat{a}}}
\def\hA{{\widehat{A}}}
\def\hb{{\widehat{b}}}
\def\hB{{\widehat{B}}}
\def\hC{{\widehat{C}}}
\def\hM{{\widehat{M}}}
\def\hdelta{{\widehat{\delta}}}

\def\hz{{\widehat{z}}}

\def\hP{{\widehat{P}}}

\def\dd{{\rm d}}
\def\la{{\langle}}
\def\ra{{\rangle}}

\def\ta{{\widetilde{a}}}
\def\tA{{\widetilde{A}}}

\def\ttheta{{\widetilde{\theta}}}

\def\tv{{\widetilde{v}}}
\def\tw{{\widetilde{w}}}

\def\tp{{\widetilde{p}}}

\def\tz{{\widetilde{z}}}

\def\atanh{{\text{atanh} \, }}
\def\E{{\mathbb{E}}}

\def\tE{{\widetilde{\mathbb{E}}}}

\def\eqd{\overset{\rm d}{=}}
\def\tod{\overset{\rm d}{\to}}
\def\oeta{\overline{\eta}}
\def\onu{\overline{\nu}}
\def\om{\overline{m}}
\def\omc{\overline{m}_{\rm c}}

\def\N{{\cal N}}

\def\thetaKS{\theta_{\rm KS}}
\def\sign{\text{sign}}

\def\ve{\varepsilon}

\def\pj{p_{\rm j}}

\begin{document}
\bibliographystyle{myunsrt}

\title{Typology of phase transitions in Bayesian inference problems}

\author{Federico Ricci-Tersenghi}
\affiliation{Diparimento di Fisica, Sapienza Universit\`a di Roma,
  and Nanotec-CNR, UOS di Roma, and INFN-Sezione di Roma 1,
  P.le Aldo Moro 5, 00185 Roma Italy}
\author{Guilhem Semerjian}
\affiliation{Laboratoire de Physique de l’Ecole normale sup\'erieure, ENS, Universit\'e PSL, CNRS, Sorbonne Universit\'e, Universit\'e Paris-Diderot, Sorbonne Paris Cit\'e, Paris, France}
\author{Lenka Zdeborov\'a}
\affiliation{Institut de Physique Th\'eorique, CNRS, CEA, Universit\'e Paris-Saclay, Gif-sur-Yvette, France}

\begin{abstract}

Many inference problems undergo phase transitions as a function of the signal-to-noise ratio, a prominent example of this phenomenon being found in the stochastic block model (SBM) that generates a random graph with a hidden community structure. Some of these phase transitions affect the information theoretically optimal (but possibly computationally expensive) estimation procedure, others concern the behavior of efficient iterative algorithms. A computational gap opens when the phase transitions for these two aspects do not coincide, leading to a hard phase in which optimal inference is computationally challenging. In this paper we refine this description in two ways. In a qualitative perspective, we emphasize the existence of more generic phase diagrams with a hybrid-hard phase, in which it is computationally easy to reach a non-trivial inference accuracy, but computationally hard to match the information theoretically optimal one. We support this discussion by quantitative expansions of the functional cavity equations that describe inference problems on sparse graphs, around their trivial solution. These expansions shed light on the existence of hybrid-hard phases, for a large class of planted constraint satisfaction problems, and on the question of the tightness of the Kesten-Stigum (KS) bound for the associated tree reconstruction problem. Our results show that the instability of the trivial fixed point is not a generic evidence for the Bayes-optimality of the message passing algorithms.
We clarify in particular the status of the symmetric SBM with 4 communities and of the tree reconstruction of the associated Potts model: in the assortative (ferromagnetic) case the KS bound is always tight, whereas in the disassortative (antiferromagnetic) case we exhibit an explicit criterion involving the degree distribution that separates a large degree regime where the KS bound is tight and a low degree regime where it is not. We also investigate the SBM with 2 communities of different sizes, a.k.a. the asymmetric Ising model, and describe quantitatively its computational gap as a function of its asymmetry, as well as a version of the SBM with two groups of $q_1$ and $q_2$ communities. We complement this study with numerical simulations of the Belief Propagation iterative algorithm, confirming that its behavior on large samples is well described by the cavity method.

\end{abstract}

\maketitle

\tableofcontents

\section{Introduction}

Problems of statistical inference where a signal is observed via noisy
measurements appear ubiquitously in situations involving data
analysis. It is intuitive that as the signal-to-noise ratio
decreases, the recovery of the signal becomes harder. 
Quantifying the optimal performance is one of the main theoretical tools to 
provide performance guarantees in practical situations. 
Numerous models of statistical inference bear formal analogies to
models of disordered systems and thanks to this connection
methods from statistical physics were applied to inference problems,
for reviews see e.g.~\cite{NishimoriBook01,MezardMontanari07,ZdKr16_review,BaPeWe18}. 
Sharp changes in behaviour, known as phase transitions, appear in many 
situations in physical systems. Due to the above connection it does not come 
as a surprise that various types of phase transitions
were also identified in inference problems. The statistical and
computational implications of the presence of phase transitions recently
attracted attention in several fields.

A number of works, including the present one, focus on inference models 
defined on sparse graphs or hypergraphs where the signal is related to a 
(hidden) labeling of the nodes, and the graph is generated via a rule that 
depends on these labels. A paradigmatic example of this class is the 
stochastic block model (SBM) where nodes belong to ``communities'' and the 
probability of observing an edge between a pair of nodes depends on the 
communities of the two nodes. There is a long history of studies of this 
model, see e.g.~\cite{ZdKr16_review,Cris_review_SBM,Abbe_review_SBM} and
references therein, particularly because of its relevance in the context
of complex networks~\cite{Fortunato10}.  
In terms of phase transitions in the SBM, 
the work of~\cite{DecelleKrzakala11,Decelle_SBM_long} was particularly
influential as it identified three phases - undetectable, hard and
easy - and located quantitatively
transitions between them. In the undetectable phase inference better
than random guessing is not possible, in the hard phase it is
information-theoretically possible, but known algorithms are not able
to perform better than random guesses (depending on parameters of the
model, this phase is sometimes missing), in the easy phase known
algorithms match the information-theoretically optimal performance 
that is in this case strictly better than random guesses.  
Another example of inference problems on sparse hyper-graphs are
the planted constraint satisfaction problems with various types of
constraints~\cite{KrzakalaZdeborova09,ZdeborovaKrzakala09,AbMo13,FePeVe15}. 
Planted constraint satisfaction problems are instrumental in studies of
average algorithmic complexity~\cite{Fe02}, hardness of finding a 
solution~\cite{BarthelHartmann02} and refutation of satisfiability formulas. 
The same three phases explained above in the SBM example have also been found 
in planted constraint satisfaction problems, as well as in many other 
``dense'' inference problems in which the signal is observed through all 
pairs of variables, not only the edges of a sparse graph 
(see~\cite{ZdKr16_review} for a review).

This paper provides a more detailed picture of these phase transitions.
We emphasize indeed the existence of hybrid-hard phases, in which
efficient inference better than
random guessing is possible, however, matching the
information-theoretically optimal performance is computationally
hard. This type of phase was identified in dense
inference problems, see e.g.~\cite{LeKrZd17}, but was missed in some 
previous studies of sparse problems~\cite{ZdeborovaKrzakala09}. We 
revisit the latter by means of the cavity method, that relates the analytic
description (in particular the optimal errors, as defined
below) of inference problems on sparse graphs and hyper-graphs to 
reconstruction problems on trees~\cite{Mossel01,JansonMossel04,MezardMontanari06,KS_tight_Ising,Sly08,Sly11} (the local limit of the graphs in the large 
size limit). These tree problems are solved by fixed point equations
whose unknown (called order parameter) are functions (more precisely 
probability distributions), at variance with their dense counterparts 
where the order parameters are finite dimensional. We will focus on problems 
where the undetectable phase exists at small signal to noise ratio,
which translates into the existence of a so-called non-informative (or trivial) 
fixed point of the corresponding cavity equations (in such cases there exists 
a close connection between the inference problems and the associated uniformly 
random models~\cite{KrzakalaMontanari06,CoKrPeZd16,CoEfJaKaKa17} through 
the notion of quiet planting). There are several reasons that motivate this
restriction of the family of inference problems we consider here: in a certain
sense these are the hardest ones, as no local information can be exploited for
the inference of the hidden signal they contain. Moreover they exhibit a
richer phenomenology in terms of phase transitions than the problems where a
local property (the degree of a vertex in the SBM for instance) is correlated
to the signal. Finally from a technical point of view the existence of a
trivial fixed point allows to set up a systematic perturbative expansion
of the cavity equations, in the neighborhood of the so-called Kesten-Stigum 
transition where the trivial fixed point looses its stability; this
perturbative computation will be our main technical contribution in this paper. Similar expansions can be found 
in~\cite{KS_tight_Ising,Sly11,CoEfJaKaKa17} but our results are either valid 
for more general models or pushed to an higher order in the perturbation 
series. This technical tool allows us to clarify several features of the
phase diagrams of inference and tree reconstruction problems.

For the symmetric stochastic block model with $q$ groups (or the tree 
reconstruction of the $q$ state Potts model), it was known that the hard phase 
does not exist for 2 and 3 groups (in other words the Kesten-Stigum bound on 
reconstruction is tight), whereas it always exists for 5 and more 
groups~\cite{Sly11}. In the intermediate case of $q=4$, whose status remained
unclear up to now, we provide an explicit condition on the degree
distribution and the ferromagnetic (or assortative) character of the model
to distinguish these two qualitatively different behaviors.

We also study the asymmetric SBM with two groups of different sizes (but equal
average degree) and derive conditions on the asymmetry that induces the 
appearance of the hard phase. We recover the critical fraction $1/2-1/\sqrt{12}$
for the size of the smaller group below which the hard phase exists,
that appeared in related studies in the dense 
regime~\cite{BaDiMaKrLeZd16,LeKrZd17} or in the limit of large 
degrees~\cite{CaLeMi16,LeMi16}; quite strikingly this critical
asymmetry does not depend on the degree distribution. This phenomenon was also
studied in the tree reconstruction perspective under the name of asymmetric 
Ising model~\cite{Mossel01,KS_tight_Ising}.

Based on this expansion, we also conclude that the hybrid-hard phase always 
exists in a very generic class of sparse inference models where Boolean 
variables are observed through $k$-uplets, for any $k$, as long as they 
preserve a global symmetry between the two possible values of the variables.
This class encompasses a large part of the Boolean occupation problems 
of~\cite{ZdeborovaKrzakala09}. More precisely, in that class of models we 
rule out the existence of a hard phase that would not be accompanied by the 
hybrid-hard phase.

Our analytical findings are confirmed by running Belief Propagation (BP) on 
large instances of the corresponding problems, thus providing a strong 
evidence that the analytical solution obtained in the thermodynamical limit 
is also of practical relevance for problems of large but finite size.

The two papers~\cite{liu2017reconstruction,liu2017tightness} that appeared
while we were finishing the present work have some overlap with ours.
The authors study indeed the reconstruction problem on regular $d$-ary trees 
and provide expansions around the Kesten-Stigum transition, for the asymmetric
Ising model and a $q$-state Potts model whose symmetry is partially broken.
While the results of~\cite{liu2017reconstruction,liu2017tightness} are rigorous
our expansions are more generic, pushed to an higher order, and put to use 
in connection with the inference problems on graphs.

The rest of the paper is organized as follows. In Sec.~\ref{sec_typology} we present in generic terms the structure describing the performance of optimal and efficient estimators in inference problems, review the well-known phase diagrams that have been deduced from it, and explain the more complicated possible scenarios that can be observed. In Sec.~\ref{sec_generic} we define a family of inference problems on sparse (hyper)graphs, introduce the cavity equations that describe them and explain their link to the tree reconstruction problem; for the convenience of the reader we then summarize our main results in Sec.~\ref{sec_mainresults}. The latter have been obtained by systematic moment expansions of the functional cavity equations, that are presented in Sec.~\ref{sec_Potts} for arbitrary discrete variables with pairwise interactions, and in Sec.~\ref{sec_Ising} for boolean variables with $k \ge 2$-wise interactions; the more technical parts of these computations are deferred to a series of Appendices. Sec.~\ref{sec_numerical} is then devoted to numerical experiments with Belief Propagation on finite size samples, and finally we draw some conclusions and perspectives in Sec.~\ref{sec_conclu}.

\section{Typology of phase transitions}
\label{sec_typology}

This first section is dedicated to the presentation of our qualitative results on the variety of possible phase diagrams in inference problems. We shall first briefly review the Bayesian perspective on inference and the way statistical mechanics handles it, staying at an abstract level of description (concrete examples will be introduced in Sec.~\ref{sec_generic}). The typology of phase transitions will then naturally follow from this discussion, as well as the quantitative computations that can be performed on concrete examples as first steps towards this classification  (which will form the core of the rest of the paper).

\subsection{Bayesian inference problems}
\label{sec_typology_Bayesian}

In a statistical inference problem a ground-truth vector $\underline s^* \in {\mathbb R}^N$ is to be inferred from some observations (or data) denoted $G$, that are correlated to $\underline s^*$. In the Bayesian setting $\underline s^*$ is a random variable whose distribution is called ``prior'', and $G$ is a random variable with a conditional law $P(G|\underline s^*)$, that correlates it to the ground-truth. An observer provided with a sample of $G$, and with the knowledge of the model that generated it (i.e. of the prior distribution and of $P(G|\underline s^*)$) must base her inference of $\underline s^*$ on the posterior distribution that follows from Bayes' theorem
\beq
P(\underline s|G) = \frac{1}{Z(G)} P(G|\underline s) \prod_{i=1}^N P_0(s_i) \ ,
\label{posterior}
\eeq
where we assumed for simplicity a prior distribution $P_0$ factorized on the components of the vector, and $Z(G)$ is a normalizing constant. The observer uses an estimator $\widehat{\underline{s}}(G)$ that should ideally be ``close'' to $\underline s^*$. Which estimator is optimal depends on the definition of ``closeness'' between $\widehat{\underline{s}}(G)$ and the true value $\underline s^*$. To clarify this point let us denote the marginal probability distributions of the posterior as 
\beq
\mu_i(s_i) = \sum_{\{s_j\}_{j\neq i}}  P(\underline s|G) \ . 
\eeq
If the distance between $\widehat{\underline{s}}(G)$ and $\underline s^*$ is measured in terms of the mean squared error ${\rm MSE}=\sum_{i}(\widehat{s}_i-s_i^*)^2/N$, then the optimal estimator (that minimizes the MSE) is simply given by the means of the marginals 
\beq
\widehat{s}_i  = \sum_{s_i} s_i \mu_i(s_i) \ , 
\eeq 
and the error achieved by this estimator is called the minimum mean-squared error (MMSE). When the variables $s_i$ belong to a discrete set $\chi$ (as will be the case in this paper) another meaningful definition of similarity between $\widehat{\underline{s}}(G)$ and $\underline s^*$ is the mean overlap ${\rm MO} = \sum_i \delta_{\hat s_i, s_i^*} /N$. The estimator that maximizes the mean overlap is given by the value of $s_i$ for which the marginal is the largest, namely
\beq
\widehat{s}_i  = {\rm argmax}\;\mu_i(s_i) \, . 
\eeq

We are interested in this paper in phase transitions, i.e. qualitative changes of the behavior of such inference problems and non-analyticities in their optimal error. These can only occur in the so-called thermodynamic limit $N\to \infty$; the dimensionality of $G$ must be scaled appropriately for this limit to be non-trivial, in such a way that the signal-to-noise ratio (SNR) (whose precise definition depends on the specific problem considered) that measures the quantity of information on $\underline s^*$ conveyed by $G$ remains finite in the limit. We shall focus on a particularly interesting and challenging class of problems for which there exists a non-trivial low-SNR phase in which the inference problem cannot be solved more precisely than by using only the prior information; we shall call this phase {\it undetectable}, as the ground-truth $\underline s^*$ has asymptotically no effect on the posterior distribution. We will call accuracy $a \ge 0$ of an estimation procedure a measure of the additional information it exploits in the posterior distribution with respect to the prior, in such a way that $a=0$ in the undetectable phase. The precise definition of the accuracy is a model-dependent problem, on which we shall come back in Sec.~\ref{sec_estimator}, with rather subtle pitfalls (see for example~\cite{Abbe_review_SBM} for a discussion of this point): for instance when the problem admits an exact symmetry the marginals of the posterior can be strictly equal to the prior distribution for arbitrary large SNR. This is for instance the case of the symmetric SBM, which is invariant under the permutation of the labels; in such a case the definition of the distance between $\underline s^*$ and $\widehat{\underline{s}}(G)$ must be adapted to break explicitly the symmetry, by considering for instance the overlap modulo an arbitrary global permutation of the labels of the vertices.

\subsection{Statistical mechanics description}

Several Bayesian inference problems have recently been studied via statistical mechanics methods, and a good part of the predictions thus obtained have been confirmed rigorously; our goal here is not to review these results (see for instance~\cite{ZdKr16_review,Cris_review_SBM,Abbe_review_SBM,BaPeWe18}) but to emphasize their common formal structure that is at the origin of the few possible universality classes of phase diagrams.

Consider, in this abstract perspective, an inference problem parametrized by a signal-to-noise ratio $c$. In the statistical mechanics approach the accuracy of the Bayes-optimal estimator is expressed as a variational problem: one derives (with the cavity or replica method) a so-called free-energy $f$, function of $c$ and of an order parameter $a$ (an object whose nature depends on the problem under study). This free-energy has to be minimized with respect to the order parameter, and the optimal accuracy is then a function of the location of this global minimum (the minimal free-energy yields instead the mutual information between the signal and the observations). In ``dense models'' (those in which there are much more than $N$ measurements, each giving a weak information on the signal), see e.g.~\cite{LeKrZd17}, the order parameter is a scalar (or a finite dimensional vector); in the sparse models studied later in this paper the order parameter is functional, yet the qualitative behavior is the same. To simplify this formal discussion let us stick to a scalar order parameter, and furthermore use the accuracy $a \ge 0$ as the order parameter itself, with the minimal value $a=0$ corresponding to the trivial, uninformative estimator (exploiting only the prior and not the observations). We shall write the free-energy as
\beq
f(a,c) = \Phi(a,c) - \frac{a^2}{2} \  , 
\label{eq:abstract_fe}
\eeq
the motivation for the subtraction of the second term will be clarified below. The location of the global minimum of $f$ with respect to $a$, denoted $a^*(c)$, is thus the prediction for the accuracy of the optimal estimator, irrespectively of its computational complexity.

The cavity method also enlightens the computational complexity of the inference problems: as a matter of fact the derivation of the free-energy is tightly linked to the analysis of an efficient iterative algorithm (called Approximate Message Passing~\cite{DoMaMa09,BaMo11} in the dense case, or Belief Propagation in the sparse regime~\cite{KschischangFrey01}), and provides a dynamical map that describes the discrete time evolution of the order parameter during the execution of the algorithm. In our abstract setting this yields an accuracy $a^{(n)}$ after $n$ steps of the algorithm, that evolves according to
\beq
a^{(n)} = \varphi(a^{(n-1)},c) \ , 
\label{eq:abstract_rec}
\eeq 
with the initial condition $a^{(0)}=\varepsilon$, where $\varepsilon>0$ is an infinitesimal positive constant: initially the algorithm only knows the prior on the signal, and at each time step uses the observation to iteratively update and improve the accuracy of its belief on the true value of the signal. We shall denote $a_{\rm alg}(c) = \lim_{n\to \infty} \, a^{(n)}$ the accuracy ultimately reached by this algorithm.

The fundamental connection between the information-theoretically optimal accuracy and the description of the iterative algorithm is expressed mathematically in our setting by the relation
\beq
\varphi(a,c) = \Phi'(a,c) \ ,
\eeq
where here and in the following the primes denote derivatives with respect to the variable $a$. This equation implies indeed a tight connection between the static description in terms of the free-energy $f(a,c)$ and the dynamical one in terms of the iteration function $\varphi(a,c)$: the stationary points of the former correspond to the fixed points of the latter. As mentioned previously we focus in this paper on problems which admit an undetectable phase; this translates in the abstract formalism employed here into the assumption that $a=0$ is, for all values of the SNR $c$, a fixed point of the iteration (\ref{eq:abstract_rec}), that we shall call the {\it trivial} or {\it non-informative} fixed point. Equivalently $a=0$ is a stationary point of the free-energy $f(a,c)$.

\subsection{Possible phase diagrams} 
\label{sec_typology_pd}

It should be clear at this point that the description of the phase transitions and the classification of the possible phase diagrams is nothing else than a bifurcation analysis. As the SNR $c$ is varied the functions $f(a,c)$ (resp. $\varphi(a,c)$) evolve in a smooth way; their stationary (resp. fixed) points also do, except at bifurcations where their number and nature can change in a singular way. The qualitative aspects of such a bifurcation diagram do not depend on the details of the function $f(a,c)$, but only on its behavior close to the bifurcations (essentially the order and sign of the first non-vanishing derivatives), which explains the high level of universality among inference problems which can have very different origins. 

Let us now describe in this perspective the phase transitions in inference problems, interpreted as bifurcations for the stationary points of $f(a,c)$; among the possibly many stationary points two (which can coincide) will play a particular role: $a^*(c)$, the global minimum of $f$, that gives the information-theoretically best possible accuracy, and $a_{\rm alg}(c)$, the fixed point reached by iterations starting infinitesimally close to the trivial one, as it corresponds to the accuracy reachable by an efficient algorithm.

As $a=0$ is assumed here to be a fixed point for all values of $c$, and as the algorithm starts infinitesimally close to it, the local stability of $a=0$ under the iterations (\ref{eq:abstract_rec}) yields a crucial information on the accuracy $a_{\rm alg}(c)$ reached by the algorithm: if $a=0$ is stable the iterations will drive the algorithm to the trivial accuracy, then $a_{\rm alg}(c)=0$, if not the dynamical system flows away from this fixed point and one reaches $a_{\rm alg}(c)>0$. This stability can be determined very easily by expanding (\ref{eq:abstract_rec}) at first order in $a$,
\beq
a^{(n)} \approx \varphi'(0,c) \, a^{(n-1)} \  ,  
\label{eq:dense_1st}
\eeq 
which shows that $a=0$ is stable if and only if $\varphi'(0,c) < 1$ (as $a^{(n)} \ge 0$ we can assume $\varphi'(0,c) \ge 0$). As $c$ represents here a signal-to-noise ratio increasing values of $c$ tends to destabilize the trivial fixed point. We shall thus define the so-called \emph{KS threshold} (the abbreviation stands for Kesten-Stigum \cite{KestenStigum66} and is elucidated below) $c_{\rm KS}$ as the largest signal-to-noise ratio $c$ for which $a=0$ is a stable fixed point; as mentioned above for $c<c_{\rm KS}$ one has $a_{\rm alg}(c)=0$, the algorithm will not be able to infer the underlying signal better than a random guess from the prior. On the other hand for $c>c_{\rm KS}$ the iterative algorithm reaches a non-trivial accuracy $a_{\rm alg}(c)>0$ and provides a strictly positive correlation between the ground truth signal and its estimator (but not necessarily as good as the optimal estimator, see the discussion below).

\begin{figure}
\centerline{
\includegraphics[width=7cm]{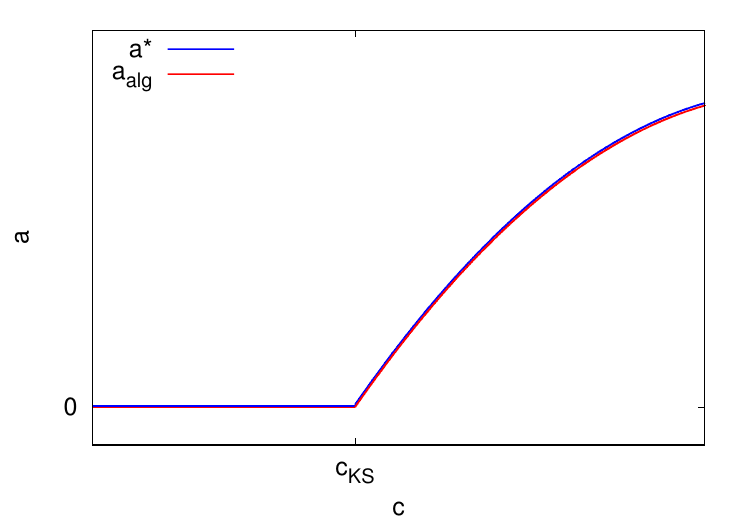}
\hspace{1cm}
\includegraphics[width=7cm]{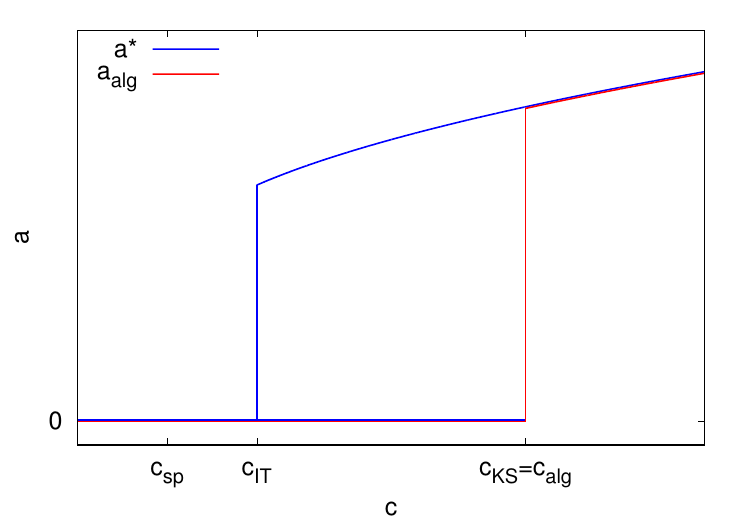}
}
\centerline{
\includegraphics[width=7cm]{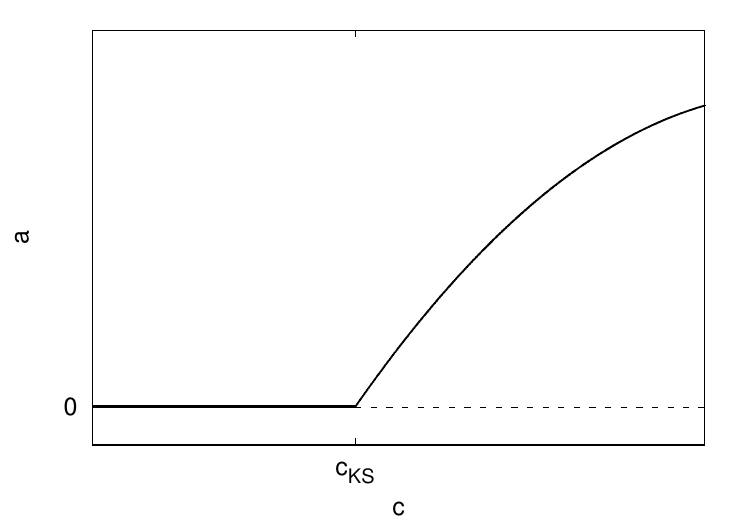}
\hspace{1cm}
\includegraphics[width=7cm]{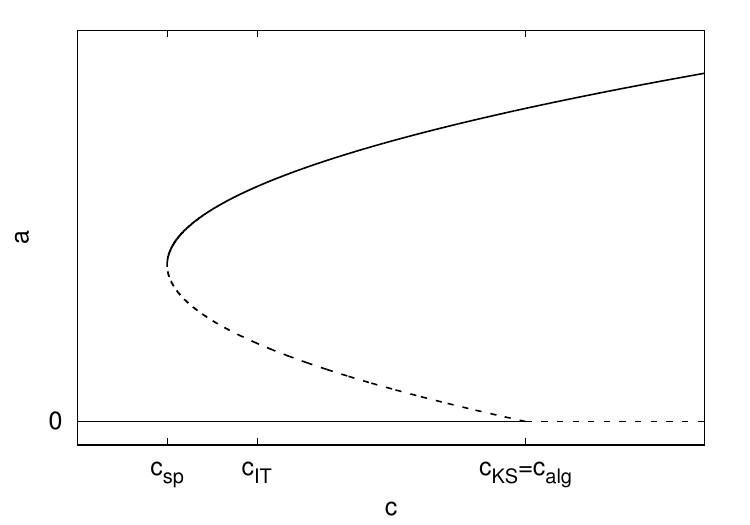}
}
\caption{This figure presents the two ``standard'' scenarios with either a second order (continuous) transition (two panels on the left) or a first order (discontinuous) transition (two panels on the right). For each of them the upper panel presents the Bayes-optimal ($a^*$, in blue) and algorithmic ($a_{\rm alg}$, in red) performance, while the bottom panel displays the fixed points of Eq.~(\ref{eq:abstract_rec}), with full (resp. dashed) lines for stable (resp. unstable) fixed points. In all panels the vertical axis is the accuracy $a$, the horizontal axis is the signal-to-noise ratio $c$. The thresholds $c_{\rm sp}$, $c_{\rm IT}$, $c_{\rm KS}$ and $c_{\rm alg}$, as defined in the text, are marked on the horizontal axes. 
} 
\label{fig:standard}
\end{figure}

A moment of thought reveals that this change of stability of the trivial fixed point at $c_{\rm KS}$ must be accompanied by a modification in the total number of fixed points, as the free-energy $f(a,c)$ is smooth. Keeping in mind the restriction $a \ge 0$, the two simplest global bifurcation diagrams that can arise are depicted in the lower panels of Fig.~\ref{fig:standard}, where all the fixed points are represented as a function of the SNR $c$, with solid (resp. dashed) lines when they are stable (resp. unstable).

The bifurcation diagram presented in the left part of Fig.~\ref{fig:standard} corresponds to a so-called continuous (2nd order) phase transition. For all values of $c$ there is a single stable fixed point, with a trivial accuracy when $c<c_{\rm KS}$ and a non-trivial one for $c>c_{\rm KS}$. The optimal accuracy $a^*(c)$ and the algorithmic one $a_{\rm alg}(c)$, depicted with red and blue lines in the upper panels of Fig.~\ref{fig:standard}, coincide for all values of the SNR $c$. The transition at $c_{\rm KS}$ thus separates an \emph{undetectable} phase (for $c<c_{\rm KS}$ no estimator can detect the signal as $a^*(c)=0$) from an \emph{easy} phase (for $c>c_{\rm KS}$ the iterative efficient algorithm matches the optimal performance as $a_{\rm alg}(c)=a^*(c)>0$).

The next possible bifurcation diagram, in the order of increasing complexity, is presented on the right of Fig.~\ref{fig:standard} and presents a discontinuous (1st order) phase transition. What happens by increasing $c$ around the KS transition is now the disappearance of an unstable non-trivial fixed point (instead of the appearance of a stable non-trivial fixed point for a continuous phase transition). There thus exists an interval of SNR, $c \in [c_{\rm sp},c_{\rm KS}]$, where two stable fixed points coexist (the trivial one $a=0$ and a non-trivial one). A bifurcation occurs at the \emph{spinodal} transition $c_{\rm sp}$, where the high-accuracy branch appears discontinuously. The consequences of this bifurcation diagram on the optimal and algorithmic accuracies $a^*(c)$ and $a_{\rm alg}(c)$ are presented in the upper right panel of Fig.~\ref{fig:standard}; the latter is only affected by the change of stability of the trivial fixed point at $c_{\rm KS}$, above which it jumps to the only stable fixed point which must thus coincide with the optimal accuracy $a^*(c)$. In the interval $c \in [c_{\rm sp},c_{\rm KS}]$ the two stable fixed points correspond to two local minima of the free-energy $f(a,c)$; their values cross each other at the \emph{information-theoretic} phase transition $c_{\rm IT} \in [c_{\rm sp},c_{\rm KS}]$, the trivial (resp. non-trivial) fixed point being the global minimum for $c \in [c_{\rm sp},c_{\rm IT}]$ (resp. for $c \in [c_{\rm IT},c_{\rm KS}]$). One concludes in this case that the undetectable phase (for $c<c_{\rm IT}$) and the easy phase (for $c>c_{\rm KS}$) are separated by an hard phase (for $c \in [c_{\rm IT},c_{\rm KS}]$) in which a non-trivial accuracy is information theoretically possible ($a^*(c)>0$) yet the iterative algorithm does not provide any correlation with the signal ($a_{\rm alg}(c)=0$).

\begin{figure}
\centerline{
\includegraphics[width=7cm]{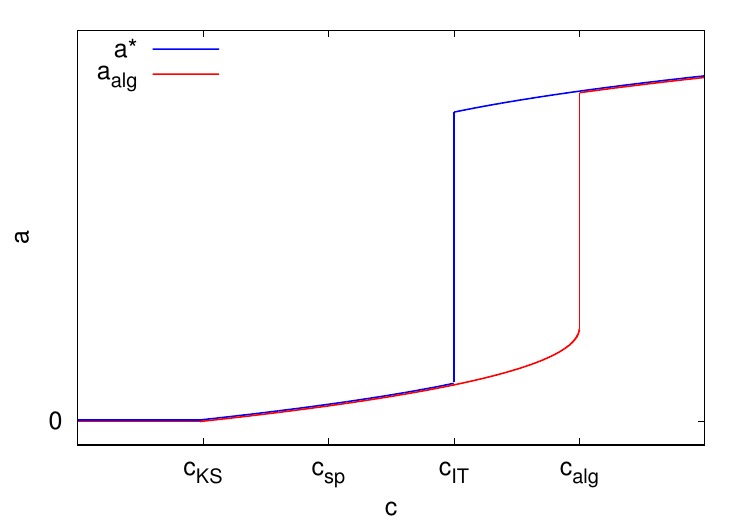}
\hspace{1cm}
\includegraphics[width=7cm]{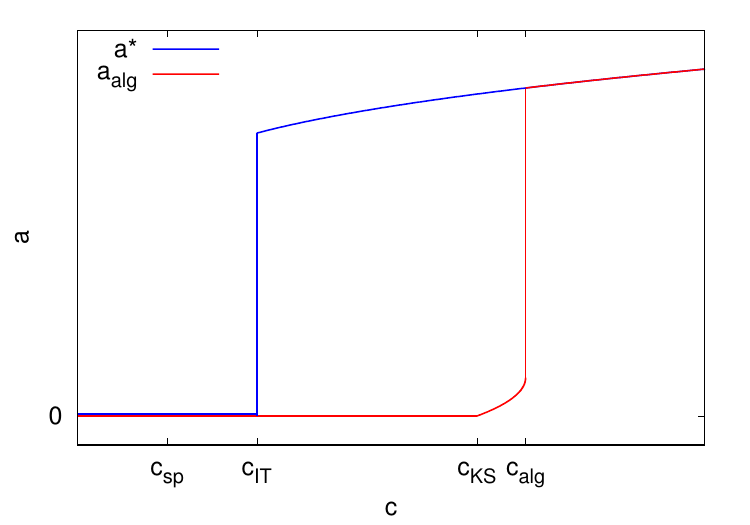}
}
\centerline{
\includegraphics[width=7cm]{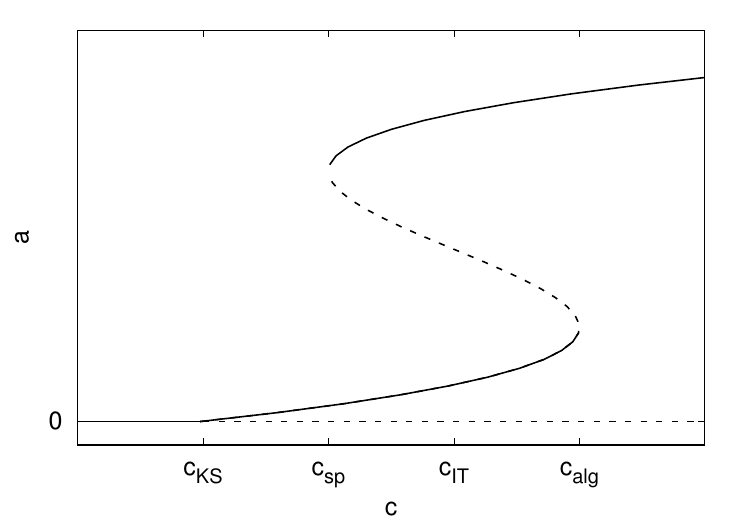}
\hspace{1cm}
\includegraphics[width=7cm]{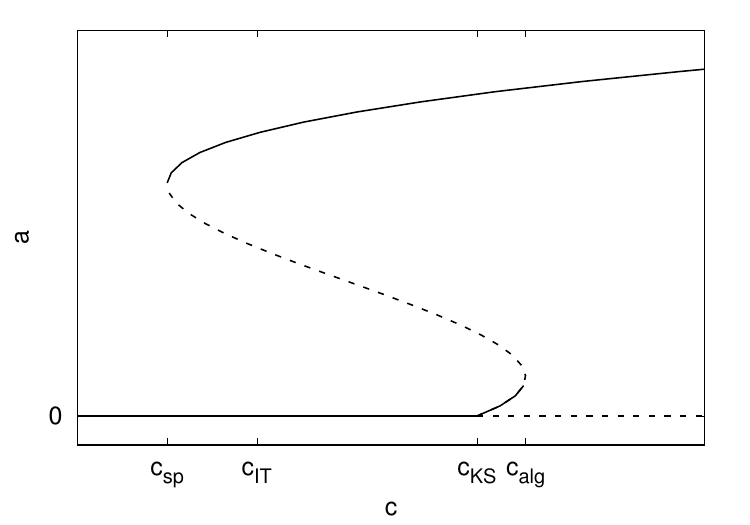}
}
\caption{This figure presents two ``new'' scenarios of phase transitions for inference problems, one on the two left panels, one on the two right panels. As in Fig.~\ref{fig:standard} the upper panels displays the Bayes-optimal ($a^*$, in blue) and algorithmic ($a_{\rm alg}$, in red) performance, while the lower panels present the fixed points of Eq.~(\ref{eq:abstract_rec}), with full (resp. dashed) lines for stable (resp. unstable) fixed points.
In all panels the accuracy $a$ is plotted as a function of the signal-to-noise ratio $c$. The thresholds $c_{\rm sp}$, $c_{\rm IT}$, $c_{\rm KS}$ and $c_{\rm alg}$, as defined in the text, are marked on the x-axes. The KS threshold $c_{\rm KS}$ here is strictly
  smaller than the algorithmic one $c_{\rm alg}$. We note that the thresholds can also be ordered as $c_{\rm sp}< c_{\rm KS}  < c_{\rm IT} < c_{\rm alg}$, in terms of $a^*$ and $a_{\rm alg}$ this still corresponds to the behavior shown in the top left panel.
}
\label{fig:new}
\end{figure}

One can easily imagine more and more complicated bifurcation diagrams that will be exhibited by free-energies with more and more stationary points; we shall content ourselves with the next possibility, depicted in Fig.~\ref{fig:new}. As shown in its lower part the branch of the non-trivial stable fixed point that appears continuously for $c>c_{\rm KS}$ undergoes a bifurcation at an \emph{algorithmic spinodal} $c_{\rm alg} >c_{\rm KS}$, while the high accuracy branch disappears discontinuously at the spinodal $c_{\rm sp} < c_{\rm alg}$ (the spinodal $c_{\rm sp}$ can occur after or before $c_{\rm KS}$, as shown in the left and right part of Fig.~\ref{fig:new}). The algorithmic accuracy $a_{\rm alg}(c)$ is in this case vanishing for $c<c_{\rm KS}$, growing continuously on the interval $[c_{\rm KS},c_{\rm alg}]$, and jumping discontinuously at $c_{\rm alg}$. As we assumed here the existence of a single stable fixed point for $c>c_{\rm alg}$ the algorithmic accuracy coincides then with $a^*(c)$ (i.e. the jump reaches the optimal value). To determine the information-theoretically optimal accuracy in the presence of two stable fixed points of the recursion one has to compare the values of the corresponding local minima of the free-energy $f(a,c)$. Let us call $c_{\rm IT}$ the information-theoretic transition at which these two free-energies cross each other, inducing a discontinuity in the accuracy $a^*(c)$ of the global minimum of the free-energy. The left and right part of Fig.~\ref{fig:new} distinguish further two different scenarios. On the left one has $c_{\rm IT} > c_{\rm KS}$, and the four regimes separated by the transitions $c_{\rm KS} < c_{\rm IT} < c_{\rm alg}$ shall be called, for increasing values of $c$: undetectable, easy, hybrid-hard, easy; on the right $c_{\rm IT} < c_{\rm KS}$ and the four phases separated by $c_{\rm IT} < c_{\rm KS} < c_{\rm alg}$  are: undetectable, hard, hybrid-hard, easy. In this terminology we define an undetectable phase by the condition $a^*(c)=a_{\rm alg}(c)=0$, an easy phase by $a^*(c)=a_{\rm alg}(c)>0$, a hard phase by  $a^*(c)>a_{\rm alg}(c)=0$, and an hybrid-hard phase by $a^*(c)>a_{\rm alg}(c)>0$. The hybrid character of this phase arises from the simultaneous easiness to beat the trivial accuracy of the uninformative estimator ($a_{\rm alg}(c)>0$), and hardness of achieving the optimal accuracy ($a^*(c)>a_{\rm alg}(c)$).

Let us make a few remarks on the classification of possible phase diagrams for inference problems we just presented:
\begin{itemize}
\item
The scenarios described in this section and more generally in this paper are the simplest ones, and they do not by far cover all possibilities: one can always devise more complicated free-energies $f(a,c)$ with more distinct stationary points and/or more bifurcations, yielding for instance several hard phases separated by easy regimes; we shall not discuss these more complicated situations as they did not arise in any of the cases we analyzed (see~\cite{MeassonMontanari05} for such examples in the related context of error-correcting codes).

\item 
We defined the computational easiness or hardness of inference with respect to one specific efficient algorithm (Approximate Message Passing or Belief Propagation), as to this day there are no better efficient (i.e. running in polynomial time) algorithms known for the problems that are NP-hard from the worst case computational complexity point of view. It remains of course a very challenging open question to prove or disprove (under some computational complexity hypothesis) the existence of more accurate efficient algorithms than these message passing ones.

\item
The two simplest scenarios presented in Fig.~\ref{fig:standard} are well known in the context of inference on sparse random graphs; for instance the study of the symmetric stochastic block model in~\cite{Decelle_SBM_long} demonstrated a continuous (left panel) phase transition for $q \le 3$ communities, and a discontinuous one (right panel) for $q \ge 5$. The more complicated phase diagrams of Fig.~\ref{fig:new} were until now (as far as we know) only discussed in the context of dense inference problems (in particular for constrained low rank matrix estimation, see figure 3 and 6 in~\cite{LeKrZd17}). However, we shall see in the rest of the paper that they do also occur naturally in sparse problems, a fact that remained previously unobserved. 

\item
It is known that for inference problems for which $a=0$ is a fixed point of (\ref{eq:abstract_rec}) there is a close relation between the inference (planted) model and the model with random uncorrelated disorder (see for instance~\cite{CoKrPeZd16}). In random optimization and constraint satisfaction problems some of the equilibrium phase transitions described in the literature correspond to phase transitions defined above. Notably, the {\it dynamical} phase transition \cite{KrzakalaMontanari06}, sometimes referred to as the {\it mode-coupling theory} transition~\cite{parisi2010mean} in the literature on the mean-field theory of structural glasses, and referred to as the {\it reconstruction threshold}~\cite{MezardMontanari06} in the theory of reconstruction on tree, corresponds to $c_{\rm d} =  {\rm min} (c_{\rm KS}, c_{\rm sp})$. This is indeed the smallest SNR for which a non-trivial fixed point does exist. The {\it condensation} phase transition from random constraint satisfaction problems \cite{KrzakalaMontanari06}, sometimes referred to as the {\it Kauzmann} ideal glass transition in the mean-field theory of structural glasses~\cite{parisi2010mean}, correspond to $c_{\rm cond}= {\rm min} (c_{\rm KS}, c_{\rm IT})$. The KS phase transition would be referred to as the de Almeida-Thouless instability in the theory of spin glasses \cite{AlmeidaThouless78} and corresponds to the point starting from which iterative message passing algorithms such as belief propagation cease to converge on random instances of the corresponding problems. This correspondence between planted and random instances does not hold for all phase transitions: for instance the satisfiability~\cite{MezardParisi02} of random constraint satisfaction problems has no counterpart in the planted/inference setting, neither does the algorithmic spinodal $c_{\rm alg} > c_{\rm KS}$ of inference problems in the random ensembles. This correspondence also breaks down for models where $a=0$ is not a fixed point of (\ref{eq:abstract_rec}), most notably satisfiability of random Boolean formulas.

\end{itemize}

\subsection{Expansions around the trivial fixed point}
\label{sec_bifurcation_scalar}

The classification of the phase diagrams presented above relies on a global bifurcation analysis, which requires the identification of all fixed points of Eq.~(\ref{eq:abstract_rec}) and the study of their domain of existence as a function of the SNR parameter~$c$. This is relatively easy to do when the order parameter is a scalar, or a finite-dimensional object, and when the recursion function $\varphi$ can be studied explicitly. This global analysis becomes much more difficult for sparse inference models, because the order parameter is infinite-dimensional in this case. The main methodological contribution of the present paper is a generalization to the sparse case of a local bifurcation analysis of the trivial fixed point in the neighborhood of the Kesten-Stigum transition. For pedagogical reasons let us first explain here how this analysis is performed in the case of a scalar order parameter (as done for instance in section IV.C.3 of~\cite{LeKrZd17}), which will help to understand the strategy followed in the functional case.

The location of the Kesten-Stigum transition $c_{\rm KS}$, defined by the condition $\varphi'(0,c_{\rm KS})=1$, was obtained through the linearization of (\ref{eq:abstract_rec}). In order to study the non-trivial fixed point in the neighborhood of this transition we shall expand the fixed-point equation $a=\varphi(a,c)$ at the next order in $a$, to obtain
\beq
a = \varphi'(0,c) \, a + \frac{1}{2} \varphi''(0,c) \, a^2 + O(a^3) \ . 
\eeq
In the neighborhood of $c_{\rm KS}$ the derivative $\varphi'(0,c)$ is close to 1, we shall hence trade the SNR $c$ for a parameter $\epsilon$ defined by $\varphi'(0,c)=1+\epsilon$, such that $\epsilon=0$ corresponds to the KS transition, $\epsilon >0$ (resp. $\epsilon <0$) to the high SNR (resp. low SNR) regime. If the second order derivative of $\varphi$ does not vanish exactly at $c_{\rm KS}$ we can rewrite this equation at lowest order as
\beq
a = a + \epsilon \, a + v \, a^2 + O(a^3) \ ,
\label{eq:dense_1st_expanded}
\eeq
where we defined $v=\varphi''(0,c_{\rm KS})/2 \neq 0$. The non-trivial solution of this equation is obviously $a = - \epsilon /v$; at this point it is crucial to remember the positivity condition $a \ge 0$ (without it the bifurcation diagrams would be qualitatively different). Indeed this requirement implies that if $v<0$ the non-trivial perturbative fixed point exists for $\epsilon >0$, in the high SNR regime; this case corresponds both to the continuous transition (left panel of Fig.~\ref{fig:standard}), and to the more complicated phase diagrams of Fig.~\ref{fig:new}. On the contrary if $v>0$ it is in the low SNR regime ($\epsilon <0$) that it exists, as in the first order transition case depicted in the right panel of Fig.~\ref{fig:standard}.

Let us emphasize the limitations of such a local study: by definition it can only provide perturbative information on the phase diagram, in the neighborhood of the KS transition and for the branch of fixed point that coalesces with the trivial one; if this is enough to distinguish the first order transition scenario (Fig.~\ref{fig:standard} right) and exclude the three other ones when $v>0$, in the opposite case one cannot decide between the second order transition (Fig.~\ref{fig:standard} left) and the scenarios of Fig.~\ref{fig:new} solely from the condition $v<0$, as non-perturbative (finite $a$) features further differentiate these cases.

We shall develop in the sparse case an expansion that goes one order further, corresponding to an equation of the form
\beq
a = a + \epsilon \, a + v \, a^2 + w \, a^3 + O(a^4) \ .
\label{eq:dense_expansion2}
\eeq
The motivations for continuing the expansion to this order are twofold. On the one hand, there are important inference problems for which $v=0$ exactly (most notably the symmetric stochastic block model for $q=4$ groups, both in the sparse~\cite{Sly11} and in the dense regime~\cite{LeKrZd17}, we shall come back in details on this point in Sec.~\ref{sec_applications_symPotts}). In that case the non-trivial solution is $a=\sqrt{-\epsilon/w}$ and it is the sign of $w$ (if $w\neq 0$) that allows to determine whether the perturbative non-trivial fixed point exist in the low or high SNR phase (the argument of the square root must be positive). On the other hand, some problems have, in addition to the SNR, another continuously varying parameter; this is in particular the case of the asymmetric 2 group stochastic block model (or asymmetric Ising model in the tree reconstruction language), where the asymmetry between the sizes of the two groups can be tuned independently of the SNR. Depending on this asymmetry parameter, that we shall denote $\om$ in the following, the type of transition changes from second to first order, which comes from a change of sign of $v$ when $\om$ crosses a critical value $\omc$. Having pushed the expansion to the third order will thus allow us to study the neighborhood of the ``tri-critical'' point $(\epsilon,\om)=(0,\omc)$ in parameter space, where the nature of the Kesten-Stigum transition changes. Taking simultaneously the limits $\epsilon \to 0$ and $\om \to \omc$ with well-chosen scalings the last three terms of (\ref{eq:dense_expansion2}) will indeed be of the same order (this will be further discussed in Sec.~\ref{sec_Ising_asym}).

\section{Cavity equations and main results}
\label{sec_generic}

In the previous section we have discussed the classification of phase transitions and phase diagrams of Bayesian inference problems, relying on the bifurcation analysis for a scalar order parameter; dense inference problems can indeed be reduced to such a scalar (or more generically finite-dimensional) representation in the large-size limit. In the rest of the paper we shall turn instead to inference problems defined on sparse graphs, for which the corresponding order parameter becomes functional. In this section we shall introduce two exemplary cases of these sparse inference problems, that cover the main applications of this paper, state in a rather generic way the cavity formalism that can be used to handle them, briefly explain its interpretation and finally present our main results, as obtained in the later sections via an expansion of the cavity equations around their trivial solution.

\subsection{Definitions of two exemplary inference problems on sparse (hyper)graphs}
\label{sec_definitions_graphs}

\subsubsection{Stochastic Block Model}
\label{sec_definitions_SBM}

The Stochastic Block Model (SBM) is a random graph ensemble which generates
networks with a community structure~\cite{fienberg1981categorical,holland1983stochastic,BoJaRi07}. It is specified by the following parameters: $N$, the number of vertices (or nodes) of the graph $G=(V,E)$, $q$ the number of communities, $\oeta=(\oeta_1,\dots,\oeta_q)$ a prior probability distribution on the communities, and $c=\{c_{\s \t}\}$ a $q \times q$ symmetric matrix (not to be confused with the SNR we also denoted $c$ in Sec.~\ref{sec_typology}). A graph of this ensemble is generated by drawing for each vertex $i\in V$ a label $\s_i \in \{1,\dots,q\}$, independently at random with probability $\oeta_{\s_i}$; this label represents the community of the vertex.  Once the labels $\us=(\s_1,\dots,\s_N)$ are chosen, each of the $N(N-1)/2$ possible edges $\la i,j \ra$ between pairs of distinct vertices is included in the set of edges $E$ of the graph with probability $c_{\s_i,\s_j}/N$. The inference problem we shall consider is the determination of the labels $\us$ (up to a potential symmetry between communities) from the mere observation of the graph $G$ and from the knowledge of the parameters $\oeta$ and $c$.

We are in particular interested in the large-size limit $N \to \infty$, taken for a fixed value of the affinity matrix $c$. In this limit the degree of a vertex
labelled $\s$ becomes a Poisson random variable with average $d_\s = \sum_{\s'} c_{\s \s'}
\oeta_{\s'}$; the finiteness of these degrees justifies the name ``sparse'' given to this type of inference problems. The phase transitions in the Bayes-optimal inference in the sparse stochastic block model were studied in detail in \cite{DecelleKrzakala11,Decelle_SBM_long}. Large part of these results were confirmed rigorously in subsequent works, see e.g. \cite{mossel2013proof,massoulie2014community,mossel2016,AbSa15,Abbe_review_SBM}.  

We shall assume in the following that  $d_\s = d$ independently of $\s$, in such a way that the degree of a vertex is uninformative of its label. This is the condition for the existence of an undetectable phase at small $d$, in which case the SBM is asymptotically contiguous to a purely random Erd\"os-R\'enyi graph of the same average degree and the optimal estimation of the labels can only rely on the prior distribution $\oeta$.

\subsubsection{Planted occupation models}
\label{sec_definitions_planted}

We consider now the family of Constraint Satisfaction Problems (CSP) in which $N$ Boolean variables, denoted $\us=(\s_1,\dots,\s_N) \in \{0,1\}^N$, are required to satisfy simultaneously $M$ constraints of the following form: the $a$-th constraint bears on a subset of $k$ variables, denoted $\da \subset \{1,\dots,N\}$, and is satisfied by $\us$ if and only if $\sum_{i \in \da} \s_i \in S$, where $S \subset \{0,1,2,\dots,k\}$ defines the type of problem. The interpretation of $\s_i=1$ (resp. 0) as a site occupied by a particle (resp. empty) justifies the name of occupation model, each constraint restricting the number of particles adjacent to it (in the bipartite graph on $N+M$ vertices with an edge put between $i$ and $a$ iff $i\in \da$) to belong to $S$. This family of CSPs contains as special cases the bicoloring of hypergraphs (when $S=\{1,\dots,k-1\}$) and XORSAT problems (also known as parity checks, when $S=\{0,2,4,\dots,\}$ or $S=\{1,3,5,\dots,\}$). Random ensembles of occupation problems, in which the neighborhoods $\da$ are drawn uniformly at random among the $\binom{N}{k}$ possible ones, have been studied e.g. in~\cite{ZdeborovaMezard08b}. 

An inference problem can be associated to these occupation models if instead of this random ensemble of CSPs one considers its planted version, in which one first chooses a configuration $\us$, drawing independently each $\s_i$ with a prior probability distribution $\oeta_{\s_i}$ (with $\oeta_0 + \oeta_1 =1$), then draws the $M$ neighborhoods $\da$ uniformly among those that are satisfied by $\us$ (i.e. such that $\sum_{i \in \da} \s_i \in S$). The inference problem is then to reconstruct the planted configuration $\us$ solely from the observation of the bipartite graph linking variables to constraints and the knowledge of the subset $S$. Special cases of planted occupation problems and their corresponding phase transition have been studied previously in~\cite{ZdeborovaKrzakala09,angelini2015spectral}.

In this paper we shall focus on ``symmetric'' planted occupation models, in the sense that $\oeta_0  = \oeta_1 = 1/2$ and the set $S$ of the number of allowed particles is invariant under the exchange of occupied and empty sites (i.e. we assume that $n \in S$ if and only if $k-n \in S$); this condition ensures the existence of an undetectable phase if the density of constraints $\alpha=M/N$ is small enough.

\subsection{Cavity equations and free-entropy functional}
\label{sec_cavity_formalism}

We shall now present a formalism, known as the cavity method~\cite{MezardParisi01,MezardMontanari07}, that allows to deal with inference problems on sparse random (hyper-)graphs. We will first state the approach in a formal way, with a level of generality that encompasses the two examples above, before discussing its interpretation and justification in the next subsection. 

The cavity equations depend on the following list of parameters:
\begin{itemize}
\item $\chi$, a discrete alphabet of spins of cardinality $q$, taken for concreteness to be $\chi=\{1,\dots,q\}$;
\item a probability distribution $\oeta$ on $\chi$, i.e. $(\oeta_1,\dots,\oeta_q)$ with $\oeta_\s \ge 0$ and $\sum_\s \oeta_\s =1$;
\item a probability law $p_\ell$ on non-negative integers representing
  the degree distribution of variables, and its size-biased version
\beq
\tp_\ell = \frac{(\ell+1)p_{\ell+1}}{\E[\ell]} \ ,
\label{eq_tpell}
\eeq
where $\E[\ell] = \sum_\ell p_\ell \ell$ is the average degree;

\item an integer $k \ge 2$;

\item a joint probability law $p_{\rm j}(\s_1,\dots,\s_k)$ on $\chi^k$, the subscript j standing for ``joint'', that we assume to be invariant under all the permutations of its arguments, and to have $\oeta$ as marginal laws for a single argument:
\beq
p_{\rm j}(\s_1,\dots,\s_k) = p_{\rm j}(\s_{\pi(1)},\dots,\s_{\pi(k)}) \ \ \forall \pi \in {\cal S}_k \ , \qquad
\sum_{\s_2,\dots,\s_k} p_{\rm j}(\s_1,\dots,\s_k) = \oeta_{\s_1} \ .
\label{eq_hypotheses_pj}
\eeq

\end{itemize}

We shall also need the conditional version of this probability law, to be denoted $p_{\rm c}(\s_1,\dots,\s_{k-1}|\s) = p_{\rm j}(\s_1,\dots,\s_{k-1},\s)\frac{1}{\oeta_\s}$, where the subscript c stands for ``conditional''. As a consequence of the hypothesis made on $p_{\rm j}$ this conditional law is invariant under the permutations of its $k-1$ first arguments, and fulfils the reversibility property:
\beq
p_{\rm c}(\s_1,\dots,\s_{k-1}|\s) \oeta_\s = p_{\rm c}(\s,\s_2,\dots,\s_{k-1}|\s_1) \oeta_{\s_1} \ .
\label{eq_reversibility}
\eeq

We denote by $\eta$ and $\nu$ probability distributions on $\chi$
(i.e. $q$-dimensional vectors of positive reals that sum to one).
We shall call conditional version of the cavity equations associated to the parameters $(q,\oeta,p_\ell,k,p_{\rm j})$ the following recursive relations on the $2 q$ distributions $\{P_\t^{(n)}(\eta) , \hP_\t^{(n)}(\nu)\}_{\t \in \chi}$:
\bea
P_\t^{(n+1)}(\eta) &=& \sum_{\ell=0}^\infty \tp_\ell 
\int \dd \hP_\t^{(n)} (\nu^1) \dots \dd \hP_\t^{(n)} (\nu^\ell) \, 
\delta(\eta - f(\nu^1,\dots,\nu^\ell)) \ , 
\label{eq_generic_conditional_hPtoP}
\\
\hP_\t^{(n)}(\nu) &=& 
\sum_{\t_1,\dots,\t_{k-1}} p_{\rm c}(\t_1,\dots,\t_{k-1}|\t)
\int \dd P_{\t_1}^{(n)}(\eta^1) \dots \dd P_{\t_{k-1}}^{(n)}(\eta^{k-1}) \, 
\delta(\nu-\hf(\eta^1,\dots,\eta^{k-1})) \ , 
\label{eq_generic_conditional_PtohP}
\eea
where here and in the following unspecified summation over spin indices $\s$ or $\t$ are understood to run over $\chi$, and $n$ should be thought of as a time index along the recursion (the initial condition for $n=0$ will be discussed below). The functions $f$ and $\hf$ appearing above are called Belief Propagations (BP) recursions and are defined as follows: $\eta = f(\nu^1,\dots,\nu^\ell)$ means
\beq
\eta_\s = \frac{1}{z(\nu^1,\dots,\nu^\ell)}\, \oeta_\s \prod_{i=1}^\ell \frac{\nu^i_\s}{\onu_\s} \ ,
\label{eq_generic_nutoeta}
\eeq
where $z$ enforces the normalization of $\eta$, and where we denoted $\onu$ the uniform distribution on $\chi$ (i.e. $\onu_\s = \frac{1}{q}$). The other recursion function $\nu=\hf(\eta^1,\dots,\eta^{k-1})$ is spelled out as
\beq
\nu_\s = \frac{1}{\hz(\eta^1,\dots,\eta^{k-1})} \, \frac{\onu_\s}{\oeta_\s} \sum_{\s_1,\dots,\s_{k-1}} p_{\rm j}(\s,\s_1,\dots,\s_{k-1}) \prod_{i=1}^{k-1} \frac{\eta^i_{\s_i}}{\oeta_{\s_i}} \ ,
\label{eq_generic_etatonu}
\eeq
with again $\hz$ a normalizing factor. One can check easily that the hypotheses (\ref{eq_hypotheses_pj}) made on the joint probability law $p_{\rm j}$ implies that $f(\onu,\dots,\onu) = \oeta$ and $\hf(\oeta,\dots,\oeta)=\onu$, hence that $P_\t^{(n)}(\eta) = \delta(\eta - \oeta)$, $\hP_\t^{(n)}(\nu) = \delta(\nu - \onu)$ is a stationary solution of the cavity equations (\ref{eq_generic_conditional_hPtoP},\ref{eq_generic_conditional_PtohP}), that will be called {\it trivial fixed point} in the following.

Let us also define what we shall call the unconditional version of the cavity equations associated to the parameters $(q,\oeta,p_\ell,k,p_{\rm j})$, that bear on the $2$ sequences of distributions $P^{(n)}(\eta)$ and $\hP^{(n)}(\nu)$:
\bea
P^{(n+1)}(\eta) &=& \sum_{\ell=0}^\infty \tp_\ell 
\int \dd \hP^{(n)} (\nu^1) \dots \dd \hP^{(n)} (\nu^\ell) \, 
\delta(\eta - f(\nu^1,\dots,\nu^\ell)) \, z(\nu^1,\dots,\nu^\ell)\ , 
\label{eq_generic_unconditional_hPtoP} \\
\hP^{(n)}(\nu) &=& 
\int \dd P^{(n)}(\eta^1) \dots \dd P^{(n)}(\eta^{k-1}) \, 
\delta(\nu-\hf(\eta^1,\dots,\eta^{k-1})) \, \hz(\eta^1,\dots,\eta^{k-1}) \ ,
\label{eq_generic_unconditional_PtohP}
\eea
where the functions $f$, $z$, $\hf$ and $\hz$ have been defined in (\ref{eq_generic_nutoeta},\ref{eq_generic_etatonu}), and the distributions $P^{(n)}(\eta)$, $\hP^{(n)}(\nu)$ are required to have mean $\oeta$ and $\onu$ respectively:
\beq
\int \dd P^{(n)}(\eta) \, \eta = \oeta \ , \qquad 
\int \dd \hP^{(n)}(\nu) \, \nu = \onu \ . 
\label{eq_average_P_hP}
\eeq
With the choice of normalization made in (\ref{eq_generic_nutoeta},\ref{eq_generic_etatonu}) one has $z(\onu,\dots,\onu)=\hz(\oeta,\dots,\oeta)=1$, hence the normalization of the distributions $P^{(n)}$ and $\hP^{(n)}$, as well as the conditions (\ref{eq_average_P_hP}) on their averages, are preserved by the recursions of Eqs.~(\ref{eq_generic_unconditional_hPtoP},\ref{eq_generic_unconditional_PtohP}). Note that the conditional and unconditional versions of the cavity equations are actually equivalent: one can go from one form to the other according to the relations
\beq
P^{(n)}(\eta) = \sum_\t \oeta_\t P_\t^{(n)}(\eta) \ , \qquad
P_\t^{(n)}(\eta) = \frac{\eta_\t}{\oeta_\t} P^{(n)}(\eta) \ ,
\label{eq_Bayes}
\eeq
for all $n$, and similarly for $\hP^{(n)}$,
\beq
\hP^{(n)}(\nu) = \sum_\t \onu_\t \hP_\t^{(n)}(\nu) \ , \qquad
\hP_\t^{(n)}(\nu) = \frac{\nu_\t}{\onu_\t} \hP^{(n)}(\nu) \ .
\eeq
In particular the trivial fixed point, in the unconditional version of the cavity equations, corresponds to $P^{(n)}(\eta) = \delta(\eta - \oeta)$, $\hP^{(n)}(\nu) = \delta(\nu - \onu)$. 

Let us conclude this formal statement of the cavity formalism by the introduction of a functional, called free-entropy, that associates in its unconditional form a real value to a pair $(P,\hP)$ of distributions that satisfy (\ref{eq_average_P_hP}) according to
\bea
\phi(P,\hP) =
&-&\E[\ell] \int \dd P(\eta) \dd \hP(\nu) \, \ze(\eta,\nu) \ln \ze(\eta,\nu) \label{eq_phiint} \nonumber \\
&+& \frac{\E[\ell]}{k} \int \dd P(\eta^1) \dots \dd P(\eta^k) \, \zc(\eta^1,\dots,\eta^k) \ln \zc(\eta^1,\dots,\eta^k) \nonumber \\
&+& \sum_{\ell=1}^\infty p_\ell \int \dd \hP(\nu^1) \dots \dd \hP(\nu^\ell)
\, \zv(\nu^1,\dots,\nu^\ell) \ln \zv(\nu^1,\dots,\nu^\ell) \ , \label{eq:phi}
\eea
where 
\bea
\ze(\eta,\nu) &=& \sum_\s \eta_\s \frac{\nu_\s}{\onu_\s} \ , \\
\zc(\eta^1,\dots,\eta^k) &=& \sum_{\s_1,\dots,\s_k} p_{\rm j}(\s_1,\dots,\s_k) \prod_{i=1}^k \frac{\eta_{\s_i}^i}{\oeta_{\s_i}} \ , \\
\zv(\nu^1,\dots,\nu^\ell)&=& \sum_\s \oeta_\s \prod_{i=1}^\ell \frac{\nu_\s^i}{\onu_\s} \ .
\eea
An equivalent form in the conditional formalism reads
\bea
\phi(\{P_\s,\hP_\s\}_{\s \in \chi}) =
&-&\E[\ell] \sum_\s \oeta_\s \int \dd P_\s(\eta) \dd \hP_\s(\nu) \ln \ze(\eta,\nu) \nonumber \\
&+& \frac{\E[\ell]}{k} \sum_{\s_1,\dots,\s_k} \pj(\s_1,\dots,\s_k) \int \dd P_{\s_1}(\eta^1) \dots \dd P_{\s_k}(\eta^k) \ln \zc(\eta^1,\dots,\eta^k) \nonumber \\
&+& \sum_{\ell=1}^\infty p_\ell \sum_\s \oeta_\s \int \dd \hP_\s(\nu^1) \dots \dd \hP_\s(\nu^\ell) \ln \zv(\nu^1,\dots,\nu^\ell) \ . \label{eq_phiint_conditional}
\eea

Note that these expressions of $\phi$ are variational, in the sense that their derivatives with respect to their arguments vanish when the latter are fixed point solutions of the cavity recursions (i.e. (\ref{eq_generic_conditional_hPtoP},\ref{eq_generic_conditional_PtohP}) in the conditional version, (\ref{eq_generic_unconditional_hPtoP},\ref{eq_generic_unconditional_PtohP}) in the unconditional one). They are, however, well-defined even if their arguments are not solutions of the cavity equations (as long as the condition (\ref{eq_average_P_hP}) is satisfied); this property has been recently exploited in~\cite{CoKrPeZd16,CoEfJaKaKa17}, we shall come back on this point in Sec.~\ref{sec_expansion_phi}. Let us also mention that our choices of normalization leads to $\phi=0$ on the trivial fixed point of the cavity equations.

\subsection{Interpretation of the cavity equations}
\label{sec:cavity_vs_inference}

The cavity equations written above arise in the context of several slightly different and delicately intertwined problems, in particular in the study of graphical models on random structures, tree reconstruction problems, and inference on sparse graphs. We shall not attempt here an exhaustive discussion of these various interpretations and refer the reader to the literature \cite{KrzakalaMontanari06,MezardMontanari06,AbMo13,ZdKr16_review,CoKrPeZd16} for more details on these various perspectives and the connections between them, and instead give a brief description of two interpretations in order to clarify the meaning of our main results to be discussed below.

\subsubsection{Reconstruction on trees}
\label{sec_reconstruction}

The first interpretation of the cavity equations we shall discuss concerns the tree reconstruction problem~\cite{Mossel01,JansonMossel04,MezardMontanari06,KS_tight_Ising,Sly08,Sly11}. Consider a rooted bipartite tree, with two types of vertices, called variable nodes and interaction nodes, the latter being all of degree $k$, the root and the leaves being variable nodes. This tree is used as an information channel, in the following way: each variable node $i$ bears a spin $\s_i \in \chi $, and an information is sent from the root to the leaves, by first drawing the spin of the root vertex with a prior probability $\oeta$ on $\chi$, and then recursively assigning the values of the spins at distance $n=1,2,\dots$ from the root. This last step is done independently for each interaction node $a$: denoting $\s_1,\dots,\s_{k-1}$ the $k-1$ variables adjacent to $a$ that are at distance $n+1$ from the root, and $\s$ the variable at distance $n$ (that has been assigned in the previous step of the induction), one draws the former conditional to the latter, with a law denoted $p_{\rm c}(\s_1,\dots,\s_{k-1}|\s)$. The tree reconstruction question is now: given the observation of the spins at distance $n$, is it possible to infer the value of the spin at the root that started this broadcast process? Thanks to the tree structure of the channel it is possible to compute easily the posterior probability $\eta$ of the root given this observation; when the distance $n$ becomes very large two possibilities can arise, depending on the level of noise in the channel: either $\eta$ becomes very close to $\oeta$, then all information on the root value is lost, or $\eta$ keeps a trace of the correlation with the root, and an inference of its value with a better chance of success than the one allowed by the prior probability $\oeta$ is possible. 

It is not too difficult to convince oneself that the distribution $P_\t^{(n)}(\eta)$ that solves the cavity equations (\ref{eq_generic_conditional_hPtoP},\ref{eq_generic_conditional_PtohP}) is precisely the one described above in this tree reconstruction problem (see e.g.~\cite{MezardMontanari06} for a detailed derivation with similar notations), conditional to the root variable being $\t$ in the broadcast process, and when the tree used as a channel is a random Galton-Watson tree with offspring degree probability $\tp_\ell$. In this light the unconditional version of the cavity equations (\ref{eq_generic_unconditional_hPtoP},\ref{eq_generic_unconditional_PtohP}) is seen to describe the distribution of the marginal probability law of the root $\eta$, conditional to the observation of the leaves, when the law of the latter is not conditioned on the value of the root in the broadcast, and the relations (\ref{eq_Bayes}) are mere traductions of the Bayes theorem.

Following this interpretation the initial condition of the induction on $n$ should be
\beq
P_\t^{(n=0)}(\eta) =  \delta(\eta - \delta_{\t,\cdot}) \ ,
\eeq
with $\delta_{\t,\cdot}$ the probability law on $\chi$ supported solely on $\t$, which expresses the perfect knowledge of the variables observed at the leaves of the tree. As mentioned above the distribution $\delta(\eta-\oeta)$ is a fixed point of the cavity recursions (\ref{eq_generic_conditional_hPtoP},\ref{eq_generic_conditional_PtohP}): the tree reconstruction problem can thus be rephrased, in this setting, as determining whether the recursion bring, for $n\to\infty$, the distributions $P_\t^{(n)}$ towards this trivial fixed point, in which case all information on the root is lost, or towards a non-trivial fixed point that contains more information on the root than the prior probability $\oeta$.

It is actually useful to consider a more generic initial condition, namely
\beq
P_\t^{(n=0)}(\eta) = \ve \, \delta(\eta - \delta_{\t,\cdot}) + (1-\ve) \, \delta(\eta-\oeta) \ , 
\label{eq_initialcondition}
\eeq
with $\ve \in [0,1]$. The iterations of (\ref{eq_generic_conditional_hPtoP},\ref{eq_generic_conditional_PtohP}), starting from this initial condition, describe now the inference problem of the tree where the spin value of each leaf is observed with probability $\ve$, and kept hidden otherwise. The usual tree reconstruction problem is recovered for $\ve=1$; the variant known as the {\it robust reconstruction} problem~\cite{JansonMossel04} corresponds to the limit $\ve \to 0$, taken after the $n \to \infty$ limit, and again asks whether the iterations of the cavity equation bring back the distributions $P_\t^{(n)}$ to the trivial fixed point (corresponding to $\ve =0$) or lead it to a non-trivial one. Robust reconstruction is thus possible if an infinitesimal amount of information on a far away boundary is amplified in the reconstruction process and yields a non-vanishing information on the root. It was shown in~\cite{JansonMossel04} that the threshold for robust reconstruction corresponds to the Kesten-Stigum condition \cite{KestenStigum66}, in other words the robust reconstruction problem is solvable if and only if the trivial fixed point is locally unstable. This provides a bound on the original reconstruction problem, which is certainly solvable whenever the robust variant is; the converse is not true in general, the tightness or not of the Kesten-Stigum bound depending on the channel will be a point that our main results will enlighten.

\subsubsection{Link with inference problems on graphs}
\label{sec_link_graphs}

Let us now turn to the second interpretation of the cavity formalism, that corresponds to the inference problems on random graphs described in Sec.~\ref{sec_definitions_graphs}. For concreteness we concentrate on the Stochastic Block Model example, parametrized by the probability law $\oeta$ and the affinity matrix $c$, that verify the condition $d_\s = \sum_{\s'} c_{\s \s'} \oeta_{\s'}=d$ independently of the label $\sigma$. Consider a graph $G$ drawn from the SBM ensemble, along with its labels $\us$, and choose uniformly at random one of its vertices. One can then show (see for instance proposition 2 in~\cite{Mossel2015} for a formal statement) that its local neighborhood of fixed radius $n$ converges in distribution, in the large size limit, to a Galton-Watson tree with an offspring distribution $\tp_\ell$ that is a Poisson law of average $d$, decorated by labels $\s_i$ that have the law described above in the tree reconstruction perspective: the label of the root is drawn with probability $\oeta$, and broadcasted along the edges of the tree with the conditional law 
\beq
p_{\rm c}(\s'|\s) = \frac{1}{d} c_{\s \s'} \oeta_{\s'} \ ,
\eeq
that verifies the reversibility property (\ref{eq_reversibility}). This simple observation unveils a link between the inference problem on a graph and its tree counterpart; of course from a rigorous point of view the interplay between these two problems is subtler, in particular because there is no explicit observation of any label in the graph problem, contrary to the tree case. Another difference is the (weak) interaction between vertices that are not linked by an edge in the posterior probability law of the graph SBM problem, which is not explicitly present in the cavity formalism but can be traced to the condition (\ref{eq_average_P_hP}). The cavity method approach leads to the following conjectures:

\begin{itemize}
\item[(i)] When the robust reconstruction tree problem is solvable, then the labels of the graph inference problem can be inferred with a better accuracy than using only the prior information $\oeta$, via a polynomial time message-passing algorithm known as Belief-Propagation, whose accuracy is described in the large size limit by the fixed point of the cavity equations reached through the robust reconstruction initial condition (i.e. with $\ve \to 0$ after $n\to\infty$ in (\ref{eq_initialcondition})).

\item[(ii)] The accuracy of the optimal estimator of the labels (that may require an exponential time to be computed) is described by the fixed point of the cavity equations (\ref{eq_generic_conditional_hPtoP},\ref{eq_generic_conditional_PtohP}) that yields the largest possible value of the free-entropy (\ref{eq_phiint_conditional}), the latter being related to the mutual information between $\us$ and $G$. In particular, if (\ref{eq_generic_conditional_hPtoP},\ref{eq_generic_conditional_PtohP}) only admit the trivial fixed point, then the hidden labels cannot be inferred better than with their prior distributions.

\item[(iii)] No polynomial time algorithm is able to beat the estimation accuracy of the Belief Propagation algorithm, hence the existence of the ``hard'' phase, where according to (ii) it is information-theoretically possible to infer the labels with a non-trivial accuracy, but this is not possible with Belief Propagation. We stress here that in problems where the hybrid-hard phase (with BP giving better estimate than random, but still being largely suboptimal) exists, the non-optimality of BP occurs for SNR above the Kesten-Stigum threshold. 

\end{itemize}

Parts of points (i) and (ii) have been established rigorously, in particular for the stochastic block model as reviewed in \cite{Abbe_review_SBM}. The most detailed results have been obtained for the symmetric 2 groups SBM~\cite{mossel2013proof,massoulie2014community}. Point (ii) is established in~\cite{CoKrPeZd16,CoEfJaKaKa17,coja2018replica}, for a large set of models; roughly speaking these works shows that the existence of probability distributions $(P,\hP)$ with $\phi(P,\hP) >0$ implies the possibility of non-trivial inference of the labels in the graph problem (a similar variational principle for the tree reconstruction problem can be found in~\cite{MezardMontanari06}). The optimality of Belief Propagation for large signal-to-noise ratio (above an unspecified constant multiplying the KS threshold) was established in \cite{mossel2016}, but remains an open problem in the low signal-to-noise ratio regime. Point (i) is very much related to theorem 2.12 in~\cite{mossel2016}, where an auxiliary algorithm provides the initial condition of BP on a finite graph that emulates the observation of a few labels in the tree robust reconstruction problem.

\subsection{Main results}
\label{sec_mainresults}

The main technical results of this paper, to be presented in the subsequent sections, are expansions of the functional cavity equations (\ref{eq_generic_conditional_hPtoP},\ref{eq_generic_conditional_PtohP}) and free-entropy (\ref{eq_phiint_conditional}) around their trivial fixed point (that exists because we consider models having an undetectable phase). These computations allow us to generalize the local bifurcation analysis sketched in Sec.~\ref{sec_bifurcation_scalar} for a finite-dimensional order parameter, to the functional infinite-dimensional case. When the parameters $(\oeta,p_{\rm j},\tp_\ell)$ of the models are varied in the neighborhood of the Kesten-Stigum transition, at which the trivial fixed point changes stability, a non-trivial fixed point bifurcates from it continuously. The local bifurcation analysis allows us to determine whether it bifurcates as an unstable fixed point in the direction of lower SNR, in which case the Kesten-Stigum threshold and the algorithmic threshold coincide (right panel of Fig.~\ref{fig:standard}) and the hard phase must exist, or if the solution bifurcates as a stable fixed point in the direction of higher SNR (left panel of Fig.~\ref{fig:standard} or Fig.~\ref{fig:new}). To be able to further distinguish between the 3 scenarios in the latter case, we have also solved the fixed point cavity equations numerically using the population dynamics technique (a presentation of this numerical algorithm and of the subtleties in its implementation can be found in Appendix~\ref{app_numerical_resolution}). For the convenience of the reader we summarize here our main findings for the stochastic block model and the planted occupation models, and defer the detailed technical explanations to Sec.~\ref{sec_Potts} and \ref{sec_Ising} respectively.

We also show the numerical confirmation that BP run on large single instances of the inference problem indeed follows the behaviour derived from the
distributional cavity equations.

\subsubsection{Symmetric SBM, the marginal case of 4 groups}
\label{sec_mainresults_sbm}

The symmetric SBM is a special case of the model introduced in Sec.~\ref{sec_definitions_SBM}, where every community has the same size, i.e. $\oeta_\sigma = 1/q$, and the probability of an edge between two vertices of labels $\s$ and $\s'$ only depends on whether $\s=\s'$ or not: $c_{\sigma \sigma} = a$, and $c_{\sigma \sigma'} = b$ for ${\sigma \neq \sigma'}$. The average degree is then $d=[a+(q-1)b]/q$, and it is convenient to parametrize the difference between the intra-community and inter-communities connectivity by defining $\theta=(a-b)/(a+(q-1)b)$. This parameter is positive in the assortative case ($a>b$), negative in the disassortative case ($a<b$). As explained in Sec.~\ref{sec:cavity_vs_inference} the study of this problem by the cavity method~\cite{DecelleKrzakala11,Decelle_SBM_long} is tightly linked to the tree reconstruction problem for the symmetric Potts model~\cite{Bleher1995,EvansKenyon00,MezardMontanari06,Sly11}, with an offspring distribution of the Galton-Watson tree $\tp_\ell$ Poissonian of average $d$; in this context $\theta>0$ (resp. $\theta<0$) is called the ferromagnetic (resp. antiferromagnetic) case.  The Kesten-Stigum transition is easily seen in the tree reconstruction perspective to occur when $d \theta^2=1$, where $d$ is the average of $\tp_\ell$, this combination of parameters can thus be regarded as a relevant signal-to-noise ratio (keeping in mind the sign of $\theta$ that disappears from this definition).

The results of our local bifurcation analysis, performed for an arbitrary distribution $\tp_\ell$, can be summarized as follows (the details can be found in Sec.~\ref{sec_applications_symPotts}):
\begin{itemize}
\item for $q \le 3$, for both ferromagnetic and antiferromagnetic models and for any $\tp_\ell$, the Kesten-Stigum transition is continuous (2nd order, as in the left panel of Fig.~\ref{fig:standard}).

\item for $q\ge 5$, for both ferromagnetic and antiferromagnetic models and for any $\tp_\ell$, the transition is discontinuous (1st order, cf. the right panel of Fig.~\ref{fig:standard}).

\item for $q=4$ the order of the phase transition depends explicitly on the degree distribution and on the ferromagnetic/antiferromagnetic character of the model, through the sign of the following quantity
\beq
w=-\frac{7}{3} \frac{\tE[\ell(\ell-1)(\ell-2)]}{\tE[\ell]^3} 
+ \left(\frac{\tE[\ell(\ell-1)]}{\tE[\ell]^2} \right)^2
\left( \frac{5}{\tE[\ell]-1} - \frac{12}{\sign(\theta) \sqrt{\tE[\ell]}-1}
\right)\ ,  \label{eq:4groups}
\eeq
where $\tE$ denotes the average with respect to the offspring degree distribution $\tp_\ell$. A 1st order phase transition is predicted to arise if $w>0$, while the transition is continuous (2nd order) for $w<0$. In the ferromagnetic/assortative case ($\sign(\theta)=1$), one can check easily that $w<0$ for all degree distributions, which yields a continuous transition (one has necessarily $d=\tE[\ell]>1$ for the problem to make sense). On the contrary in the antiferromagnetic/disassortative case the sign of $w$ changes according to the degree distribution. For instance when $\tp_\ell$ is a Poisson law with average $d$, as in the application to the graph SBM problem, there is a critical degree $d_{\rm c}\approx 22.2694$ such that $w>0$ for small degrees $d < d_{\rm c}$, leading to a 1st order phase transition, while for large degrees $d > d_{\rm c}$ the coefficient $w$ becomes negative and hence the transition is of 2nd order type.
\end{itemize}

This problem was extensively studied in the literature from various perspectives, notably tree reconstruction~\cite{Bleher1995,EvansKenyon00,MezardMontanari06,Sly11}, stochastic block model~\cite{DecelleKrzakala11,Decelle_SBM_long}, Potts model on random graphs~\cite{KrzakalaZdeborova07}, and part of this characterization was already known (in particular the tightness of the KS transition for $q=2$ was established in~\cite{Bleher1995,EvansKenyon00}, and~\cite{Sly11} proved the non-tightness of the KS transition for $q\ge 5$ for all degrees, and its tightness for $q=3$ and large enough degrees). However, the situation of the $q=4$ case remained rather obscure up to now (for instance Fig.~2 of \cite{KrzakalaZdeborova07} missed the crossover towards a continuous transition at large degrees), even if it could have been anticipated that the ferromagnetic and antiferromagnetic models must behave in the same way in the large degree limit (see for instance~\cite{Sly11,lesieur2015mmse} and the discussion in Sec.~\ref{sec_large_degree}). The explicit condition (\ref{eq:4groups}) that characterizes sharply the degree distributions for which the antiferromagnetic 4 groups Potts model (or disassortative symmetric SBM) has a gap between the information-theoretical performance and the best known algorithmic one is thus one of our main original results.

\begin{figure}[ht]
\centerline{
\includegraphics[width=8cm]{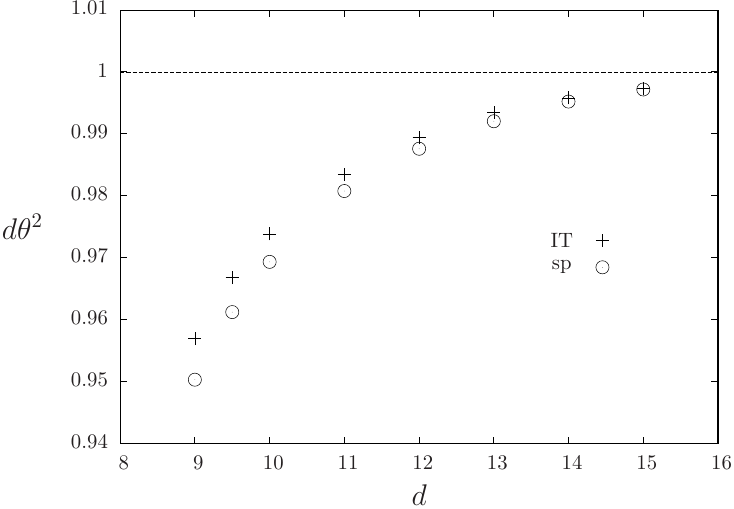}
\hspace{6mm}
\includegraphics[width=9cm]{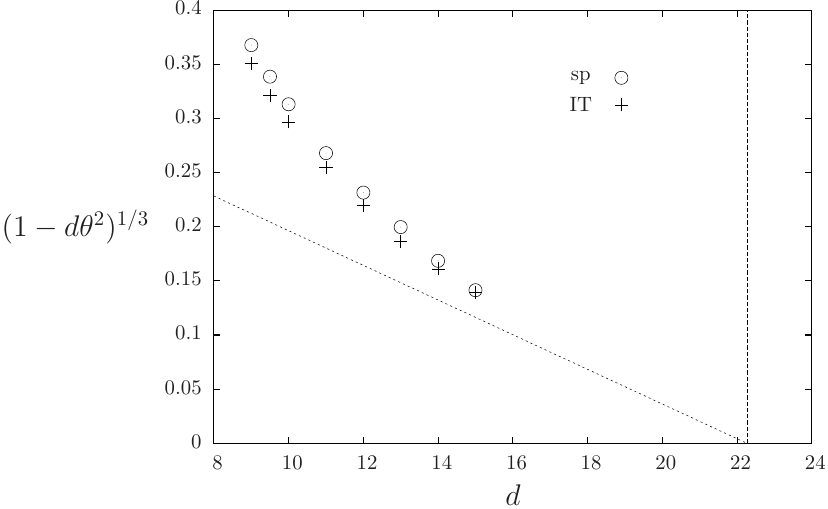}
}
\caption{The phase diagram of the 4 groups disassortative symmetric SBM (equivalently the reconstruction of antiferromagnetic $q=4$ Potts model on Poissonian Galton-Watson trees), as a function of the average degree $d$. On the left panel the spinodal and IT transitions (see the text and Fig.~\ref{fig:standard} for their definitions) are plotted in terms of the signal-to-noise parameter $d \theta^2$, which equals 1 at the Kesten-Stigum transition (horizontal dashed line). On the right panel the same data is drawn with a change of variable on the vertical axis to better appreciate the crossover from 1st to 2nd order transition as $d$ reaches $d_{\rm c}$; as explained in the text and with more details in Sec.~\ref{sec_applications_symPotts} we expect the curves on the right panel to vanish linearly when $d \to d_{\rm c}^-$. The vertical line marks our analytical prediction for $d_{\rm c}\approx 22.2694$, the other line is only here as a guide to the eye.}
\label{fig:4groups}
\end{figure}

In order to test this prediction we performed a numerical resolution of the cavity equations (see Appendix~\ref{app_numerical_resolution} for more details) for the $q=4$ antiferromagnet on Poissonian Galton-Watson trees. In the left panel of Fig.~\ref{fig:4groups} one sees clearly that for low average degrees the spinodal and information theoretic transition are below the Kesten-Stigum bound, and that they get closer to it as $d$ grows. A simple scaling argument, to be explained in Sec.~\ref{sec_applications_symPotts}, predicts that the signal-to-noise ratio $d \theta^2$ of the spinodal and IT transitions reaches the Kesten-Stigum value $1$ with a correction scaling as $(d_{\rm c}- d)^3$ as $d \to d_{\rm c}^-$. In the right panel of Fig.~\ref{fig:4groups} we thus plotted the same phase diagram with the change of variable $d\theta^2 \to (1-d\theta^2)^{1/3}$, in such a way that the deviation to the KS bound should vanish linearly when $d \to d_{\rm c}^-$ in these units. It is fair to say that our numerical data are compatible with our analytical prediction of $d_{\rm c}$, without bringing an independent confirmation of it. As a matter of fact, the largest degrees for which we could estimate reliably the location of the spinodal and IT transitions are still rather far from our prediction of $d_{\rm c}$; it is indeed very difficult to distinguish numerically a continuous transition from a very weakly discontinuous one, and the relatively large exponent 3 in the behavior of the signal to noise ratio is unfavourable for this numerical study. As a consequence we cannot perform a linear fit from the plot on the right of Fig.~\ref{fig:4groups}, the line we drew as a guide to the eye looks however as a possible interpolation towards our analytically computed value for $d_{\rm c}$.

\subsubsection{Asymmetric balanced two groups SBM}

We have also dedicated a specific attention, detailed in section \ref{sec_Ising_asym}, to the case of the $q=2$ communities SBM with different communities sizes ($\oeta_1\neq \oeta_2$), yet with an affinity matrix such that the average degree of the vertices in the two communities is the same (hence the degree of a vertex is uninformative of its label, and the model exhibits an undetectable phase at low degrees). We will parametrize this model by a ``magnetization'' parameter $\om \in [0,1]$ such that the size of the smaller group is $\oeta_1 = \frac{1-\om}{2}$, the symmetric case being recovered for $\om=0$. With the condition of both groups having average degree $d$ we are left with one more free parameter, we will call it $\theta$ and define it as $\theta = 1 - \frac{c_{12}}{d}$ where $c_{12}=c_{21}$ is $N$ times the probability that two nodes in different groups are connected by an edge. 

\begin{figure}
\centerline{
\includegraphics[width=8cm]{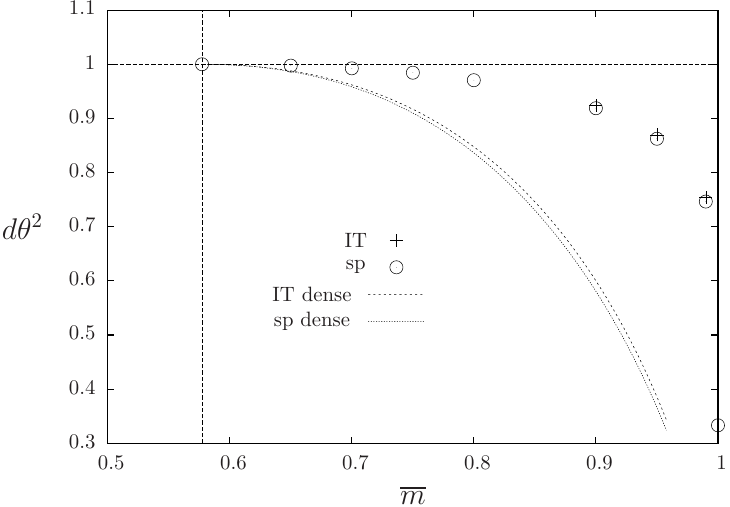}
\hspace{6mm}
\includegraphics[width=9cm]{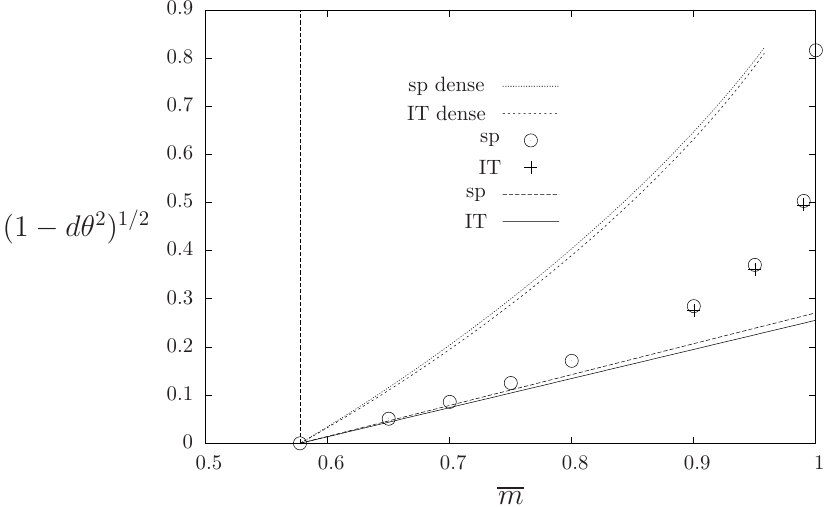}
}
\caption{The phase diagram of the ferromagnetic ($\theta>0$) asymmetric Ising model on a regular tree with offspring degree distribution $\tp_{\ell} = \delta_{\ell,d}$, $d=3$. The magnetization parameter $\om$ encodes the size of the smaller community, in the SBM interpretation, as $(1-\om)/2$. On the left panel the spinodal and IT transitions are plotted in terms of the signal-to-noise parameter $d \theta^2$, which equals 1 at the Kesten-Stigum transition (horizontal dashed line). Within our numerical accuracy we cannot distinguish the spinodal and IT points for $\om$ smaller than $0.8$, we used a single symbol in this case. The two lines corresponds to the large degree limit, as computed for the dense models in~\cite{BaDiMaKrLeZd16,CaLeMi16,LeKrZd17}. On the right panel the same data is drawn in rescaled units to better appreciate the crossover from 1st to 2nd order transition as $\om$ reaches its critical value $\om_{\rm c}=1/\sqrt{3} \approx 0.577$, marked as a vertical line. The two straight lines are the analytical predictions of Eqs.~(\ref{eq_Ising_line_td},\ref{eq_Ising_line_tc}) for the leading order behavior of the spinodal and IT transitions as $\om \to \om_{\rm c}^+$, for the sparse case $\tp_{\ell} = \delta_{\ell,3}$.} \label{fig:asymmetric}
\end{figure}

The cavity method relates this model to a tree reconstruction problem known as the asymmetric Ising model~\cite{Mossel01,KS_tight_Ising,liu2017reconstruction}, for which the KS transition occurs at $d \theta^2 =1$, with $d=\tE[\ell]$ the average offspring degree. Our expansions of the cavity equations around this point reveals a striking universality phenomenon: for all degree distributions, and for both ferromagnetic and antiferromagnetic models, the qualitative nature of the transition changes when the parameter $\om$ that quantifies the asymmetry of the model crosses the same value $\om_{\rm c}=1/\sqrt{3}$. The transition is indeed second order (left panel of Fig.~\ref{fig:standard}) for $\om < \om_{\rm c}$, and first order (right panel of Fig.~\ref{fig:standard}) for $\om > \om_{\rm c}$. Until very recently it was only known from~\cite{KS_tight_Ising} that the Kesten-Stigum bound was tight (i.e. the transition second order) for small enough asymmetries, and that it was not tight for large enough asymmetries~\cite{Mossel01}, but these papers did not estimate the critical asymmetry separating these two regimes. The value $\om_{\rm c}=1/\sqrt{3}$ was deduced independently in~\cite{liu2017reconstruction} which provide a rigorous proof of the non-tightness of the Kesten-Stigum bound for $\om > \om_{\rm c}$ on regular trees, via moment expansions similar to ours. Our (non-rigorous) computations presented in Sec.~\ref{sec_Ising_asym} are more generic, as they encompass arbitrary offspring distributions $\tp_\ell$, but more importantly have been pushed to an higher order in the expansion. This allows us to predict not only the critical asymmetry $\om_{\rm c}$ above which the Kesten-Stigum bound is not tight, but also the leading order behavior of the spinodal (reconstruction) and IT transition lines in the regime $\om \to \om_{\rm c}^+$ (which is not universal and depends on the degree distribution and on the ferromagnetic/antiferromagnetic character of the model). As an illustration we present in Fig.~\ref{fig:asymmetric} the phase diagram for the ferromagnetic Ising model on a regular tree, obtained by a numerical resolution of the cavity equations. One sees indeed on the left panel that the spinodal and information theoretic transition occur before the KS threshold for asymmetries larger than $\om_{\rm c}$, and that this discontinuity of the transition vanishes when $\om \to \om_{\rm c}^+$. In the right panel we present the same data in rescaled units, chosen such that the approach to $\om_{\rm c}$ is linear; the two straight lines of the plot are the results of our analytical predictions to be explained in Sec.~\ref{sec_Ising_asym}, and are in agreement with the numerical results within the accuracy we could reach.

Note that the independence of $\om_{\rm c}$ on the degree distribution explains why the same phenomenology, and the same critical asymmetry, was already obtained in the corresponding dense inference problems~\cite{BaDiMaKrLeZd16,LeKrZd17}, or in the large degree limit~\cite{CaLeMi16,LeMi16}. The community detection problem in networks with unequal groups was also recently investigated in~\cite{Montanari2015,zhang2016community}, but these works treat the case where the degree of a vertex is correlated to its label; on the contrary in all the problems we treated in this paper the degree of a vertex is not informative of its label.

\subsubsection{$q_1+q_2$ SBM}
\label{sec:main_q1q2}

\begin{figure}
\centerline{
\includegraphics[width=5cm]{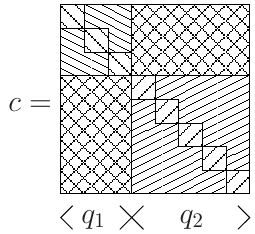}
}
\caption{The shape of the connectivity matrix in the $q_1+q_2$ Potts model, matrix elements 
filled with the same pattern
are equal.}
\label{fig_matrix_q1q2}
\end{figure}

As we have seen in the two cases above the qualitative properties of the phase diagram of the SBM strongly depends on the structure of the prior distribution $\oeta$ and of the affinity matrix $c$, in particular on their symmetry properties. It is of course impossible to investigate explicitly all the possible ways to break the full permutation symmetry between the $q$ labels, for arbitrary $q$, and to determine the typology of the phase transitions of the associated models. We have studied a class of SBM that generalizes and encompasses both the symmetric SBM model for $q$ arbitrary and the asymmetric $q=2$ case, by dividing the $q$ possible labels in two ``super-groups'' $G_1$ and $G_2$ of cardinality $q_1$ and $q_2$ (with $q=q_1+q_2$) and keeping the permutation symmetry inside each of these two sets. This leads to a prior $\oeta_\s$ that is constant inside $G_1$ and $G_2$, and an affinity matrix $c_{\s \t}$ which depends only on the super-groups $\s$ and $\t$ belong to, and on the fact that $\s=\t$ or not (see Fig.~\ref{fig_matrix_q1q2} for an illustration). One can view the graph generated in this way as a superposition of two symmetric SBMs on the vertices in the groups $G_1$ and $G_2$, and an asymmetric Ising model that coarse-grain all the labels in a group $G_i$ as a single symbol. Once $q_1$ and $q_2$ is chosen the model is defined by 5 parameters: the average degree of a vertex (assumed to be independent of its label), the fraction of vertices in the two super-groups, and the three SNR's quantifying the information provided by the edges of each SBM in the decomposition explained above. Depending on the choice of parameters the Kesten-Stigum transition can arise from any of these three subgraphs; in section~\ref{sec:q1+q2} we present a classification of the typology of the corresponding bifurcation. Moreover we will argue there that for well-chosen parameters the more complicated bifurcation depicted in Fig.~\ref{fig:new} must occur in this model.

Note that a special case of this problem (with an additional symmetry between the two sub-groups) was recently studied in~\cite{liu2017tightness,liu2018}.

\subsubsection{The hybrid-hard phase in the planted occupation models}
\label{sec_mainresults_occ}

In Sec.~\ref{sec_Ising} we shall study in detail the cavity equations for Ising spin ($q=2$) variables, denoting the spin alphabet $\chi=\{-1,+1\}$, with $k$-wise interactions, for $k\ge 2$ arbitrary. We concentrate on models that have a global spin-flip symmetry, i.e. such that $\oeta_+ = \oeta_- = 1/2$ and $p_{\rm j}(\s_1,\dots,\s_k)=p_{\rm j}(-\s_1,\dots,-\s_k)$ in the notations of Sec.~\ref{sec_cavity_formalism}. Quite strikingly we shall prove that all the problems fulfilling this symmetry property have, around their Kesten-Stigum threshold, a bifurcating non-trivial solution on the large signal-to-noise ratio side of the KS transition. In other words none of these models can exhibit a first order transition as in the right panel of Fig.~\ref{fig:standard}; depending on the model the phase diagram is either of the second order type (cf. left panel of Fig.~\ref{fig:standard}), or exhibits an hybrid-hard phase as in Fig.~\ref{fig:new}.

\begin{figure}
\includegraphics[width=5.5cm]{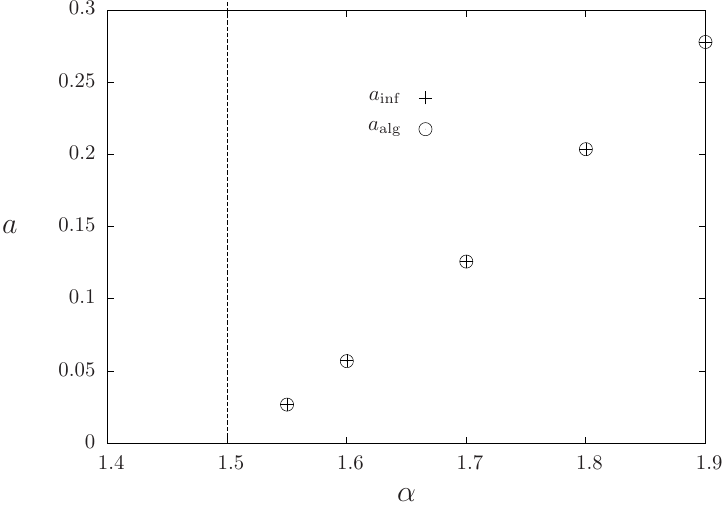}
\includegraphics[width=5.5cm]{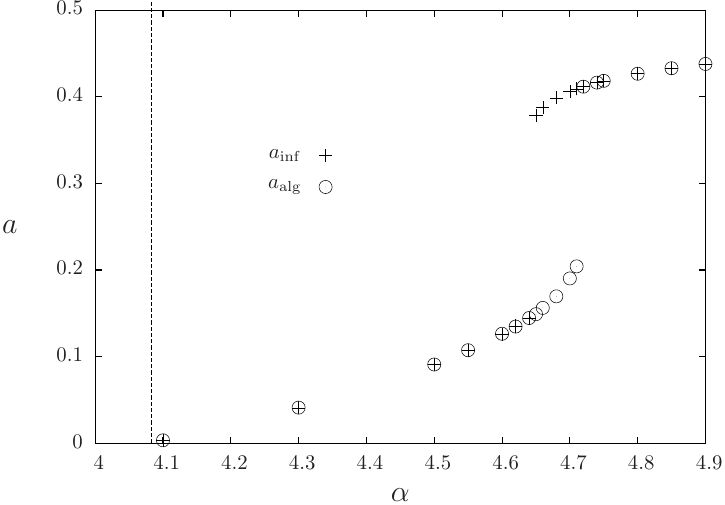}
\includegraphics[width=5.5cm]{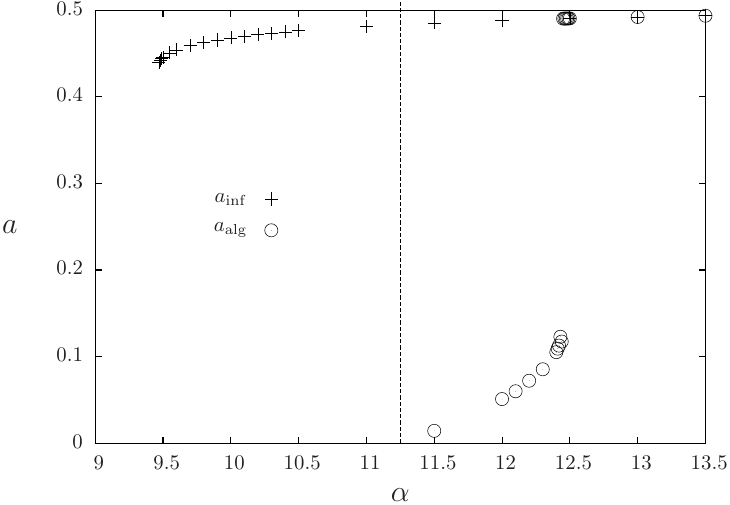}
\caption{Stable fixed points of the cavity equations in terms of their accuracies for the planted bicoloring of $k$-uniform hypergraphs, with $k=3,4,5$ from left to right. The accuracy is defined here as the difference between the probability that a sample from the posterior marginal outputs the correct label and the same probability when only the prior is used, see Sec.~\ref{sec_estimator} for more details. The vertical line marks the KS threshold. These curves have been obtained by a numerical resolution of the cavity equations (\ref{eq_generic_conditional_hPtoP},\ref{eq_generic_conditional_PtohP}), $a_{\rm inf}$ corresponds to the fixed point reached from an informative initialization (corresponding to the reconstruction problem, $\varepsilon=1$ in Eq.~(\ref{eq_initialcondition})), $a_{\rm alg}$ was obtained from an uninformative initialization (corresponding to the robust reconstruction problem, $\varepsilon \to 0$ in Eq.~(\ref{eq_initialcondition})). The parameter $\alpha$ controls the offspring degree distribution $\tp_\ell$, which is a Poisson law of average $\alpha k$. The optimal accuracy $a^*$ coincides with $a_{\rm inf}$ above the IT transition, whose location can be found in Table \ref{tab_bicoloring_main}.}
\label{fig_bicoloring_q_main}
\end{figure}

As an illustrating example we present in Fig.~\ref{fig_bicoloring_q_main} the bifurcation diagrams for the planted $k$-uniform hypergraph bicoloring, as defined in Sec.~\ref{sec_definitions_planted}, in which each hyperedge forbids the $k$ variables around it to be all in the same state. The random version of this Constraint Satisfaction Problem has been studied in some details (see for instance~\cite{gabrie2017phase} and references therein), but we are not aware of previous studies of its planted version for small values of $k$. The plots of Fig.~\ref{fig_bicoloring_q_main} show that for $k=3$ the transition is second order, while $k=4$ and $k=5$ realizes the two scenarios sketched in the left and right panel of Fig.~\ref{fig:new}, respectively, and whose interpretation was discussed in general terms in Sec.~\ref{sec_typology_pd}. In these latter two cases the hybrid-hard phase occupies a sizeable part of the phase diagram and can be easily identified numerically; we expect all $k \ge 5$ to behave qualitatively in the same way, and checked this for $k=6$. We report in Table~\ref{tab_bicoloring_main} the corresponding values of the thresholds $\alpha_{\rm KS}$ for the Kesten-Stigum transition, $\alpha_{\rm sp}$ and $\alpha_{\rm alg}$ for the spinodals of the high and low accuracy branches, respectively, and $\alpha_{\rm IT}$, corresponding to the crossings of the free-entropies of these two fixed-points (the corresponding numerical data will be shown in Sec.~\ref{sec_Ising}).

As far as we know the hybrid-hard phase in which Belief Propagation is able to reach a non-trivial accuracy, yet cannot reach the optimal one, was never identified previously in sparse inference problems that have an undetectable phase. This can sound surprising in view of our statement above: hybrid-hard phases do appear as soon as the phase diagram is not a purely second order transition. The resolution of this apparent paradox is that the hybrid-hard phase is often very narrow: for instance the so-called 2-in-4 planted SAT problem was reported, wrongly, to have a first order transition in~\cite{ZdeborovaKrzakala09,angelini2015spectral}. As we show in Sec.~\ref{sec_Ising} the hybrid-hard phase does exist in this model, but its width is so tiny that it was easily missed in previous numerical studies.

\begin{table}
\begin{tabular}{|c||c|c|c|c|c|c|}
\hline
$k$ & $\alpha_{\rm KS}$ & $\alpha_{\rm sp}$  & $\alpha_{\rm IT}$ & $\alpha_{\rm alg}$  \\
\hline
3   & 1.5                         & --                & --         &  -- \\
\hline
4   & 4.083                     & 4.64             & 4.673    & 4.71  \\
\hline
5   & 11.25                     & 9.46             & 10.296  & 12.44  \\
\hline
6   & 32.03                   & 18.08            & 21.425    & 34.5  \\
\hline
\end{tabular}
\caption{Thresholds for the bicoloring on Poissonian random hyper-graphs, see the text and Fig.~\ref{fig:new} for their definitions.}
\label{tab_bicoloring_main}
\end{table}

\subsubsection{Numerical experiments with Belief Propagation}

Our main technical results are based on numerical resolutions and 
analytic expansion of the distributional cavity equations
(\ref{eq_generic_unconditional_hPtoP}-\ref{eq_generic_unconditional_PtohP}). 
These equations describe directly the tree reconstruction problem, but their 
interpretations in terms of the inference problems on graphs and hypergraphs
relies on conjectured connections summarized in Sec.~\ref{sec_link_graphs}. 
We shall
therefore devote section~\ref{sec_numerical} to a numerical test of this
connection, and show that the behavior of the BP algorithm on large
but finite size samples is well described by the (infinite size) tree 
reconstruction problems. As a striking example let us anticipate that the 
curves of Fig.~\ref{fig_bicoloring_q_main} 
shall be reproduced in Fig.~\ref{fig:bicol_ms} from the output of
single sample experiments, including the coexistence of non-trivial branches
in the hybrid-hard phases.

\section{Moment expansions for pairwise interacting Potts variables}
\label{sec_Potts}

This section is devoted to the first of the two specializations of the formalism introduced in Sec.~\ref{sec_generic}: we shall only study here pairwise interacting models (i.e. $k=2$), in other words we consider graphs instead of hypergraphs, but keep an arbitrary $q$-state variable alphabet (in Sec.~\ref{sec_Ising} we shall instead turn to $k$-wise interactions with arbitrary $k$ but restrict ourselves to Ising spin variables).

\subsection{Cavity equations and free-entropy functional}

Let us start by rewriting the cavity equations and free-entropy functional introduced in Sec.~\ref{sec_generic} in a simplified form, exploiting the pairwise character of the interactions considered here. Our problem is now defined in terms of a probability distribution $\oeta$ on $\chi=\{1,\dots,q\}$, a $q \times q$ transition matrix $M_{\s \s'}=p_{\rm c}(\s'|\s)$, and the degree distributions $p_\ell$ and $\tp_\ell$ related by the size-bias (\ref{eq_tpell}). The two equations (\ref{eq_generic_conditional_hPtoP},\ref{eq_generic_conditional_PtohP}) can be joined into a single recursion on the conditional distribution:
\beq
P^{(n+1)}_\t(\eta) = \sum_{\ell=0}^\infty \tp_\ell \sum_{\t_1,\dots,\t_\ell} M_{\t \t_1} \dots M_{\t \t_\ell} \int \dd P_{\t_1}^{(n)}(\eta^1) \dots  \dd P_{\t_\ell}^{(n)}(\eta^\ell) \, \delta(\eta - f(\eta^1,\dots,\eta^\ell)) \ ,
\label{eq_Potts_recursion_Pt}
\eeq
where the Belief Propagation recursion function $f$ is obtained by concatenating (\ref{eq_generic_nutoeta}) and (\ref{eq_generic_etatonu}). In an explicit form,
$\eta=f(\eta^1,\dots,\eta^\ell)$ means that for all labels $\s \in \{1,\dots,q\}$,
\beq
\eta_\s = \frac{z_\s(\eta^1,\dots,\eta^\ell)}{z(\eta^1,\dots,\eta^\ell)} \ , \quad z(\eta^1,\dots,\eta^\ell) = \sum_\g z_\g(\eta^1,\dots,\eta^\ell) \ , \quad z_\s(\eta^1,\dots,\eta^\ell) = \oeta_\s \prod_{i=1}^\ell \sum_{\s'} \hM_{\s \s'} \eta^i_{\s'} \ ,
\label{eq_BP}
\eeq
where we introduced the notation
\beq
\hM_{\s \s'}=M_{\s \s'} \frac{1}{\oeta_{\s'}} \ .
\label{eq_def_hM}
\eeq
An equivalent and more compact form of (\ref{eq_Potts_recursion_Pt}) can be written as an equality in distribution of random variables,
\beq
\eta^{(n+1,\t)} \eqd f(\eta^{(n,\t_1)},\dots,\eta^{(n,\t_\ell)}) \ ,
\eeq
where all random variables are independent, $\ell$ has the distribution $\tp_\ell$, $\eta^{(n,\t)}$ stands for a random draw from $P^{(n)}_\t$, and the $\t_i$'s of the r.h.s. are generated with probability $M_{\t \t_i}$.

From the equations (\ref{eq_generic_unconditional_hPtoP},\ref{eq_generic_unconditional_PtohP}) one obtains the following recursion for the unconditional distribution,
\beq
P^{(n+1)}(\eta) = \sum_{\ell=0}^\infty \tp_\ell \int \dd P^{(n)}(\eta^1) \dots  \dd P^{(n)}(\eta^\ell) \, \delta(\eta - f(\eta^1,\dots,\eta^\ell)) \, z(\eta^1,\dots,\eta^\ell) \ ,
\label{eq_Potts_recursion_P}
\eeq
where $z$ is the quantity defined in (\ref{eq_BP}).

We gave in Sec.~\ref{sec_generic} an expression of the free-entropy for generic random factor graph models. Specializing it to the case of pairwise interactions one obtains:
\bea
\phi(\{P_\t\}) &=& \sum_{\ell=1}^\infty p_\ell \sum_\t \oeta_\t \sum_{\t_1,\dots,\t_\ell} M_{\t \t_1} \dots M_{\t \t_\ell} \int \dd P_{\t_1}(\eta^1) \dots \dd P_{\t_\ell}(\eta^\ell) \ln \zv(\eta^1,\dots,\eta^\ell) \nonumber \\ && - \frac{1}{2} \E[\ell] \sum_{\t_1,\t_2} \oeta_{\t_1} M_{\t_1 \t_2} \int \dd P_{\t_1}(\eta^1) \, \dd P_{\t_2}(\eta^2) \ln \ze(\eta^1,\eta^2) \ ,
\label{eq_Potts_phi_cond}
\eea
where $\E[\ell] = \sum_\ell p_\ell \ell$ is the average degree. The variable contribution $\zv$ reads exactly as $z$ in (\ref{eq_BP}), while the edge contribution is
\beq
\ze(\eta^1,\eta^2) = \sum_{\s_1,\s_2} \eta^1_{\s_1} \hM_{\s_1 \s_2} \eta^2_{\s_2} \ .
\label{eq_Potts_ze}
\eeq
An equivalent form in terms of the unconditional distribution is
\bea
\phi(P) &=& \sum_{\ell=1}^\infty p_\ell \int \dd P(\eta^1) \dots \dd P(\eta^\ell) \, \zv(\eta^1,\dots,\eta^\ell) \ln \zv(\eta^1,\dots,\eta^\ell) \nonumber \\ && - \frac{1}{2} \E[\ell] \int \dd P(\eta^1) \, \dd P(\eta^2) \, \ze(\eta^1,\eta^2) \ln \ze(\eta^1,\eta^2) \ .
\label{eq_Potts_phi_uncond}
\eea

\subsection{Properties of the Markov matrix $M$}
\label{sec_Potts_diagM}

Let us review the properties of the transition matrix $M_{\s \s'}=p_{\rm c}(\s'|\s)$ that follows from the hypothesis on $p_{\rm c}$ made in Sec.~\ref{sec_generic}.
\begin{itemize}
\item
The normalization of $p_{\rm c}$, that ensures the conservation of probabilities in the broadcast process, implies that $M$ is a stochastic matrix, $\sum_{\s'} M_{\s \s'} =1$ for all $\s$.
\item
The assumption (\ref{eq_reversibility}) yields the reversibility of $M$ with respect to $\oeta$ (the detailed balance condition in physics jargon), $\oeta_\s M_{\s \s'}=\oeta_{\s'} M_{\s' \s}$. 
\item
This reversibility is a sufficient condition for $\oeta$ to be stationary for the Markov chain with transition probability matrix $M$, i.e. $\sum_{\s} \oeta_\s M_{\s \s'} = \oeta_{\s'}$ for all $\s'$. Hence $M$ has an eigenvalue $\theta_1=1$, with right eigenvector $(1,\dots,1)$ and left eigenvector $\oeta$. 
\end{itemize}

Assuming that $M$ is irreducible and aperiodic the Perron-Frobenius theorem ensures that all its other eigenvalues are strictly smaller in absolute value, we will order them as $1 > |\theta_2| \ge \dots \ge |\theta_q|$. The reversibility assumption simplifies the study of the eigen-decomposition of $M$: consider indeed the matrix $W_{\s \s'}= \oeta_\s ^{1/2} M_{\s \s'} \oeta_{\s'}^{- 1/2}$. It is easily seen that $W$ is real and symmetric, hence diagonalizable in an orthonormal basis, and that $M$ has the same eigenvalues as $W$. Let us write the eigen-decomposition of $W$ as
\beq
W_{\s \s'} = \sum_{j=1}^q \theta_j v_\s^{(j)} v_{\s'}^{(j)} \ , 
\eeq
where the vectors $v$ of the orthonormal basis verify
\beq
\sum_\s  v_\s^{(j)} v_\s^{(k)} = \delta_{j,k} \ , \qquad
\sum_j  v_\s^{(j)} v_{\s'}^{(j)} = \delta_{\s,\s'} \ ,
\eeq
with here and in the following $\delta_{j,k}$ the Kronecker symbol, equal to 1 if the two arguments are equal, 0 otherwise. The Perron eigenvector of $W$ reads in this basis:
\beq
\theta_1 = 1 \ , \ \ \ v^{(1)}_\s = \oeta_\s^{1/2} \ .
\eeq
Transforming back from $W$ to $M$ we introduce the basis of the left and right eigenvectors of $M$,
\beq
r_\s^{(j)} = \oeta_{\s}^{- 1/2} v_\s^{(j)} \ , \qquad
l_\s^{(j)} = \oeta_{\s}^{1/2} v_\s^{(j)} \ , 
\label{eq_def_lr}
\eeq
which obey the following relations:
\beq
\sum_\s  l_\s^{(j)} r_\s^{(k)} = \delta_{j,k} \ , \qquad 
\sum_j  l_\s^{(j)} r_{\s'}^{(j)} = \delta_{\s,\s'} \ , \qquad 
r_\s^{(j)} = \frac{1}{\oeta_\s} l_\s^{(j)} \ , \qquad 
l_\s^{(1)} = \oeta_\s \ , \qquad r_\s^{(1)} = 1 \ .
\eeq
The matrix $M$ can thus be expressed as:
\beq
M_{\s \s'} = \sum_{j=1}^q \theta_j r_\s^{(j)} l_{\s'}^{(j)}
= \oeta_{\s'}+\sum_{j=2}^q \theta_j r_\s^{(j)} l_{\s'}^{(j)}\ .
\eeq
The matrix $\hM$ introduced in (\ref{eq_def_hM}) is symmetric thanks to the reversibility condition, and it can be expressed in terms of the eigenvector decomposition as
\beq
\hM_{\s \s'} = \sum_{j=1}^q \theta_j r_\s^{(j)} r_{\s'}^{(j)}
= 1 +\sum_{j=2}^q \theta_j r_\s^{(j)} r_{\s'}^{(j)} \ . 
\label{eq_Potts_defhM}
\eeq

\subsection{Symmetry properties}

Let us recall that we stated in (\ref{eq_Bayes}) some consequences of the Bayes theorem that bridge the conditional and unconditional distributions. These will play an important role in the following, hence we first spell out some consequences of these symmetry relations. We will denote $\E_\t^{(n)}$ and $\E^{(n)}$ the averages with respect to $P_\t^{(n)}$ and $P^{(n)}$. One has $\E^{(n)}[\eta]=\oeta$ for all $n$. For any function $g$, the relationship between the densities of $P_\t^{(n)}$ and $P^{(n)}$ given in (\ref{eq_Bayes}) yields:
\beq
\oeta_\t \E_\t^{(n)}[g(\eta)] = \E^{(n)}[\eta_\t g(\eta)] \ .
\label{eq_Bayes_Potts}
\eeq
Applying this identity with $g(\eta) = 1$ gives back the already mentioned property $\E^{(n)}[\eta]=\oeta$. If one uses instead $g(\eta) = \eta_\s$, this yields
\beq
\oeta_\t \E_\t^{(n)}[\eta_\s] = \E^{(n)}[\eta_\t \eta_\s] \ .
\label{eq_moment_Potts_0}
\eeq
Let us define a $q$-dimensional vector $\delta$ as the difference $\delta = \eta - \oeta$ between a message and the stationary value. One has in this way $\E^{(n)}[\delta]=0$, and from (\ref{eq_moment_Potts_0}) one obtains
\beq
\oeta_\t \E_\t^{(n)}[\delta_\s] = \E^{(n)}[\eta_\t \eta_\s] - \oeta_\t \oeta_\s = \E^{(n)}[\delta_\t \delta_\s] \ .
\label{eq_moment_Potts_1}
\eeq

\subsection{Stability analysis of the trivial fixed-point via moment expansions}
\label{sec_expansion_Potts}

By construction the probability distribution $\oeta$ is a fixed point of the BP equation (\ref{eq_BP}), i.e. $f(\oeta,\dots,\oeta) = \oeta$, reflecting the stationarity of $\oeta$ under the Markov chain generated by $M$. This implies that $P_\t(\eta) = \delta(\eta-\oeta)$ is a fixed point of the cavity recursion (\ref{eq_Potts_recursion_Pt}) (and similarly $P(\eta) = \delta(\eta-\oeta)$ is invariant under the recursion (\ref{eq_Potts_recursion_P}) for the unconditional distributions). We shall now study the stability of this trivial fixed point and determine perturbatively the non-trivial fixed points that arise in the neighborhood of the limit of stability. This bifurcation analysis will follow the steps explained in Sec.~\ref{sec_bifurcation_scalar} for scalar recursions. The additional complication due to the functional character of the recursions (\ref{eq_Potts_recursion_Pt},\ref{eq_Potts_recursion_P}) will be dealt with by concentrating on the low order moments of the distributions $P_\t$, thereby reducing the infinite-dimensional bifurcation analysis to a finite-dimensional one.

\subsubsection{Linear analysis}

We denote $\delta = \eta-\oeta$ the ($q$-dimensional vector) difference between the BP message and the trivial fixed point. As both $\eta$ and $\oeta$ are normalized one has $\sum_\s \delta_\s = 0$. The BP equation (\ref{eq_BP}) can be expressed as
\beq
\delta_\s = \frac{\oeta_\s \, \overset{\ell}{\underset{i=1}{\prod}}(1+\hdelta_\s^i)}{\underset{\g}{\sum} \oeta_\g \, \overset{\ell}{\underset{i=1}{\prod}}(1+\hdelta_\g^i)
} - \oeta_\s \ ,
\label{eq_BP_delta}
\eeq
where we defined
\beq
\hdelta_\s \equiv \sum_{\s'} \hM_{\s \s'} \delta_{\s'}
\label{eq_BP_def_hdelta}
\eeq
as a linearly transformed version of $\delta$. The normalization condition becomes, under this transformation, $\sum_\s \oeta_\s \hdelta_\s = 0$. For this reason the denominator of (\ref{eq_BP_delta}) is equal to 1 up to quadratic corrections in $\delta$, hence the linearization of the BP equation gives
\beq
\delta_\s \approx \oeta_\s \sum_{i=1}^\ell \hdelta^i_{\s} 
\ .
\eeq
Taking averages according to Eq.~(\ref{eq_Potts_recursion_Pt}) we thus obtain the evolution of the first moments of the distributions $P_\t^{(n)}$, within this linearized approximation, as
\beq
\E_\t^{(n+1)}[\delta_\s] = \tE[\ell] \oeta_\s \sum_{\t'} M_{\t \t'}  \E_{\t'}^{(n)}[\hdelta_\s] 
\ ,
\label{eq_Potts_linear}
\eeq
where $\tE[\ell] = \sum_\ell \tp_\ell \ell$ is the average offspring degree; note that in (\ref{eq_Potts_linear}), and in many equations that will follow, we keep implicit higher order correction terms in the right hand side for the sake of readability. To put this evolution equation under a more convenient form let us first exploit the moment identities to write
\beq
\sum_{\t'} M_{\t \t'} \E_{\t'}^{(n)}[\hdelta_\s] = \sum_{\t'} \hM_{\t \t'} \oeta_{\t'} \E_{\t'}^{(n)}[\hdelta_\s] = \sum_{\t'} \hM_{\t \t'}  \E^{(n)}[\hdelta_\s \eta_{\t'}] \ ,
\eeq
where in the first step we used the definition (\ref{eq_Potts_defhM}) of $\hM$ and in the second one we exploited the relation (\ref{eq_Bayes_Potts}) between conditional and unconditional averages. Noting that $\sum_{\t'} \hM_{\t \t'} \oeta_{\t'}=1$ and that $\E^{(n)}[\hdelta_\s]=0$ we obtain the identity:
\beq
\sum_{\t'} M_{\t \t'} \E_{\t'}^{(n)}[\hdelta_\s] = \E^{(n)}[\hdelta_\s \hdelta_\t] \ ,
\label{eq_Potts_id}
\eeq
Multiplying the linearized evolution equation (\ref{eq_Potts_linear}) by $\oeta_\t$ and using (\ref{eq_moment_Potts_1}) to transform the left hand side and (\ref{eq_Potts_id}) to treat the right hand side we obtain, without making further approximations,
\beq
\E^{(n+1)}[\delta_\s \delta_\t] 
= \tE[\ell] \oeta_\s \oeta_\t \E^{(n)}[\hdelta_\s \hdelta_\t] \ .
\label{eq_Potts_linear2}
\eeq
We shall call $A^{(n)}_{\s \t} = \E^{(n)}[\delta_\s \delta_\t]$; it can be viewed as a matrix $A^{(n)}$ of size $q \times q$, which is symmetric positive semi-definite with vanishing row sums. We also denote
\beq
\hA^{(n)}_{\s \t} = \E^{(n)}[\hdelta_\s \hdelta_\t] = \sum_{\s',\t'} \hM_{\s \s'} \hM_{\t \t'} A^{(n)}_{\s' \t'} \ ,
\label{eq_Potts_defhA}
\eeq
which defines a matrix $\hA$, symmetric positive semi-definite with $\sum_\s \oeta_\s \hA^{(n)}_{\s \t} =0$. With these notations the linearized recursion (\ref{eq_Potts_linear2}) becomes
\beq
A^{(n+1)}_{\s \t} = \tE[\ell] \oeta_\s \oeta_\t \hA^{(n)}_{\s \t} \ ,
\label{eq_Potts_linear_A}
\eeq
which, considering now $A$ as a $q^2$-dimensional vector, can be viewed as a matrix multiplication:
\beq
A^{(n+1)} = N A^{(n)} \ , \qquad N_{\s \t , \s' \t'} = \tE[\ell] \oeta_\s \oeta_\t \hM_{\s \s'} \hM_{\t \t'} = \tE[\ell] \oeta_\s M_{\s \s'} \frac{1}{\oeta_{\s'}} \oeta_\t M_{\t \t'} \frac{1}{\oeta_{\t'}} \ .
\eeq
In this equation $N$ is a $q^2 \times q^2$ symmetric matrix; its
eigenvalues are easily computed, exploiting its tensor product form, 
and seen to be $\{\tE[\ell] \theta_i \theta_j\}_{i,j=1,\dots,q}$,
where the $\theta$'s are the eigenvalues of $M$ (see
Sec.~\ref{sec_Potts_diagM}). As $A^{(n)}$ must obey the normalization
condition $\sum_{\s} A^{(n)}_{\s \t}=0$ the relevant eigenvalues of
$N$ are only those with $i,j\ge 2$. The stability of $A=0$ under its
multiplication by $N$ is equivalent to the largest absolute value
among the (relevant) eigenvalues of $N$ being smaller than 1: we have
thus recovered, as expected, the well-known Kesten-Stigum~\cite{KestenStigum66} condition
$\tE[\ell] \theta_2^2 < 1$ for the trivial fixed point $P(\eta) =
\delta(\eta-\oeta)$ to be stable under small perturbations.

\subsubsection{Second order expansion}

Our goal now is to characterize the non-trivial fixed points of the functional recursions (\ref{eq_Potts_recursion_Pt},\ref{eq_Potts_recursion_P}) that bifurcate continuously from the trivial one at the Kesten-Stigum transition; we shall hence drop the iteration index $n$ in the following, and denote $\E_\t[\bullet]$ and $\E[\bullet]$ as the averages with respect to $P_\t$ and $P$ fixed-point solutions of (\ref{eq_Potts_recursion_Pt}) and (\ref{eq_Potts_recursion_P}) respectively. It is rather intuitive, and can be easily seen in the scalar case treated in Sec.~\ref{sec_bifurcation_scalar}, that this task will rely on an expansion of (\ref{eq_BP_delta}) beyond the linear order. However, the functional character of the fixed point equation complicates the determination of the relevant non-linear terms in this expansion. In other words one has to specify in which precise sense the sought-for fixed point is ``close'' to the trivial one. Let us consider the unconditional distribution $P(\eta)$ and discuss the relative scaling of its centered moments. By definition the first one $\E[\delta_\s]$ must vanish; the second one, $A_{\s \t} = \E[\delta_\s \delta_\t ]$ must be non-zero for the distribution to be non-trivial, and we assume that it is small, of the order of some parameter $\kappa \ll 1$. We shall make the following ansatz for the higher-order moments,
\beq
B_{\s \t \g}=\E[\delta_\s \delta_\t \delta_\g] = O(\kappa^2) \ , \qquad
C_{\s \t \g \b}=\E[\delta_\s \delta_\t \delta_\g \delta_\b] = O(\kappa^2) \ ,
\qquad \dots  \qquad
\E[\delta_{\s_1} \dots \delta_{\s_p}] = O( \kappa^{\lceil p/2 \rceil}) \ .
\label{eq_ansatz}
\eeq
This ansatz was inspired by the numerical resolution of the functional equations, and we refer the reader to the Appendix \ref{sec_app_expansion_Potts} for a proof of its self-consistency.

We shall now proceed with these assumptions, and compute the leading behavior of $A_{\s \t}$ by considering the following truncated expansion of (\ref{eq_BP_delta}):
\bea
\delta_\s &=& \oeta_\s \sum_{i=1}^\ell  \hdelta^i_\s \nonumber \\ &+& 
\oeta_\s \sum_{1\le i<j \le \ell}\left[
 \hdelta^i_\s \hdelta^j_\s 
- \sum_\g \oeta_\g \hdelta^i_\g \hdelta^j_\g  
- \sum_\g \oeta_\g (\hdelta^i_\s + \hdelta^j_\s) \hdelta^i_\g \hdelta^j_\g
- \sum_\g \oeta_\g \hdelta^i_\s \hdelta^i_\g  \hdelta^j_\s \hdelta^j_\g 
+  \sum_{\g,\b} \oeta_\g \oeta_\b \hdelta^i_\g  \hdelta^i_\b  \hdelta^j_\g  \hdelta^j_\b 
\right] \ ;
\eea
we have included non-linear terms that are of order 2,3 and 4 in terms of $\delta$, but that will all be of order $\kappa^2$ once averaged, according to the ansatz (\ref{eq_ansatz}). Taking averages with respects to the fixed point conditional distributions yields, without further approximations,
\bea
\E_\t[\delta_\s] &=& \tE[\ell] \oeta_\s \sum_{\t'} M_{\t \t'}  \E_{\t'}[\hdelta_\s] \nonumber \\ &+& \frac{1}{2} \tE[\ell (\ell-1)]
\oeta_\s \sum_{\t',\t''} M_{\t \t'} M_{\t \t''}
\left[ \E_{\t'}[\hdelta_\s] \E_{\t''}[\hdelta_\s]
-\sum_\g \oeta_\g \E_{\t'}[\hdelta_\g] \E_{\t''}[\hdelta_\g]
-2\sum_\g \oeta_\g \E_{\t'}[\hdelta_\s \hdelta_\g] \E_{\t''}[\hdelta_\g] \right. \nonumber \\
&& \left.
- \sum_\g \oeta_\g \E_{\t'}[\hdelta_\s \hdelta_\g] \E_{\t''}[\hdelta_\s \hdelta_\g] 
+ \sum_{\g,\b} \oeta_\g \oeta_\b \E_{\t'}[\hdelta_\g \hdelta_\b] \E_{\t''}[\hdelta_\g \hdelta_\b] 
\right] \ .
\label{eq_Potts_dev2}
\eea
In order to simplify this equation we first state an identity similar to (\ref{eq_Potts_id}):
\beq
\sum_{\t'} M_{\t \t'} \E_{\t'}[\hdelta_\s \hdelta_\g] = 
\sum_{\t'} \hM_{\t \t'} \E[\hdelta_\s \hdelta_\g \eta_{\t'}]
=  \E[\hdelta_\s \hdelta_\g ] + \E[\hdelta_\s \hdelta_\g \hdelta_\t] \ ;
\label{eq_Potts_id2}
\eeq
note that according to our ansatz the second term is negligible with respect to the first one (it is of order $\kappa^2 \ll \kappa$). 

We shall now multiply (\ref{eq_Potts_dev2}) by $\oeta_\t$, in such a way that its left hand side becomes $A_{\s \t}$; the right hand side can be simplified by using the identities (\ref{eq_Potts_id},\ref{eq_Potts_id2}). Keeping only the terms of order $\kappa$ and $\kappa^2$ 
yields
\bea
A_{\s \t} &=& \tE[\ell] \oeta_\s \oeta_\t \hA_{\s \t}
+ \frac{1}{2}\tE[\ell(\ell-1)] \oeta_\s \oeta_\t \left[ 
(\hA_{\s \t} )^2 
- \sum_\g \oeta_\g (\hA_{\s \g} + \hA_{\g \t})^2
+  \sum_{\g \b} \oeta_\g \oeta_\b (\hA_{\g \b})^2
\right] \ ,
\label{eq_Potts_order2} \\
\hA_{\s \t} &=& \sum_{\s',\t'} \hM_{\s \s'} \hM_{\t \t'} A_{\s' \t'} \ ,
\eea
where we recalled the equation linking $A$ and $\hA$ in the second line.

An equivalent form of these equations, which will turn out to be more convenient in some cases, can be given in the basis that diagonalizes the matrix $M$ defined in Sec.~\ref{sec_Potts_diagM}. Using the  eigenvectors defined in (\ref{eq_def_lr}) we can trade the matrix $A$ for a matrix $A'$, now indexed by these basis vectors, in an invertible way:
\beq
A_{\s \t} = \sum_{j k} l_\s^{(j)} l_\t^{(k)} A'_{jk} 
\quad \leftrightarrow \quad
A'_{jk} = \sum_{\s \t} r_\s^{(j)} r_\t^{(k)} A_{\s \t}  \ .
\label{eq_Potts_def_Ap}
\eeq
The normalization condition that implied the vanishing of the row and column sums of $A$ shows up now as $A'_{jk}=0$ whenever $j=1$ or $k=1$. The transformation from $A$ to $\hA$, i.e. the matrix multiplication by $\hM$ for each of the two indices, becomes for $A'$ a simple multiplication by the eigenvalues $\theta_j$ of the matrix $M$. The equation (\ref{eq_Potts_order2}) thus becomes in this basis:
\beq
A'_{jk} = \tE[\ell] \theta_j \theta_k A'_{jk} + 
\frac{1}{2} \tE[\ell (\ell-1)] \left[\sum_{j_1 j_2 k_1 k_2} f_{j j_1 j_2} f_{k k_1 k_2} \theta_{j_1} \theta_{j_2} \theta_{k_1} \theta_{k_2} A'_{j_1 k_1} A'_{j_2 k_2} - 2 \theta_j \theta_k \sum_l A'_{jl} A'_{lk} \theta_l^2 \right] \ ,
\label{eq_Potts_order2_Ap}
\eeq
where we defined
\beq
f_{j_1 j_2 j_3} = \sum_\s \oeta_\s r_\s^{(j_1)} r_\s^{(j_2)} r_\s^{(j_3)} \ .
\label{eq_def_tensorf}
\eeq

We have achieved here our first technical goal: this quadratic equation (\ref{eq_Potts_order2}) on the matrix $A$ (or equivalently (\ref{eq_Potts_order2_Ap}) on the matrix $A'$) constitutes the equivalent of (\ref{eq:dense_1st_expanded}) for the scalar bifurcation analysis, but now for a generic model with pairwise interaction between Potts variables in the sparse regime. In particular imposing the positive definiteness of $A$ will allow us to discriminate between different bifurcation scenarios, as explained in Sec.~\ref{sec_bifurcation_scalar} in the simpler scalar case. We shall see in the following various applications of this formula, for different choices of the matrix $M$ and stationary distribution $\oeta$.

\subsubsection{Third order expansion}
\label{sec_Potts_third_order}

Before turning to these applications let us first state the results of this moment expansion at the next order (which in some cases will be crucial to discriminate between different bifurcation scenarios). According to our ansatz (\ref{eq_ansatz}) we have now to determine the centered moments of $P(\eta)$ up to the fourth moment, we thus define
\beq
A_{\s \t}=\E[\delta_\s \delta_\t] \ , \qquad
B_{\s \t \g}=\E[\delta_\s \delta_\t \delta_\g] \ , \qquad
C_{\s \t \g \b}=\E[\delta_\s \delta_\t \delta_\g \delta_\b] \ .
\label{eq_Potts_def_ABC}
\eeq
Generalizing the derivation of (\ref{eq_Potts_order2}) above by including higher order terms we have obtained the following set of equations on $A$, $B$ and $C$ (see Appendix~\ref{sec_app_expansion_Potts} for a justification through a formal expansion at all orders):
\bea
A_{\s \t} &=& \tE[\ell] \oeta_\s \oeta_\t \hA_{\s \t}
+ \frac{1}{2}\tE[\ell(\ell-1)] \oeta_\s \oeta_\t \left[ 
(\hA_{\s \t})^2 
- \sum_\g \oeta_\g (\hA_{\s \g} + \hA_{\g \t})^2
+  \sum_{\g \b} \oeta_\g \oeta_\b (\hA_{\g \b})^2
\right]  \label{eq_Potts_cavity_A} \\
&+& \tE[\ell(\ell-1)] \oeta_\s \oeta_\t \left[
-\sum_\g \oeta_\g \hB_{\s \t \g}(\hA_{\s \g} + \hA_{\t \g})
+\sum_{\g \b} \oeta_\g \oeta_\b (\hB_{\s \g \b} + \hB_{\t \g \b}) \hA_{\g \b}
+\sum_{\g \b} \oeta_\g \oeta_\b \hC_{\s \t \g \b} \hA_{\g \b}
\right] \nonumber \\
&+& \tE[\ell(\ell-1)(\ell-2)] \oeta_\s \oeta_\t \left[
\frac{1}{6} (\hA_{\s \t})^3 - \frac{1}{6} \sum_\g \oeta_\g (\hA_{\s \g} + \hA_{ \t \g})^3 - \frac{1}{2}  \hA_{\s \t} \sum_\g \oeta_\g (\hA_{\s \g} + \hA_{ \t \g})^2 + \frac{1}{6} \sum_{\g \b} \oeta_\g \oeta_\b (\hA_{\g \b})^3 \right. 
 \nonumber \\ && \hspace{1cm} \left. 
+ \frac{1}{2}  \sum_{\g \b} \oeta_\g \oeta_\b (\hA_{\g \b})^2
(\hA_{\s \t} + \hA_{\s \g} + \hA_{\t \g} + \hA_{\s \b} + \hA_{\t \b} )
+  \sum_{\g \b} \oeta_\g \oeta_\b \hA_{\g \b}
(\hA_{\s \g} + \hA_{\t \g})(\hA_{\s \b} + \hA_{\t \b} )
\right. \nonumber \\ && \hspace{1cm} \left.
- \sum_{\g \b \a} \oeta_\g \oeta_\b \oeta_\a \hA_{\g \b} \hA_{\b \a} \hA_{\a \g} \right] \ , \nonumber \\
B_{\s \t \g} &=& \tE[\ell] \oeta_\s \oeta_\t \oeta_\g \hB_{\s \t \g} 
\label{eq_Potts_cavity_B}
\\ &+& \tE[\ell (\ell-1)] \oeta_\s \oeta_\t \oeta_\g \left[
\hA_{\s \t} \hA_{\t \g} + \hA_{\s \g} \hA_{\g \t} + \hA_{\t \s} \hA_{\s \g}  
- \sum_\b \oeta_\b (\hA_{\s \b} \hA_{\b \g} + \hA_{\s \b} \hA_{\b \t} + \hA_{\t \b} \hA_{\b \g}  )
\right] \ , \nonumber \\
C_{\s \t \g \b} &=& \tE[\ell] \oeta_\s \oeta_\t \oeta_\g \oeta_\b  \hC_{\s \t \g \b} + \tE[\ell (\ell-1)] \oeta_\s \oeta_\t \oeta_\g \oeta_\b 
(\hA_{\s \t} \hA_{\g \b} + \hA_{\s \g} \hA_{\t \b} + \hA_{\s \b} \hA_{\t \g})
\ , \label{eq_Potts_cavity_C} \\
\hA_{\s \t} &=& \sum_{\s' \t'} \hM_{\s \s'} \hM_{\t \t'} A_{\s' \t'} \ , \label{eq_Potts_cavity_hA}\\
\hB_{\s \t \g} &=& \sum_{\s' \t' \g'} \hM_{\s \s'} \hM_{\t \t'} \hM_{\g \g'} B_{\s' \t' \g'} \ , \label{eq_Potts_cavity_hB}\\
\hC_{\s \t \g \b} &=& \sum_{\s' \t' \g' \b'} \hM_{\s \s'} \hM_{\t \t'} \hM_{\g \g'} \hM_{\b \b'} C_{\s' \t' \g' \b'} \ .\label{eq_Potts_cavity_hC}
\eea

\subsection{Expansion of the free-entropy}
\label{sec_expansion_phi}

Let us now focus on the second fundamental object introduced besides the cavity equation recursions, namely the free-entropy functional, in particular the expressions (\ref{eq_Potts_phi_cond},\ref{eq_Potts_phi_uncond}) for the case of pairwise interacting Potts variables. With the conventions we used the value of $\phi$ for the trivial fixed point $P_{\rm triv}(\eta)=\delta(\eta-\oeta)$ is $\phi(P_{\rm triv})=0$. We have computed an expansion of $\phi(P)$ when $P$ is close to $P_{\rm triv}$, i.e. when the centered moments of $P$ are small (imposing the condition $\int \dd P(\eta) \, \eta = \oeta$). Denoting again $A$, $B$ and $C$ the tensors encoding these centered moments of $P$ of order 2, 3 and 4 (as defined in (\ref{eq_Potts_def_ABC})), and $\hA$, $\hB$ and $\hC$ their transformed versions according to (\ref{eq_Potts_cavity_hA}-\ref{eq_Potts_cavity_hC}), we have found (see Appendix~\ref{sec_app_expansion_Potts_phi} for the details of the derivation):
\bea
\frac{1}{\E[\ell]} \phi(P) &=& -\frac{1}{4} \sum_{\s \t} A_{\s \t} \hA_{\s \t}
+ \frac{1}{12} \sum_{\s \t \g} B_{\s \t \g} \hB_{\s \t \g}
- \frac{1}{24} \sum_{\s \t \g \b} C_{\s \t \g \b} \hC_{\s \t \g \b} 
\label{eq_Potts_phi} \\
&& + \frac{1}{4} \tE[\ell] \sum_{\s \t} \oeta_\s \oeta_\t (\hA_{\s \t})^2
- \frac{1}{12} \tE[\ell] \sum_{\s \t \g} \oeta_\s \oeta_\t \oeta_\g (\hB_{\s \t \g})^2
+ \frac{1}{24} \tE[\ell] \sum_{\s \t \g \b} \oeta_\s \oeta_\t \oeta_\g \oeta_\b (\hC_{\s \t \g \b})^2 \nonumber \\
&& + \frac{1}{12} \tE[\ell(\ell-1)] \left[\sum_{\s \t} \oeta_\s \oeta_\t (\hA_{\s \t})^3 - 2 \sum_{\s \t \g} \oeta_\s \oeta_\t \oeta_\g \hA_{\s \t} \hA_{\t \g} \hA_{\g \s} \right] \nonumber \\
&& - \frac{1}{2} \tE[\ell(\ell-1)] \sum_{\s \t \g} \oeta_\s \oeta_\t \oeta_\g \hB_{\s \t \g} \hA_{\s \t} \hA_{\s \g}
+ \frac{1}{4} \tE[\ell(\ell-1)] \sum_{\s \t \g \b} \oeta_\s \oeta_\t \oeta_\g \oeta_\b \hC_{\s \t \g \b} \hA_{\s \t} \hA_{\g \b} \nonumber \\
&& + \frac{1}{48} \tE[\ell(\ell-1)(\ell-2)] \left[
\sum_{\s \t} \oeta_\s \oeta_\t (\hA_{\s \t})^4
- 6 \sum_{\s \t \g} \oeta_\s \oeta_\t \oeta_\g (\hA_{\s \t})^2 (\hA_{\s \g})^2
-12  \sum_{\s \t \g} \oeta_\s \oeta_\t \oeta_\g (\hA_{\s \t})^2 \hA_{\s \g} \hA_{\t \g} \right. \nonumber \\ && \left. \hspace{4cm}
+ 3 \left(\sum_{\s \t} \oeta_\s \oeta_\t (\hA_{\s \t})^2\right)^2
+ 12 \sum_{\s \t \g \b} \oeta_\s \oeta_\t \oeta_\g \oeta_\b \hA_{\s \t} \hA_{\t \g} \hA_{\g \b} \hA_{\b \s} \right] \ , \nonumber
\eea
where of course this expression is a truncation of an infinite series involving moments of all orders. Let us make a few comments on this result:
\begin{itemize}
\item One can rewrite this expansion in the eigenbasis of $M$ (i.e. in terms of the quantities $A'_{jk}$ instead of $A_{\s \t}$, as defined in (\ref{eq_Potts_def_Ap})), the corresponding expression is given in Appendix~\ref{sec_app_expansion_Potts_phi}.

\item The expression (\ref{eq_Potts_phi_uncond}) of the free-entropy $\phi(P)$ is variational, in the sense that its variation with respect to $P$ vanishes when $P$ is a fixed point solution of the recursion equation (\ref{eq_Potts_recursion_P}). Accordingly, the derivatives of the truncated expansion (\ref{eq_Potts_phi}) with respect to $A$, $B$ and $C$ vanish when the latter are solutions of the moment equations (\ref{eq_Potts_cavity_A}-\ref{eq_Potts_cavity_hC}).

\item If the centered moments of $P$ verify the scaling ansatz (\ref{eq_ansatz}) that we argued was the correct one for fixed points of the cavity recursion, then the expansion (\ref{eq_Potts_phi}) contains all the terms of order $\kappa^2$, $\kappa^3$ and $\kappa^4$.

\item However, the free-entropy functional $\phi(P)$ of equation (\ref{eq_Potts_phi_uncond}) is well-defined even if $P$ is not a fixed point of the recursion equation (\ref{eq_Potts_recursion_P}) (as long as it satisfies $\int \dd P(\eta) \, \eta = \oeta$), and accordingly the expansion (\ref{eq_Potts_phi}) can be used even if $A$, $B$ and $C$ do not satisfy (\ref{eq_Potts_cavity_A}-\ref{eq_Potts_cavity_hC}). As a matter of fact, such an expansion was performed in~\cite{CoEfJaKaKa17}, let us briefly mention the similarities and differences with respect to ours. In the context of the models studied in this section, one of the results of~\cite{CoEfJaKaKa17} (i.e. their theorem 2.3) can be rephrased as: above the Kesten-Stigum transition, i.e. whenever $\tE[\ell]\theta_2^2 > 1$ in such a way that the trivial fixed point is unstable under the cavity recursions, there exists a distribution $P$ with $\phi(P)>0$, hence the Kesten-Stigum transition is an upper-bound for the information-theoretic transition (or condensation transition in the associated non-planted random ensemle). Let us recover this statement from our computations, defining the probability distributions $P_\lambda$ by
\beq
P_\lambda(\eta) = \frac{1}{2} \, \delta(\eta - (\oeta + \lambda \, l^{(2)}))
+ \frac{1}{2} \, \delta(\eta - (\oeta - \lambda \, l^{(2)})) \ ,
\eeq
where $l^{(2)}$ is the left eigenvector of $M$ associated to the eigenvalue $\theta_2$, and $\lambda >0$ should be sufficiently small for $\oeta \pm \lambda \, l^{(2)}$ to remain inside the polytope of probability distributions. The average of $P_\lambda$ is $\oeta$ for all $\lambda$, and the first centered moments of $P_\lambda$ are easily seen to be
\beq
A_{\s \t} = \lambda^2 l^{(2)}_\s l^{(2)}_\t \ , \quad 
B_{\s \t \g} =0 \ , \quad
C_{\s \t \g \b} = \lambda^4 l^{(2)}_\s l^{(2)}_\t l^{(2)}_\g l^{(2)}_\b \ .
\eeq
More generically the odd centered moments vanish and the $2p$-th one is $\lambda^{2p}$ times the $2p$-th tensor power of $l^{(2)}$. Computing the corresponding value of $\hA_{\s \t}$ and inserting in (\ref{eq_Potts_phi}) one obtains
\beq
\frac{1}{\E[\ell]} \phi(P_\lambda) = \frac{1}{4} \theta_2^2 (\tE[\ell]\theta_2^2 - 1) \lambda^4 + O(\lambda^6) \ , 
\eeq
which can indeed be made strictly positive as soon as $\tE[\ell]\theta_2^2 > 1$ by taking a sufficiently small $\lambda$. Note that the centered moments of $P_\lambda$ do not satisfy our ansatz (\ref{eq_ansatz}), which is not contradictory as $P_\lambda$ has no reason to be a fixed point of the cavity recursion (incidentally, there might be no perturbative non-trivial fixed point in such a situation, see the right panel of Fig.~\ref{fig:standard}). Only two terms of (\ref{eq_Potts_phi}) contribute to $\phi(P_\lambda)$ at order $\lambda^4$ (the first ones in the first and second line), hence our expansion is more detailed than the one in~\cite{CoEfJaKaKa17}, which on the other hand deals with a much more general and complicated setting ($k$-wise interaction of Potts variables with quenched disorder) and is performed in a mathematically rigorous way.

\end{itemize}

\subsection{Large degree limit}
\label{sec_large_degree}

We have underlined in the introduction of this paper the existence of two families of inference problems, the dense ones which can be described by a finite-dimensional order parameter, and the sparse ones that require the use of a distributional, infinite-dimensional, description. The focus of this paper is on the latter family, but it is important to realize that the two types of problems are not completely disjoint: as a matter of fact the sparse models reduce effectively to the simpler dense ones if one takes the limit of large degrees, after the thermodynamic limit. This is a well-known fact that has been derived rigorously in several papers (see for instance~\cite{Sly11,CaLeMi16,LeMi16}), for the sake of self-containedness and to be able to contrast the behavior of the sparse and dense models we reproduce here this derivation. 

We shall consider the recursion for the unconditional distribution (\ref{eq_Potts_recursion_P}) and simplify it in the large degree limit $\tE[\ell] \to \infty$. We will show that in this limit the evolution of the distribution $P^{(n)}$ can be tracked exactly by following a finite-dimensional object, namely the covariance matrix $A^{(n)}_{\s \t} = \E^{(n)}[\delta_\s \delta_\t]$, that obeys a recursion equation of the form $A^{(n+1)}=F(A^{(n)})$.

The main idea of the derivation is to exploit a central limit theorem, we shall thus rewrite (\ref{eq_Potts_recursion_P}) in a way that makes apparent a sum of a large number of random variables. To reach this goal we will define two functions $\psi(L)$ and $\tz(L)$ whose arguments are $q$-dimensional real vectors $L=(L_1,\dots,L_q)$, that are mapped by $\psi$ and $\tz$ to normalized probability distributions and positive reals, respectively, according to
\beq
\psi(L)_\s = \frac{e^{L_\s}}{\tz(L)} \ , \qquad
\tz(L) = \sum_{\gamma=1}^q e^{L_\gamma} \ .
\eeq
One can translate the unconditional recursion relation (\ref{eq_Potts_recursion_P}) as
\beq
\E^{(n+1)}[g(\eta)] = \E^{(n)}[g(\psi(L)) \tz(L)] \ ,
\eeq
for an arbitrary function $g$, where in the right hand side the vector $L$ has the distribution:
\beq
(L)_{\s=1,\dots,q} \eqd  (\ln z_\s(\eta^1,\dots,\eta^\ell) ) _{\s=1,\dots,q} \ ,
\eeq
with $\ell$ drawn from $\tp_\ell$ and the $\eta^i$'s drawn from $P^{(n)}$. From the expression of $z_\s$ given in (\ref{eq_BP}) one realizes that
\beq
L_\s = \ln \oeta_\s + \sum_{i=1}^\ell \ln \left(\sum_{\s'} \hM_{\s \s'} \eta^i_{\s'} \right) \ .
\eeq
We now recall that if $Y\eqd \sum_{i=1}^n X^i$ is a sum of i.i.d. random vectors, the number of summands $n$ being also random, the two first moments of the sum are given by
\bea
(\E[Y])_\s &=& \E[n] \E[X_\s] \ , \\
(\text{Cov} \, [Y])_{\s \t} &=& \E[n] \E[X_\s X_\t] + (\text{Var} \, [n] - \E[n])  \E[X_\s] \E[X_\t] \ . \label{eq_largedegree_cov}
\eea
Moreover, the large degree limit $\tE[\ell] \to \infty$ yields a non-trivial result only if the non-trivial eigenvalues of $M$, $\theta_2, \dots,\theta_q$ vanish in the limit. More precisely, they have to be of the form $\theta_i = \ttheta_i / \sqrt{\tE[\ell]}$ with $\ttheta_i$ finite. We thus have
\bea
X_\s &=& \ln \left(\sum_{\s'} \hM_{\s \s'} \eta_{\s'} \right) =
\ln\left(1+ \frac{1}{\sqrt{\tE[\ell]}} \sum_{j=2}^q \ttheta_j r_\s^{(j)}\sum_{\s'} r_{\s'}^{(j)} \delta_{\s'} \right) \\
&=& \frac{1}{\sqrt{\tE[\ell]}} \sum_{j=2}^q \ttheta_j r_\s^{(j)}\sum_{\s'} r_{\s'}^{(j)} \delta_{\s'} - \frac{1}{2} \frac{1}{\tE[\ell]} \left(\sum_{j=2}^q \ttheta_j r_\s^{(j)}\sum_{\s'} r_{\s'}^{(j)} \delta_{\s'}\right)^2 + o\left(\frac{1}{\tE[\ell]} \right) \ .
\eea
Combining these various observations, and assuming that the offspring degree distribution is sufficiently well-behaved for the second term in (\ref{eq_largedegree_cov}) to be negligible (it vanishes exactly for a Poisson distribution), we find that in the large degree limit the promised recursion $A^{(n+1)}=F(A^{(n)})$ can be decomposed in two steps:
\begin{itemize}
\item from $A^{(n)}$ compute $\tA^{(n)}$ with
\beq
\tA^{(n)}_{\s \t} = \lim \left( \tE[\ell] \sum_{\s' \t'} \hM_{\s \s'} \hM_{\t \t'}  A^{(n)}_{\s' \t'} \right) = \sum_{j,k=2}^q \ttheta_j \ttheta_k r_\s^{(j)} r_\t^{(k)}  \sum_{\s' \t'} r_{\s'}^{(j)}r_{\t'}^{(k)} A^{(n)}_{\s' \t'} \ .
\label{eq_largedegree_tA}
\eeq
\item then compute the new covariance matrix $A^{(n+1)}$ from
\beq
A^{(n+1)}_{\s \t} = \E\left[ (\psi(L)_\s - \oeta_\s)(\psi(L)_\t - \oeta_\t) \tz(L)\right]=
\E\left[ \frac{(e^{L_\s} - \oeta_\s \tz(L))(e^{L_\t} - \oeta_\t \tz(L))}{\tz(L)}
\right] \ ,
\label{eq_largedegree_recursion}
\eeq
with $L$ a Gaussian vector characterized by its first two moments
\bea
(\E[L])_\s &=& \ln \oeta_\s - \frac{1}{2} \tA^{(n)}_{\s \s} \ , \nonumber \\
(\text{Cov} \, [L])_{\s \t} &=& \tA^{(n)}_{\s \t} \ .
\label{eq_largedegree_L}
\eea
\end{itemize}

As emphasized before this finite-dimensional recursion $A^{(n+1)}=F(A^{(n)})$ is an exact description of the functional recursion on $P^{(n)}$ in the large degree limit. The trivial covariance $A=0$ is a fixed point of $F$, that reproduces the stationarity of $P(\eta)=\delta(\eta - \oeta)$ under the functional recursion. It is instructive to expand the equation $A=F(A)$ around $A=0$, in order to compare this expansion with the one we performed previously on the full functional equation. Exploiting the properties of Gaussian random variables it is relatively easy to obtain from (\ref{eq_largedegree_recursion}):
\bea
A_{\s \t} &=&
\oeta_\s \oeta_\t \tA_{\s \t}
+ \frac{1}{2} \oeta_\s \oeta_\t \left[ 
(\tA_{\s \t})^2 
- \sum_\g \oeta_\g (\tA_{\s \g} + \tA_{\g \t})^2
+  \sum_{\g \b} \oeta_\g \oeta_\b (\tA_{\g \b})^2
\right] \\
&+& \oeta_\s \oeta_\t \left[
\frac{1}{6} (\tA_{\s \t})^3 - \frac{1}{6} \sum_\g \oeta_\g (\tA_{\s \g} + \tA_{\t \g})^3 - \frac{1}{2}  \tA_{\s \t} \sum_\g \oeta_\g (\tA_{\s \g} + \tA_{ \t \g})^2 + \frac{1}{6} \sum_{\g \b} \oeta_\g \oeta_\b (\tA_{\g \b})^3 \right. 
 \nonumber \\ && \hspace{1cm} \left. 
+ \frac{1}{2}  \sum_{\g \b} \oeta_\g \oeta_\b (\tA_{\g \b})^2
(\tA_{\s \t} + \tA_{\s \g} + \tA_{\t \g} + \tA_{\s \b} + \tA_{\t \b} )
+  \sum_{\g \b} \oeta_\g \oeta_\b \tA_{\g \b}
(\tA_{\s \g} + \tA_{\t \g})(\tA_{\s \b} + \tA_{\t \b} )
\right. \nonumber \\ && \hspace{1cm} \left.
- \sum_{\g \b \a} \oeta_\g \oeta_\b \oeta_\a \tA_{\g \b} \tA_{\b \a} \tA_{\a \g} \right] + O(A^4) \ .
\eea
This is, at it should, consistent with the expansion performed in the sparse case and presented in Eq.~(\ref{eq_Potts_cavity_A}). Of course the equations in the sparse regime are richer, the effects of the third and fourth order moments $B$ and $C$ being washed out in the large degree limit by the Gaussian character of the distribution.

For completeness let us also state the large degree limit for the free-entropy, that becomes a function of a covariance matrix $A$,
\beq
\phi(A) = \E[\tz(L) \ln \tz(L)] - \frac{1}{4} \sum_{\s \t} A_{\s \t} \tA_{\s \t} \ ,
\eeq
where $\tA$ is obtained from $A$ according to (\ref{eq_largedegree_tA}), and in the first term $L$ is a Gaussian vector with the two first moments indicated in (\ref{eq_largedegree_L}).

In the remaining of this section we shall investigate the consequences
of our expansions, that were performed for an arbitrary choice of $q$,
$\oeta$ and $M$, in several specific cases. In particular we have to
analyze the bifurcation of the quadratic equation
(\ref{eq_Potts_order2}) on the covariance matrix $A$ in the
neighborhood of the Kesten-Stigum transition, and use it to
discriminate between the possible scenarios depicted in
Fig.~\ref{fig:standard} and \ref{fig:new} by imposing the positive definiteness of $A$.

\subsection{Application 1: the non-degenerate case}
\label{sec_applications_nondegenerate}

Let us first consider the case of a simple second eigenvalue of $M$ at the Kesten-Stigum transition. More explicitly, we consider that the various parameters of the model (degree distribution, $\oeta$, $M$) are functions of a single control parameter $\epsilon$, defined in such a way that $\tE[\ell] \theta_2^2 = 1+ \epsilon$, with $\epsilon=0$ at the Kesten-Stigum transition, and we assume that $\tE[\ell] \theta_j^2$ are bounded away from $1$ for $j=3,\dots,q$ in a neighborhood of $\epsilon=0$.

Consider now the set of quadratic equations (\ref{eq_Potts_order2_Ap}) on $A'$, the covariance matrix expressed in the basis of eigenvectors of $M$. With the assumption of simplicity of $\theta_2$ the only coefficient of the linear terms in the right hand side that crosses 1 at the Kesten-Stigum transition corresponds to $j=k=2$; the bifurcating solution will thus have $|A'_{22}| \gg |A'_{jk}|$ for $(j,k) \neq (2,2)$ in the neighborhood of $\epsilon=0$. We can thus close the equation on this leading term as
\beq
A'_{22} = \tE[\ell] \theta_2^2 A'_{22}
+ \frac{1}{2} \tE[\ell(\ell-1)] \theta_2^4 (A'_{22})^2 ((f_{222})^2-2) \ ,
\label{eq_Ap_nondegenerate}
\eeq
where the expression of $f_{222}$ is given in (\ref{eq_def_tensorf}). At the leading order in $\epsilon$ we have
\beq
0 = \epsilon + \frac{1}{2} \tE[\ell(\ell-1)] \theta_2^4 ((f_{222})^2-2) A'_{22} \ ,
\eeq
hence $A'_{22}$ varies linearly with $\epsilon$ around $\epsilon=0$, the other matrix elements of $A'$ being at least of order $\epsilon^2$. The crucial point is now to remember that $A'$, as a covariance matrix, must be positive definite, and as a consequence $A'_{22}$ must be a positive real. Depending on the sign of $(f_{222})^2-2$ this happens for $\epsilon > 0$ (resp. $\epsilon < 0$), corresponding for instance to the scenario sketched on the left (resp. right) of Fig.~\ref{fig:standard}. Spelling out the definition of $f_{222}$ in terms of the eigenvectors of $M$, we have obtained that the first case occurs when
\beq
\left(\sum_\s \oeta_\s (r_\s^{(2)})^3\right)^2 < 2 \ .
\label{eq_criterion_nondegenerate}
\eeq

Let us underline the striking similarity between this criterion and the one obtained in~\cite{LeKrZd17} in the context of matrix factorization. This is a dense inference problem with continuous variables, for which the condition of existence of a bifurcating solution above the Kesten-Stigum transition was found to be $\la x_0^3 \ra^2 < 2 \la x_0^2 \ra^3$, the average being here on the prior distribution of a real valued variable $x_0$ (see equation (205) in~\cite{LeKrZd17}). Using the properties of the eigenvalue decomposition of $M$ explained in Sec.~\ref{sec_Potts_diagM} one can rewrite (\ref{eq_criterion_nondegenerate}) as
\beq
\left(\sum_\s \oeta_\s (r_\s^{(2)})^3\right)^2 < 2 \left(\sum_\s \oeta_\s (r_\s^{(2)})^2\right)^3 \ ,
\eeq
a formally equivalent expression with $\oeta_\s$ replacing the prior distribution and the eigenvector $r_\s^{(2)}$ the continuous variable $x_0$.

\subsection{Application 2: the symmetric $q$ state case}
\label{sec_applications_symPotts}

We turn now to a case which can be viewed as the opposite of the previous one, with a maximal degeneracy of the second eigenvalue of $M$. Consider indeed the symmetric $q$ state Potts model, with a stationary distribution $\oeta_\s = \frac{1}{q}$ for all $\s$, and a matrix $M$ which is invariant under all permutations of its rows and columns. As $M$ is stochastic it must be of the form
\beq
M_{\s \t} = \frac{1}{q} + \theta \, K_{\s \t} \ , \qquad
\text{with} \ \ \  K_{\s \t} = \delta_{\s,\t} - \frac{1}{q} \ ;
\label{eq_Potts_sym_def}
\eeq
$\theta$ is the non-trivial eigenvalue of $M$, with a degeneracy $q-1$. The matrix elements of $M$ must be non-negative, the authorized range of $\theta$ is thus $\left]-\frac{1}{q-1},1\right[$. The model is said assortative (or ferromagnetic) if $\theta>0$ and disassortative (or antiferromagnetic) if $\theta<0$. The matrix elements of $\hM$ are $\hM_{\s \t}=1 + q\theta K_{\s \t}$.

The recursion equation (\ref{eq_Potts_recursion_P}), for this choice of $\oeta$ and $M$, will respect the permutation symmetry between the $q$ values of $\s$. If the initial condition is symmetric this will also be the case for $P^{(n)}$, for all $n$. In particular the covariance matrix $A^{(n)}$ will be invariant under all row and column permutations; as its row sums must also vanish it is necessarily of the form $A^{(n)} = a^{(n)} K$, with $a^{(n)}$ a real number and $K$ the matrix defined in (\ref{eq_Potts_sym_def}). The eigenvalues of $K$ being 0 and $1$, the positivity constraint on $A^{(n)}$ thus translates into $a^{(n)} \ge 0$. One finds easily that with this form of $A$ and $M$ that $\hA^{(n)} = (q \theta)^2 a^{(n)} K$. Plugging these forms of $A$ and $\hA$ in the linearized evolution equation (\ref{eq_Potts_linear_A}) yields $a^{(n+1)} = \tE[\ell] \theta^2 a^{(n)}$, hence the Kesten-Stigum condition $\tE[\ell] \theta^2=1$ for the limit of stability of the trivial solution $a=0$, as expected.

Looking for a fixed point solution of the form $A_{\s \t} = a K_{\s \t}$ in the neighborhood of the Kesten-Stigum transition, we obtain from (\ref{eq_Potts_order2}) after a brief computation:
\beq
a = \tE[\ell] \theta^2 a +  
\tE[\ell (\ell-1)] \theta^4 \frac{1}{2} q(q-4) a^2 \ .
\label{eq_Potts_sym_a}
\eeq
Denoting $\tE[\ell] \theta^2 = 1+\epsilon$, and keeping in mind the crucial positivity condition $a \ge 0$, one realizes that for $q<4$ the non-trivial perturbative fixed point exists for $\epsilon >0$, whereas if $q > 4$ it exists for $\epsilon <0$, which implies that the Kesten-Stigum bound is not tight for the tree reconstruction problem in the latter case. This fact was proven rigorously in~\cite{Sly11}, the above equation corresponding to (1.1) in~\cite{Sly11} (with a linear change of variable between our $a^{(n)}$ and $x_n$ of~\cite{Sly11}). Let us emphasize the necessity of imposing the permutation symmetry of $A$ in order to reach this conclusion; the bifurcation equations (\ref{eq_Potts_order2}) admit indeed, besides the symmetric solution studied above, non-symmetric spurious solutions $A$ even when $\oeta$ and $M$ are invariant under permutations.

We have thus justified the statements made in Sec.~\ref{sec_mainresults_sbm} in the cases $q\le 3$ and $q \ge 5$. When $q=4$ the coefficient of $a^2$ in the bifurcation equation vanishes exactly and one cannot conclude from this lowest-order expansion on the type of transition encountered at the Kesten-Stigum threshold. Thanks to the next order expansion presented in Sec.~\ref{sec_Potts_third_order} we shall now be able to elucidate this case, and justify the efforts put in this generalization of the expansion. In order to solve the equations (\ref{eq_Potts_cavity_A}-\ref{eq_Potts_cavity_hC}) on $A$, $B$ and $C$, we first note that $B$ and $C$ vanish when summed over any of their indices; from the expression of $\hM$ in the case under study we can thus conclude that $\hB_{\s \t \g} = (q \theta)^3 B_{\s \t \g}$ and $\hC_{\s \t \g \b} = (q \theta)^4 C_{\s \t \g \b}$. Inserting these expressions in (\ref{eq_Potts_cavity_B},\ref{eq_Potts_cavity_C}) yields an explicit solution for $B$ and $C$ in terms of $\hA$,
\bea
B_{\s \t \g} &=& \frac{\tE[\ell (\ell-1)]}{1-\tE[\ell]\theta^3} \frac{1}{q^3} 
\left[
\hA_{\s \t} \hA_{\t \g} + \hA_{\s \g} \hA_{\g \t} + \hA_{\t \s} \hA_{\s \g}  
- \frac{1}{q} \sum_\b  (\hA_{\s \b} \hA_{\b \g} + \hA_{\s \b} \hA_{\b \t} + \hA_{\t \b} \hA_{\b \g}  )
\right] \ , \\
C_{\s \t \g \b} &=& \frac{\tE[\ell (\ell-1)]}{1-\tE[\ell]\theta^4} \frac{1}{q^4}
(\hA_{\s \t} \hA_{\g \b} + \hA_{\s \g} \hA_{\t \b} + \hA_{\s \b} \hA_{\t \g}) 
\ .
\eea
Furthermore we have argued above by symmetry arguments that $A=a K$ and $\hA = a (q \theta)^2 K$, we can thus determine completely $B$ and $C$ modulo the still unknown real $a$. Plugging these expressions in (\ref{eq_Potts_cavity_A}) gives, after some combinatorial evaluations, the following equation on $a$ that generalizes (\ref{eq_Potts_sym_a}):
\beq
a = \tE[\ell] \theta^2 a +  
\tE[\ell (\ell-1)] \theta^4 \frac{1}{2} q(q-4) a^2 + w \, a ^3 \ ,
\label{eq_Potts_sym_a_third}
\eeq
where the coefficient $w$ reads
\beq
w= \frac{1}{6} q^2(q^2 -18 q +42) \tE[\ell(\ell-1)(\ell-2)] \theta^6
- \frac{6 q^2(q-2) \tE[\ell(\ell-1)]^2 \theta^9}{1 - \tE[\ell]\theta^3}
+ \frac{q^2(q+1) \tE[\ell(\ell-1)]^2 \theta^{10}}{1 - \tE[\ell]\theta^4} \ .
\eeq
In the most interesting case $q=4$ where the coefficient of $a^2$ vanishes, one finds for $w$ (substituting $\theta = \sign(\theta)/\sqrt{\tE[\ell]}$ at lowest order around the Kesten-Stigum transition):
\beq
w=16 \left(-\frac{7}{3} \frac{\tE[\ell(\ell-1)(\ell-2)]}{\tE[\ell]^3} 
+ \left(\frac{\tE[\ell(\ell-1)]}{\tE[\ell]^2} \right)^2
\left( \frac{5}{\tE[\ell]-1} - \frac{12}{\sign(\theta) \sqrt{\tE[\ell]}-1}
\right)
\right) \ ,
\label{eq_wq4}
\eeq
which justifies the expression (\ref{eq:4groups}) announced in Sec.~\ref{sec_mainresults_sbm} (the factor 16 between (\ref{eq:4groups}) and (\ref{eq_wq4}) is irrelevant for the study of the sign of $w$). Depending on the degree distribution $\tp_\ell$ and on the sign of $\theta$ this coefficient can be positive or negative; as the equation on $a$ reduces here to $0=\epsilon + w \, a^2$, with $\epsilon = \tE[\ell]\theta^2 -1$, the sign of $w$ controls the sign of $\epsilon$ for which a non-trivial solution of the cavity equations exists in a neighborhood of the trivial one, hence the type of bifurcation scenario as sketched in Fig.~\ref{fig:standard}. As mentioned in Sec.~\ref{sec_mainresults_sbm} one has $w<0$ in the ferromagnetic case for all degree distributions, which yields a second order transition; it is indeed easy to check numerically that
\beq
\frac{5}{d-1} - \frac{12}{\sqrt{d}-1} < 0 
\eeq
for all average offspring degrees $d=\tE[\ell]>1$.

In the antiferromagnetic case the sign of $w$ depends on the degree distribution. For instance when $\tp_\ell$ is a Poisson law of parameter $d$ the expression of $w$ above can be simplified to
\beq
w=16\left(-\frac{7}{3} + \frac{5}{d-1} + \frac{12}{\sqrt{d}+1} \right) \ .
\label{eq_wq4_Poisson}
\eeq
A numerical study of this function reveals the existence of a critical degree $d_{\rm c}\approx 22.2694$ such that $w>0$ for $d \in (1,d_{\rm c})$, hence a discontinuous bifurcation scenario similar to the case $q>4$, while for $d>d_{\rm c}$ one has $w<0$, yielding the scenario of $q<4$. We have also evaluated the sign of $w$ in the regular case, i.e. for an offspring degree distribution $\tp_\ell=\delta_{\ell ,d}$. Similarly to the Poisson case one finds (for the antiferromagnetic model) that $w<0$ for $d \le 24$ and $w>0$ for $d \ge 25$.

Note that for large enough degrees ferromagnetic and antiferromagnetic models behave in the same way; this could be anticipated from the study of the large degree limit recalled in Sec.~\ref{sec_large_degree}, as in this limit the degree distribution and $\theta$ only appear in the combination $\tE[\ell]\theta^2$, which is clearly independent of the sign of $\theta$. Actually an expansion to the same order than in (\ref{eq_Potts_sym_a_third}) can be found in equation (4.9) of~\cite{Sly11}, but in the large degree limit and not in the sparse regime, which did not allow to deduce the modification of the type of phase transition according to the degree distribution.

Let us finally justify the scaling exponent we used in the plot presented in the right panel of Fig.~\ref{fig:4groups}. If we had included in the expansion of the cavity equations around the trivial fixed point one more term, one would have found, for $q=4$, an equation on $a$ of the form
\beq
0 = \epsilon + w(\tp) \, a^2 + x \, a^3 \ ,
\eeq
where a priori $x \neq 0$, and we emphasized the dependency of $w$ on the degree distribution. Let us consider for concreteness $\tp$ to be Poissonian with average $d$, in such a way that $w(d)$ is given explicitly by (\ref{eq_wq4_Poisson}). In the neighborhood of the critical degree $d_{\rm c}$ where $w$ changes sign we have $w(d) \approx w' (d-d_{\rm c})$, with $w'$ some constant. Inserting this form in the equation above gives
\beq
0 = \epsilon + w' (d-d_{\rm c}) \, a^2 + x \, a^3 \ ;
\label{eq_q4_scaling}
\eeq
the scaling of $\theta_{\rm sp}$, which corresponds to the limit of existence of a solution of this equation, must be such that the three terms in (\ref{eq_q4_scaling}) are of the same order. This is easily seen to imply that $\epsilon$ is of the order of $(d-d_{\rm c})^3$, which is the reason of our choice of exponent in the right panel of Fig.~\ref{fig:4groups}: the spinodal and IT curves should, in this rescaled units, behave linearly in the neighborhood of $d_{\rm c}$.

\subsection{Application 3: the asymmetric Ising ($q=2$) case}
\label{sec_Ising_asym}

We consider now the case of binary variables ($q=2$), that for convenience we denote as Ising spins or sign variables, $\s\in\{-1,1\}=\{-,+\}$. We parametrize the stationary probability distribution $\oeta$ as $\oeta_\s=\frac{1+\s \om}{2}$, with $\om\in[-1,1]$. This bias parameter can be interpreted as a magnetization in the perspective of a physical Ising model~\cite{KS_tight_Ising}, or in the context of the SBM as controlling the relative size of the two communities of vertices which are $\oeta_+=\frac{1+\om}{2}$ and $\oeta_-=\frac{1-\om}{2}$. The symmetric case, as studied in Sec.~\ref{sec_applications_symPotts}, is recovered for $\om=0$. We write the matrix $M$ as
\beq
M = \begin{pmatrix} M_{++} & M_{+-} \\ M_{-+} & M_{--} \end{pmatrix} 
= \frac{1}{2}
\begin{pmatrix} 
1+\om & 1-\om \\ 
1+\om & 1-\om \end{pmatrix} + \theta \frac{1}{2}
\begin{pmatrix} 
1-\om    & -1+\om \\ 
-1 -\om& 1+\om
\end{pmatrix} \ , \quad \text{i.e.} \ \ \ \ 
M_{\s \t} = \frac{1+\t \om}{2} + \s \t \theta \frac{1-\s \om}{2} \ .
\eeq
This is indeed the only Markov matrix reversible with respect to
$\oeta$ with second eigenvalue $\theta$. To connect with the
definition of the SBM in section \ref{sec_definitions_SBM} the
parameter $\theta$ is defined as $\theta=1-c_{-+}/d$ where $c_{-+}=c_{+-}$ is
$N$ times the probability that nodes in different groups are
connected, and $d$ is the average degree. Given the group sizes $\oeta_\pm$, the condition of equal
average degree in the two groups then implies uniquely the
parameters $c_{--}$ and $c_{++}$. In the context of the SBM the degree distributions $p_\ell$ and $\tp_\ell$ are Poissonian with average $d$, we shall, however, consider generic degree distributions, as they are relevant in the tree reconstruction perspective.

The range of allowed values of the parameters $(m,\theta)$ is a subset of $[-1,1]^2$: imposing the non-negativity of the matrix elements of $M$ implies indeed that
\beq
\theta \ge -\frac{1-|\om|}{1+|\om|} \ ,
\label{eq_Ising_bound_theta}
\eeq
which is always fulfilled for $\theta \ge 0$ but restrict the range of allowed biases $\om$ when $\theta<0$.

The transformed version $\hM$ of $M$ defined in (\ref{eq_Potts_defhM}) reads thus:
\beq
\hM_{\s \t} = M_{\s \t} \frac{1}{\oeta_\t} = 1 + \theta \s \t \frac{1 - \s \om}{1+\t \om } \ .
\label{eq_Ising_hM}
\eeq

Thanks to the binary nature of the variables the probability laws $\eta$ can be parametrized by a single real, and the recursion equation~(\ref{eq_BP}) can be rewritten in a simpler form with this parametrization. This expression, along with some more technical details on the numerical resolution of the cavity equations, can be found in Appendix~\ref{app_numerical_resolution}.

\subsubsection{The critical asymmetry}

Let us now apply the generic results of the moment expansions presented in Sec.~\ref{sec_expansion_Potts} to this specific case. Because of the normalization condition $\delta_+ + \delta_- =0$ the deviation $\delta_\s=\eta_\s - \oeta_\s$ is equal to $\s$ multiplied by a scalar random variable. We can thus write the matrix $A^{(n)}_{\s \t}=\E^{(n)}[\delta_\s \delta_\t]$ as $A^{(n)}_{\s \t}= a^{(n)} \frac{1}{2} \s \t$, where $a^{(n)}$ is a scalar; the factor $1/2$ is chosen in such a way that $\frac{1}{2} \s \t$ coincides with the matrix $K_{\s \t}$ of (\ref{eq_Potts_sym_def}), in the case $q=2$. Once again we have a positivity requirement on this covariance, $a^{(n)}\ge 0$. From the expression of $\hM$ given in (\ref{eq_Ising_hM}) one sees that
\beq
\sum_{\s \s'} \hM_{\s \s'} \s' = \theta \frac{1}{\oeta_\s} \s \ ,
\eeq
hence with the relation (\ref{eq_Potts_defhA}) one obtains
\beq
\hA^{(n)}_{\s \t} = \theta^2 a^{(n)} \frac{1}{\oeta_\s \oeta_\t} \frac{1}{2}\s \t \ .
\eeq
The linear evolution of (\ref{eq_Potts_linear_A}) thus becomes
\beq
a^{(n+1)} \frac{1}{2} \s \t  = \tE[\ell] \theta^2 a^{(n)} \frac{1}{2} \s \t \ ,
\qquad \text{i.e.} \ \ \ a^{(n+1)} = \tE[\ell] \theta^2 a^{(n)} \ ,
\eeq
which reproduces the expected Kesten-Stigum threshold at $\tE[\ell] \theta^2=1$. 

We now look for a fixed-point solution in the neighborhood of the Kesten-Stigum transition, by injecting the above forms of $A$ and $\hA$ in equation (\ref{eq_Potts_order2}). This yields after a short computation
\beq
a = \tE[\ell] \theta^2 a
+ \tE[\ell (\ell-1)] \theta^4 \frac{2 (3 \om^2 -1) }{(1-\om^2)^2} a^2 \ .
\label{eq_Ising_lowest}
\eeq
A simple consistency check is provided by the coincidence of (\ref{eq_Potts_sym_a}) and (\ref{eq_Ising_lowest}) when $q=2$ and $\om=0$. Moreover when $q=2$ the non-trivial eigenvalue of $M$ is unique, hence certainly non-degenerate; one can check that (\ref{eq_Ising_lowest}) is also a consequence of the equation (\ref{eq_Ap_nondegenerate}) derived in Sec.~\ref{sec_applications_nondegenerate} under this non-degeneracy assumption, for all $\om$.

The crucial property of (\ref{eq_Ising_lowest}) we would like to emphasize is the change of sign of the coefficient of $a^2$ depending on whether $|\om|$ is larger or smaller than $\om_{\rm c}=1/\sqrt{3}$. Because of the condition $a \ge 0$ this implies that the non-trivial perturbative solution exists when $\tE[\ell] \theta^2 >1$ at small asymmetry ($|\om| < \om_{\rm c}$), and when $\tE[\ell] \theta^2  < 1$ at large asymmetry ($\om > \om_{\rm c}$). In the latter case one has, in the tree reconstruction language, a reconstructible phase below the Kesten-Stigum threshold, which is not tight in this case. Quite strikingly the critical asymmetry $\om_{\rm c}$ does not depend on the degree distribution; this explains why this value $1/\sqrt{3}$ was also obtained previously in the large degree limit~\cite{CaLeMi16,LeMi16} and in related dense inference problems~\cite{BaDiMaKrLeZd16,LeKrZd17}. Very recently this critical asymmetry was also discovered for the reconstruction of the Ising model on regular trees~\cite{liu2017reconstruction}; this paper proved rigorously the non-tightness of the Kesten-Stigum bound for $\om > \om_{\rm c}$, and its tightness for $\om < \om_{\rm c}$ at large enough degrees.

\begin{figure}
\includegraphics[width=8cm]{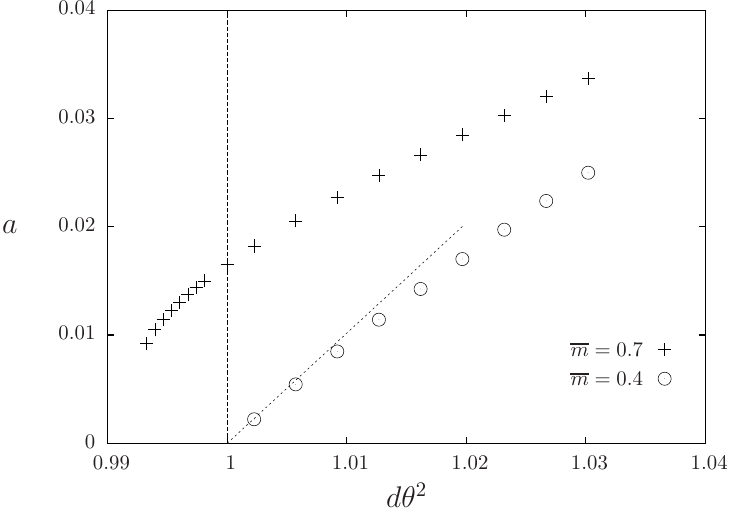}
\caption{The value of $a$ in the fixed-point solution of the cavity equations reached from an informative initial condition, as a function of the signal-to-noise ratio, for the ferromagnetic ($\theta>0$) asymmetric Ising model. We used a regular degree distribution, $\tp_{\ell}=\delta_{\ell,d}$ with $d=3$, each point correspond to a different value of $\theta >0$. The two set of symbols correspond to  $\om=0.4 < \omc$ and $\om=0.7 > \omc$. The vertical dashed line indicates the location of the Kesten-Stigum transition, while the inclined line is our analytical prediction from (\ref{eq_Ising_lowest}) for $\om=0.4$.}
\label{fig_Ising_qEAoftheta}
\end{figure}

As an illustration of this phenomenon we present in Fig.~\ref{fig_Ising_qEAoftheta} the results of a numerical resolution of the cavity equations, for two asymmetries below and above $\om_{\rm c}$. One clearly sees that in the latter case reconstruction is possible below the Kesten-Stigum threshold, and that in the former case our analytical prediction (\ref{eq_Ising_lowest}) is in agreement with the numerical results.

\subsubsection{Further expansions around the critical asymmetry}

Let us now further describe the phase diagram of the problem in the
$(\theta,\om)$ plane (for simplicity of the discussion we assume the
degree distribution to be held fixed), see Fig.~\ref{fig:asymmetric} for an example. 
In the low asymmetry ($|\om| < \om_{\rm c}$) situation there are only two phases, the tree
reconstruction being possible if and only $\tE[\ell] \theta^2 >1$. This tightness of the Kesten-Stigum bound at small non-zero asymmetry was proven in~\cite{KS_tight_Ising}, but this paper did not estimate the value of $\om_{\rm c}$. In terms of the SBM inference problem the Kesten-Stigum transition separates an easy phase from an information theoretically impossible phase; this was proven in the symmetric ($\om=0$) case in~\cite{massoulie2014community,mossel2013proof}.

The large asymmetry ($|\om| > \om_{\rm c}$) part of the
phase diagram is richer. In terms of the tree reconstruction problem
there will be two lines of transition $\theta_{{\rm sp},\pm}(\om)$ such
that reconstruction is possible for $\theta < \theta_{{\rm
    sp},-}(\om)<0$ and $\theta > \theta_{{\rm sp},+}(\om)>0$, with
$\tE[\ell]\theta_{{\rm sp},\pm}(\om)^2 <1 $; the subscript $\pm$
indicates the ferromagnetic ($\theta >0$) or antiferromagnetic
($\theta<0$) part of the phase diagram. Note that the average offspring degree $\tE[\ell]$ has to be larger than $(2+\sqrt{3})/(2-\sqrt{3})\approx 13.928$ for the large asymmetry antiferromagnetic part of the phase diagram to be non-empty, because of the condition (\ref{eq_Ising_bound_theta}). This discontinuous transition
of the tree reconstruction problem, depicted in the bottom right part of
Fig.~\ref{fig:standard}, is not directly relevant for the SBM graph
inference problem. The latter has an easy phase for $\tE[\ell]
\theta^2 >1$, an hard phase for $\theta < \theta_{{\rm IT},-}(\om)<0$
and $\theta > \theta_{{\rm IT},+}(\om)>0$, with $\tE[\ell]\theta_{{\rm
    IT},\pm}(\om)^2 \in [\tE[\ell]\theta_{{\rm sp},\pm}(\om)^2,1]$,
while the inference is information theoretically (IT) impossible for $\theta \in [\theta_{{\rm IT},-}(\om),\theta_{{\rm IT},+}(\om)]$. This ``IT'' line is defined by the vanishing of the free-entropy computed in the non-trivial solution of the cavity equations.

In the remaining of this section we shall describe more precisely the neighborhood of the (ferromagnetic and antiferromagnetic when it exists) points $(\theta,\om)=(\thetaKS,\omc)$ of the phase diagram, and in particular present an analytic description of the lines $\theta_{{\rm sp},\pm}(\om)$ and $\theta_{{\rm IT},\pm}(\om)$ when $|\om| \to \omc^+$. To reach this goal we shall exploit the next order in our generic moment expansions, as summarized in (\ref{eq_Potts_cavity_A}-\ref{eq_Potts_cavity_hC}). We first note that for symmetry reasons the tensors $B$ and $C$ depends on their spin indices as $B_{\s \t \g} = b \, \s \t \g$ and $C_{\s \t \g \b} = c \, \s \t \g \b$, with $b$ and $c$ two reals. The form of the matrix $\hM$ given in (\ref{eq_Ising_hM}) leads to
\beq
\hB_{\s \t \g} = \theta^3 b \frac{1}{\oeta_\s \oeta_\t \oeta_\g} \s \t \g \ , \qquad 
\hC_{\s \t \g \b} = \theta^4 c \frac{1}{\oeta_\s \oeta_\t \oeta_\g \oeta_\b} \s \t \g \b \ .
\eeq
Plugging these expressions into (\ref{eq_Potts_cavity_A}-\ref{eq_Potts_cavity_C}) leads after a short computation to a set of equations on $a$, $b$ and $c$, namely:
\bea
a &=& \tE[\ell]\theta^2 a + 
\tE[\ell(\ell-1)] \frac{2(3 \om^2 -1)}{(1-\om^2)^2} \theta^4 a^2 
+ \tE[\ell(\ell-1)] \frac{32 \om}{(1-\om^2)^2} \theta^5 a b + 
\tE[\ell(\ell-1)] \frac{16}{(1-\om^2)^2} \theta^6 a c  
\nonumber \\ &&  
+ \tE[\ell(\ell-1)(\ell-2)]  
\frac{4}{3} \frac{5-42\om^2+45\om^4}{(1-\om^2)^4} \theta^6  a^3 \ , 
\label{eq_Ising_a}
\\
b &=& \tE[\ell]\theta^3 b 
- \tE[\ell(\ell-1)]  \frac{3 \om}{1-\om^2} \theta^4 a^2 \ , 
\label{eq_Ising_b}
\\
c &=& \tE[\ell]\theta^4 c + \tE[\ell(\ell-1)] \frac{3}{4} \theta^4 a^2 \ .
\label{eq_Ising_c}
\eea
Alternatively these equations can be obtained via the computation of the free-entropy (\ref{eq_Potts_phi}), which is found to be
\bea
\frac{1}{\E[\ell]} \phi(a,b,c) &=& 
 \frac{\tE[\ell]\theta^2-1}{(1-\om^2)^2} \theta^2 a^2 
- \frac{16}{3} \frac{\tE[\ell]\theta^3-1}{(1-\om^2)^3} \theta^3 b^2 
+ \frac{32}{3} \frac{\tE[\ell]\theta^4-1}{(1-\om^2)^4} \theta^4 c^2 
\label{eq_phi_Ising} \\
&+&  \tE[\ell(\ell-1)] \frac{4}{3} \frac{3 \om^2 -1}{(1-\om^2)^4} \theta^6 a^3 
+ \tE[\ell(\ell-1)] \frac{32 \om}{(1-\om^2)^4} \theta^7 a^2 b 
+ \tE[\ell(\ell-1)] \frac{16}{(1-\om^2)^4} \theta^8 a^2 c \nonumber \\
&+&  \tE[\ell(\ell-1)(\ell-2)]
\frac{2}{3}  \frac{5-42\om^2+45\om^4}{(1-\om^2)^6} \theta^8 a^4 \ ; \nonumber
\eea
its derivatives with respect to $a$, $b$ and $c$ do indeed vanish when the equations (\ref{eq_Ising_a}-\ref{eq_Ising_c}) are fulfilled.

The equations (\ref{eq_Ising_b}) and (\ref{eq_Ising_c}) can be immediately solved to obtain $b$ and $c$ as a function of $a$; re-injecting these results in (\ref{eq_Ising_a}) one obtains a quadratic equation on $a$:
\bea
0 &=& u + v \, a + w \, a^2 \ , \qquad \text{with} \label{eq_Ising_a_quadratic} \\
u(\theta) &=& \tE[\ell]\theta^2 - 1\ ,  \\
v(\theta,\om) &=& 
\tE[\ell(\ell-1)]  \frac{2(3 \om^2 -1)}{(1-\om^2)^2} \theta^4\ ,  \\
w(\theta,\om) &=& \tE[\ell(\ell-1)(\ell-2)]  
\frac{4}{3} \frac{5-42\om^2+45\om^4}{(1-\om^2)^4}\theta^6 \\ && 
+ \tE[\ell(\ell-1)]^2 \frac{12 \theta^9}{(1-\om^2)^2} \left(-\frac{8
    \om^2}{1-\om^2} \frac{1}{1-\E[\ell]\theta^3} +
  \frac{\theta}{1-\E[\ell]\theta^4} \right)\ .
\eea
We want to study this equation in the neighborhood of the point $(\thetaKS,\omc)$, taking simultaneously the limits $\theta \to \tKS$ and $\om \to \omc$; we denote as before $\epsilon=\tE[\ell]\theta^2 - 1$. The coefficients $u=\epsilon$ and $v$ will thus be both small in this regime, while $\tw=w(\tKS,\omc) \neq 0$. In order to have the three terms in (\ref{eq_Ising_a_quadratic}) of the same order, one realizes that the two simultaneous limits must be taken with $\epsilon = t (\om-\omc)^2$, where $t$ is finite and constitutes the relevant control parameter in this scaling regime. Then, denoting $v(\tKS,\om) = \tv (\om-\omc) + O((\om-\omc)^2)$, we reduce (\ref{eq_Ising_a_quadratic}) at lowest order to
\beq
0 = t \, (\om-\omc)^2 + \tv \, (\om-\omc) \, a + \tw \, a^2 \ .
\eeq
Defining a reduced unknown $\ta = \frac{a}{(\om-\omc)}$, which will be finite in the limit, we obtain finally
\beq
0 = t + \tv \, \ta + \tw \, \ta^2 \quad \text{hence} \ \ 
\ta = - \frac{\tv}{2 \tw} \pm \frac{1}{2 \tw} \sqrt{\tv^2 -4 \tw t} \ . 
\label{eq_Ising_eqtX}
\eeq
The coefficients $\tv$ and $\tw$ only depend on the degree distribution; to determine them at dominating order one replaces $\theta$ by its value at the Kesten-Stigum transition, i.e. $\sign(\theta)/\sqrt{\tE[\ell]}$, and obtain:
\bea
\tv &=& 9 \sqrt{3} \frac{\tE[\ell (\ell-1)]}{\tE[\ell]^2} \ ,\\
\tw &=& -27 \frac{\tE[\ell (\ell-1)(\ell-2)]}{\tE[\ell]^3}
- 27 \left(\frac{\tE[\ell (\ell-1)]}{\tE[\ell]^2} \right)^2
\left( \frac{4}{\sign (\theta)\sqrt{\tE[\ell]}-1} - \frac{1}{\tE[\ell]-1} \right)
\ .
\label{eq_Ising_tw}
\eea
It is rather easy to check that $\tw<0$ in the ferromagnetic case ($\sign (\theta) >0$) for all degree distributions with $\tE(\ell) > 1$ (a necessary condition for the Galton-Watson tree to be infinite with positive probability, and for the Kesten-Stigum transition to exists); we believe that $\tw<0$ also in the antiferromagnetic case whenever the point $(-1/\sqrt{\tE[\ell]},\om_{\rm c})$ belongs to the authorized domain of parameters according to the condition (\ref{eq_Ising_bound_theta}) for all offspring degree distributions, but could only check it explicitly in the regular and Poisson cases. Note that in the large degree limit both coefficients $\tv$ and $\tw$ remains finite, namely $\tv \to 9\sqrt{3}$ and $\tw \to -27$.

As a consequence of $\tw < 0$ the real solutions of (\ref{eq_Ising_eqtX}) exist for $t \ge t_{\rm sp} = \tv^2/(4 \tw)$. This is precisely the condition that defines the line $\theta_{\rm sp}$; translating back from rescaled units yields the lowest order expansion of $\theta_{\rm sp}(\om)$ in the limit $\om \to \omc^+$:
\beq
\theta_{\rm sp}(\om) = \tKS \left( 1 -  \frac{\tv^2}{8 |\tw|} (\om-\omc)^2 + o((\om-\omc)^2)\right) \ ;
\label{eq_Ising_line_td}
\eeq
note that $\tw$ depends on $\sign (\theta)$, giving two distinct expansions for the ferromagnetic and antiferromagnetic lines $\theta_{{\rm sp},+}$ and $\theta_{{\rm sp},-}$.

Using the expression (\ref{eq_phi_Ising}) of $\phi$ one can also expand the IT line $\tc$ around $(\theta,\om)=(\thetaKS,\omc)$. Indeed in the scaling regime described above one finds that
\beq
\phi = \frac{9}{2} \frac{\E[\ell]}{\tE[\ell]}  (\om-\omc)^4 \left(
\frac{1}{2} t \, \ta^2 + \frac{1}{3} \tv \, \ta^3 + \frac{1}{4} \tw \, \ta^4 
\right) \ .
\label{eq_Ising_phi_rescaled}
\eeq
The IT line corresponds to $\phi=0$ (this is the value on the trivial fixed point), hence the IT threshold corresponds to the solution of
\beq
\begin{cases} 
0 = t + \tv \, \ta + \tw \, \ta^2 \\
0 = \frac{1}{2} t + \frac{1}{3} \tv \, \ta + \frac{1}{4} \tw \, \ta^2
\end{cases}
\qquad \Rightarrow \qquad
\ta = -\frac{2 \tv}{3 \tw} \ , \quad t=\frac{2 \tv^2}{9 \tw} \ .
\eeq
The IT threshold $t_{\rm IT} = \frac{2 \tv^2}{9 \tw}$ is thus distinct at this order from the spinodal one $t_{\rm sp} = \frac{\tv^2}{4 \tw}$ (even though the difference in the coefficient is small). Translating back in terms of the parameters $(\theta,\om)$ gives
\beq
\theta_{\rm IT}(\om) = \tKS \left( 1 -  \frac{\tv^2}{9 |\tw|} (\om-\omc)^2 + o((\om-\omc)^2) \right) \ ,
\label{eq_Ising_line_tc}
\eeq
with again two different expressions for the ferromagnetic and antiferromagnetic transitions.

\begin{figure}
\includegraphics[width=8cm]{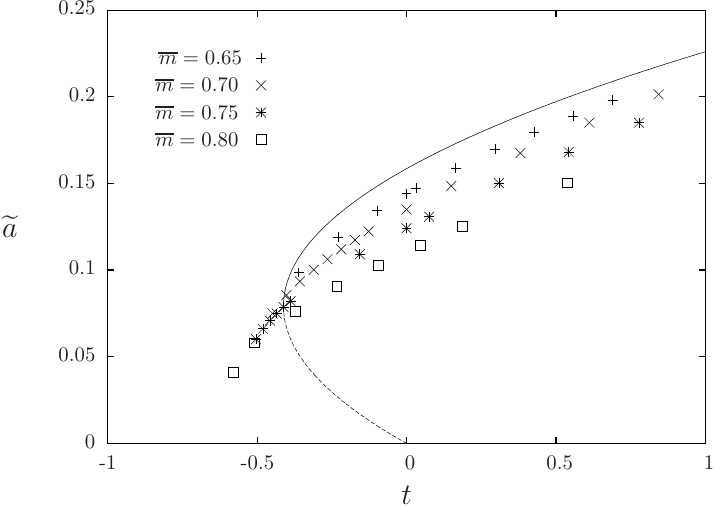}
\hspace{1cm}
\includegraphics[width=8cm]{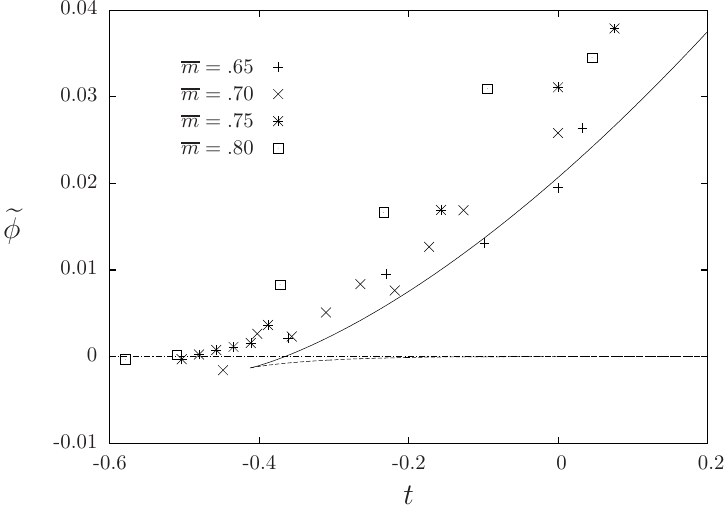}
\caption{Study of the scaling regime in the neighborhood of the point $(\tKS,\omc)$ for the ferromagnetic asymmetric Ising model with degree distribution $\tp_{\ell}=\delta_{\ell,d}$, $d=3$. The rescaled signal-to-noise ratio parameter is $t=(d\theta^2-1)/(\om-\omc)^2$, the Kesten-Stigum transition thus corresponds to $t=0$. Left panel: the rescaled accuracy $\ta=a/(\om-\omc)$ as a function of $t$; the lines are the analytic predictions for the stable (solid) and unstable (dashed) branches of fixed-points obtained in Eq.~(\ref{eq_Ising_eqtX}), the symbols are numerical results that approach the analytic line when $\om \to \omc^+$. Right panel: the rescaled free-entropy $\widetilde{\phi}=\phi/(\om-\omc)^4$ as a function of $t$; the analytic lines have been obtained by inserting in (\ref{eq_Ising_phi_rescaled}) the two branches of $\ta$ from (\ref{eq_Ising_eqtX}), the threshold $t_{\rm IT}$ corresponds to the crossing with the horizontal line $\widetilde{\phi}=0$.}
\label{fig_Ising_rescaled}
\end{figure}

These two expansions (\ref{eq_Ising_line_td},\ref{eq_Ising_line_tc}) on the behavior the spinodal and IT lines in the neighborhood of the critical asymmetry where the transition crosses over from second to first order constitute our main new results for the asymmetric Ising model. We have shown in the right panel of Fig.~\ref{fig:asymmetric} that they are in agreement with our numerical data, within the accuracy we could reach. As a further illustration we show in Fig.~\ref{fig_Ising_rescaled} the numerical determination of $a$ and $\phi$ in the scaling regime and compare it to our analytical formulae.

\subsection{Application 4: the $q_1+q_2$ case}
\label{sec:q1+q2}

We shall now turn our attention to the model introduced in Sec.~\ref{sec:main_q1q2}, that breaks the symmetry between $q$ labels in a minimal way, generalizing the two previous cases (i.e. $q$ symbols with maximal symmetry studied in Sec.~\ref{sec_applications_symPotts} and $q=2$ without any symmetry in Sec.~\ref{sec_Ising_asym}).

Let us consider indeed an alphabet of $q$ labels, $\chi=\{1,\dots,q\}$, divided into two ``super-groups'' (or ``super-communities'') $G_1$ and $G_2$ containing respectively $q_1$ and $q_2$ symbols (in such a way that $q=q_1+q_2$), namely $G_1=\{1,\dots,q_1\}$ and $G_2 = \{q_1+1,\dots,q\}$. We break the symmetry among the $q$ labels in a minimal way, according to this sub-division in two groups, by taking $\oeta_\s$ constant in $G_1$ and $G_2$: $\oeta_\s = \oeta_1$ if $\s \in G_1$ and $\oeta_\s = \oeta_q$ if $\s \in G_2$, with the normalization condition $q_1 \oeta_1 + q_2 \oeta_q =1$. We parametrize the fraction of vertices in $G_1$ and $G_2$ by $\om \in [-1,1]$, according to
\beq
\oeta_\s = \begin{cases} 
\frac{1}{q_1} \frac{1+\om}{2} & \text{if} \ \ \s \in G_1  \\
\frac{1}{q_2} \frac{1-\om}{2} & \text{if} \ \ \s \in G_2  \\
\end{cases} \ ;
\label{eq_q1q2_def_om}
\eeq
this notation is reminiscent of the asymmetric Ising case, which would be obtained by coarse-graining the labels $\s \in G_1$ (resp. $\s \in G_2$) as $\s=+$ (resp. $\s=-$). We also perform this minimal symmetry breaking on the matrix $M$, assuming that its matrix elements $M_{\s \t}$ only depend on the group of appartenance of $\s$ and $\t$, and on whether $\s=\t$ or not (the shape of the associated connectivity matrix in the SBM interpretation is sketched in Fig.~\ref{fig_matrix_q1q2}).

Imposing in addition the reversibility of the stochastic matrix $M$ with respect to $\oeta$ (this condition corresponds to the equality of the average degrees in $G_1$ and $G_2$ in the SBM interpretation) one realizes that such a matrix $M$ is necessarily of the form:
\bea
M_{\s \t} &=& \frac{1+\om}{2 q_1} \ind(\t \in G_1 ) + \frac{1-\om}{2 q_2} \ind(\t \in G_2 ) + 
\mu_1 \, \ind(\s,\t \in G_1 ) 
\left( \delta_{\s,\t} - \frac{1}{q_1} \right)
+ \mu_2 \, \ind(\s,\t \in G_2 ) 
\left( \delta_{\s,\t} - \frac{1}{q_2} \right)
\nonumber \\
&+& \mu_0 \left( \frac{1-\om}{2} \ind(\s \in G_1) - \frac{1+\om}{2} \ind(\s \in G_2) \right)
\left(\frac{1}{q_1} \ind(\t \in G_1) - \frac{1}{q_2} \ind(\t \in G_2) \right)
\ .
\label{eq_M_q1q2}
\eea
One can check that $M$ admits 1 as an eigenvalue, with eigenvectors $\oeta$ on the left and constant vector on the right; the parameters $\mu_0$, $\mu_1$ and $\mu_2$ introduced in (\ref{eq_M_q1q2}) correspond to the non-trivial eigenvalues of $M$. There are $q_1-1$ (resp. $q_2-1$) degenerate eigenvalues $\mu_1$ (resp. $\mu_2$) with eigenvectors supported on $G_1$ (resp. $G_2$) and a simple eigenvalue $\mu_0$ whose eigenvector is constant inside each of the two groups. For a given choice of $\om \in [-1,1]$ the requirement of positivity of the matrix elements of $M$ restricts the allowed domain of the parameters $(\mu_0,\mu_1,\mu_2)$ to the subset of $\left[-\frac{1-|\om|}{1+|\om|} ,1\right] \times \left[-\frac{1}{q_1-1},1\right] \times \left[-\frac{1}{q_2-1},1\right]$ that fulfils the condition
\beq
\mu_0 \ge \max\left( \frac{2 \mu_1-1-\om}{1-\om},\frac{-2(q_1-1) \mu_1-1-\om}{1-\om} ,
\frac{2 \mu_2-1+\om}{1+\om},\frac{-2(q_2-1) \mu_2-1+\om}{1+\om} \right) \ .
\eeq
One sees that the relevant eigenvalue for the Kesten-Stigum transition, i.e. the maximal one in absolute value, can be either $\mu_1,\mu_2$ or $\mu_0$ depending on the choice of the parameters. In other words the signal on the labels transmitted through the edges of the SBM can be stronger inside $G_1$, inside $G_2$ or in the difference of behavior between $G_1$ and $G_2$.

We shall now specialize the equations of Sec.~\ref{sec_expansion_Potts} to this particular case. We have first to describe the matrices $A_{\s \t}=\E[\delta_\s \delta_\t]$. Because of the permutation invariance inside each of the two groups of labels the matrix element $A_{\s \t}$ should only depend on whether $\s=\t$ or not and on the group of the two communities $\s$ and $\t$. In addition the matrix $A$ has to be symmetric and the sum of every row or column has to vanish. A moment of thought reveals that the vector space of such matrices is 3 dimensional, and is spanned by the three matrices $K_1$, $K_2$ and $K_0$ defined as
\bea
(K_1)_{\s \t} &=& \ind(\s,\t \in G_1 ) 
\left(\delta_{\s,\t} - \frac{1}{q_1} \right) \ , \qquad
(K_2)_{\s \t} = \ind(\s,\t \in G_2 ) 
\left(\delta_{\s,\t} - \frac{1}{q_2} \right) \ , \\
(K_0)_{\s \t} &=& \frac{1}{q q_1 q_2} 
\left( q_2 \ind(\s \in G_1 ) - q_1 \ind(\s \in G_2 )  \right)
\left( q_2 \ind(\t \in G_1 ) - q_1 \ind(\t \in G_2 )  \right) \ .
\eea
These matrices, which are defined similarly to the matrix $K$ of the symmetric Potts model introduced in (\ref{eq_Potts_sym_def}), are linearly independent and obey in addition the simple algebra
\beq
K_1^2 = K_1 \ , \qquad K_2^2 = K_2 \ , \qquad K_0^2 = K_0 \ , \qquad
K_i K_j =0 \ \ \ \text{if} \ \ i \neq j \ .
\eeq
We can thus parametrize $A^{(n)}$ as $a_1^{(n)} K_1 + a_2^{(n)} K_2 + a_0^{(n)} K_0$; the non-zero eigenvalues of $A^{(n)}$ are then found to be $a_1^{(n)}$, $a_2^{(n)}$ and $a_0^{(n)}$. These three reals must thus be non-negative for $A^{(n)}$ to be semi-positive definite.

The matrix $\hM$ defined in (\ref{eq_Potts_defhM}) can then be expressed as
\beq
\hM_{\s \t} = 1 + \mu_1 \frac{1}{\oeta_1} (K_1)_{\s \t} + \mu_2 \frac{1}{\oeta_q} (K_2)_{\s \t} + \mu_0 \, q \oeta_1 \oeta_q \, \frac{1}{\oeta_\s} (K_0)_{\s \t} \frac{1}{\oeta_\t} \ ,
\eeq
and a simple computation based on the algebraic properties stated above reveals that
\beq
\oeta_\s \oeta_\t \hA^{(n)}_{\s \t} = (\mu_1^2 a_1^{(n)} K_1 + \mu_2^2 a_2^{(n)} K_2 + \mu_0^2 a_0^{(n)} K_0)_{\s \t} \ .
\eeq
Hence the linearized evolution equation (\ref{eq_Potts_linear_A}) yields 
\beq
a_i^{(n+1)} = \tE[\ell] \mu_i^2 a_i^{(n)} \quad \text{for} \ \ i = 0,1,2 \ .
\eeq
The trivial fixed point $a_1=a_2=a_0=0$ becomes unstable as soon as one of the three coefficients $a_i$ grows under these iterations, hence we recover the Kesten-Stigum criterion $\tE[\ell] \max(\mu_1^2 , \mu_2^2 , \mu_0^2) =1 $ at the limit of stability of the trivial fixed point, the parameters $\mu_i$ being the non-trivial eigenvalues of $M$.

Following our usual program we incorporate the next order correction and look for a perturbative non-trivial fixed point around the Kesten-Stigum transition. Inserting the form given above for the matrices $A$ and $\hA$ into the generic equation (\ref{eq_Potts_order2}) yields, after some computations, the following system of quadratic equations for $a_0$, $a_1$ and $a_2$:
\bea
a_1 &=& \tE[\ell] \mu_1^2 a_1 + \tE[\ell(\ell-1)] \frac{2 q_1}{(1+\om)^2} \left( (q_1 - 3 - \om) \mu_1^4 a_1^2 + \frac{2 q_2}{q} \mu_0^2 \mu_1^2 a_0 a_1 \right) \ , \label{eq_q1q2_a1} \\
a_2 &=& \tE[\ell] \mu_2^2 a_2 + \tE[\ell(\ell-1)]\frac{2
  q_2}{(1-\om)^2} \left( (q_2 - 3 + \om) \mu_2^4 a_2^2 + \frac{2
    q_1}{q} \mu_0^2 \mu_2^2 a_0 a_2 \right) \ ,  \label{eq_q1q2_a2} \\
a_0 &=& \tE[\ell] \mu_0^2 a_0 \label{eq_q1q2_a0} \\ &+& \tE[\ell(\ell-1)] 
\left( \frac{4 q_1 q_2}{q} \frac{3 \om^2 -1}{(1-\om^2)^2} \mu_0^4 a_0^2
+ \frac{q q_1 (q_1-1)}{2 q_2}\frac{(1-\om)^2}{(1+\om)^2} \mu_1^4 a_1^2
+ \frac{q q_2 (q_2-1)}{2 q_1}\frac{ (1+\om)^2}{(1-\om)^2} \mu_2^4 a_2^2
\right) \ . \nonumber
\eea
From this system of equations one can recover, as a consistency check, the equations (\ref{eq_Potts_sym_a}) of the symmetric $q$-state model and (\ref{eq_Ising_lowest}) of the asymmetric Ising case. The latter is indeed obtained with $q_1=q_2=1$, in which case $K_1=K_2=0$, the only parameter is $a_0$ which obeys indeed (\ref{eq_Ising_lowest}) with the identification $\mu_0=\theta$. To recover the former case one takes $\mu_1=\mu_2=\mu_0=\theta$, $a_1=a_2=a_0$ and $\om=(q_1-q_2)/2$; then the three equations (\ref{eq_q1q2_a1}-\ref{eq_q1q2_a0}) reduces to (\ref{eq_Potts_sym_a}). Note also that these equations coincide, in the special case $q_1=q_2$, $\mu_1=\mu_2$, $\om=0$, for which $a_1=a_2$, to the moment recursions derived in~\cite{liu2017tightness}.

Let us now come back to an arbitrary choice of parameters in (\ref{eq_q1q2_a1}-\ref{eq_q1q2_a0}), and discuss the bifurcation of the non-trivial solution of this system of equations at the Kesten-Stigum transition. This discussion must be divided according to which eigenvalue $\mu_i$ causes the transition by fulfilling the condition $\tE[\ell] \mu_i^2 = 1$ (we shall assume for simplicity that only one among $\mu_1$, $\mu_2$ and $\mu_0$ becomes critical).

\subsubsection{Bifurcation driven by $\mu_0$}
Let us first consider the case where the bifurcation is driven by $\mu_0$, i.e. where $\tE[\ell] \mu_0^2 = 1 + \epsilon$ with $\epsilon$ small, while $\tE[\ell] \mu_1^2 < 1$ and $\E[\ell] \mu_2^2 < 1$. The dominant direction of the bifurcation being $a_0$ one can simplify the system (\ref{eq_q1q2_a1}-\ref{eq_q1q2_a0}) at lowest order into a single equation on $a_0$:
\beq
0 = \epsilon +  
\tE[\ell(\ell-1)] \frac{4 q_1 q_2}{q} \frac{3 \om^2 -1}{(1-\om^2)^2} \mu_0^4 a_0
\ .
\eeq
Remembering the positivity condition $a_0 \ge 0$, one realizes that the bifurcating solution exists above the Kesten-Stigum transition (i.e. for $\epsilon >0$) if and only if the asymmetry between the two groups of labels is small enough, namely if $\om < \omc = 1/\sqrt{3}$, as in the asymmetric Ising case. As in the right hand sides of (\ref{eq_q1q2_a1}-\ref{eq_q1q2_a2}) there are no terms proportional to a power of $a_0$ (without further multiplication by $a_1$ or $a_2$), the non-bifurcating unknowns $a_1$ and $a_2$ remain strictly equal to 0 in this solution.

\subsubsection{Bifurcation driven by $\mu_1$}

Suppose now that $\mu_1$ is the critical eigenvalue (the case where $\mu_2$ becomes critical can be deduced from this one by exchanging the two groups), i.e. that $\tE[\ell] \mu_1^2 = 1 + \epsilon$ with $\epsilon$ small, while $\tE[\ell] \mu_2^2 < 1$ and $\tE[\ell] \mu_0^2 < 1$. The lowest order equation on the bifurcating direction $a_1$ thus becomes
\beq
0 = \epsilon + \tE[\ell(\ell-1)] \frac{2 q_1}{(1+\om)^2} (q_1 - 3 - \om) \mu_1^4 a_1 \ ,
\label{eq_q1q2_a1driven}
\eeq
with the sub-dominant coefficient $a_0$ being of order $O(a_1^2)=O(\epsilon^2)$, while $a_2$ remains strictly zero (within the system (\ref{eq_q1q2_a1}-\ref{eq_q1q2_a0})). The sign of $\epsilon$ for which the non-trivial solution satisfies the constraint $a_1 \ge 0$ is thus the one of $-(q_1 - 3 - \om)$. To analyze the sign of this quantity we can exclude the cases where $\om = \pm 1$, as these reduce to purely symmetric models with either $q_1$ or $q_2$ labels. The type of bifurcation thus depends on the (integer) value of $q_1$ as follows:
\begin{itemize}
\item if $q_1 \ge 4$, for any value of $\om$, the bifurcating solution exists for $\epsilon <0$, leading to the non-tightness of the Kesten-Stigum bound for the reconstruction. This is the conclusion reached in~\cite{liu2017tightness}, in the special case $q_1=q_2$ and $\om=0$.
\item if $q_1 \in \{1,2\}$, for any value of $\om$, the continuous solution is present above the Kesten-Stigum transition (for $\epsilon >0$).
\item if $q_1 =3$ the scenario depends on the asymmetry parameter $\om$: for $\om<0$ the bifurcating solution exists for $\epsilon <0$, yielding a 1st order transition with the non-tightness of the Kesten-Stigum bound. On the other hand if $\om >0$ the non-trivial solution appears continuously in the large SNR regime, above the Kesten-Stigum transition.
\end{itemize}

Note that this classification is independent of  the number $q_2$ of labels in the group which does not become critical at the Kesten-Stigum transition.

\subsubsection{Higher-order terms and the existence of algorithmic spinodals}

We have computed the next order in the expansion of (\ref{eq_q1q2_a1}-\ref{eq_q1q2_a0}), computing the coefficients of the terms that are cubic in the $a_i$'s, by specializing the generic equations (\ref{eq_Potts_cavity_A}-\ref{eq_Potts_cavity_C}) to the symmetry pattern of the $q_1+q_2$ model. This requires in particular the determination of the most generic symmetric tensors with 3 and 4 indices, $B_{\s \t \g}$ and $C_{\s \t \g \b}$, that are invariant under permutations of the labels inside $G_1$ and $G_2$; the resulting equations are rather long, we shall hence not write them completely but concentrate on the additional predictions they led us to.

As long as the coefficient of the quadratic term in the equation for the bifurcating $a_i$ is non-zero in the right hand side of (\ref{eq_q1q2_a1}-\ref{eq_q1q2_a0}), these higher order terms can only affect the solution quantitatively, but not qualitatively. Suppose first that the bifurcation is driven by $\mu_0$, and that $\om=\om_{\rm c}=1/\sqrt{3}$, in such a way that the first non-linear term in $a_0$ vanishes. In this case $a_0$ is solution of a quadratic equation of the form $0=\epsilon + w \, a_0^2$; our computation yields explicitly the value of this coefficient $w$, which turns out to be proportional (with a positive constant depending only on $q_1$ and $q_2$) to the corresponding expression found in the asymmetric Ising case and written in (\ref{eq_Ising_tw}), with $\mu_1$ playing the role of $\theta$. We argued this coefficient to be always negative for all degree distributions and sign of $\mu_1$, this case thus brings no novelty with respect to the situations investigated previously.

Suppose now that the bifurcation is driven by $\mu_1$, and that $q_1=3$, and $\om=0$, in such a way that the coefficient of $a_1^2$ in (\ref{eq_q1q2_a1}) vanishes. In this much more interesting case the leading order bifurcation equation becomes $0=\epsilon + w \, a_1^2$, with the following expression for the coefficient $w$ (with $\tE[\ell] \mu_1^2 = 1$ at lowest order):
\beq
w = 36 \left[-3 \frac{\tE[\ell (\ell-1)(\ell-2)]}{\tE[\ell]^3}
+ \left(\frac{\tE[\ell (\ell-1)]}{\tE[\ell]^2} \right)^2
\left( 
\frac{4}{\tE[\ell]-1} 
- \frac{12}{\sign (\mu_1)\sqrt{\tE[\ell]}-1} 
- \frac{2\mu_0}{1-\mu_0}
+ \frac{\tE[\ell]\mu_0^2}{1-\tE[\ell]\mu_0^2}  
\right)
\right] \ .
\label{eq_w_q1q2}
\eeq
The crucial point we want to emphasize is that $w$ can be made positive for well-chosen values of the parameters of the model, in particular when $|\mu_0|$ is close to $|\mu_1|$ (but still strictly smaller for the bifurcation to be driven by $\mu_1$). In such a case the non-trivial solution $a_1$ of the equation exists for $\epsilon<0$, hence the Kesten-Stigum bound is not tight. Let us now argue that the situation depicted in Fig.~\ref{fig:new} must occur in some part of the phase diagram of the $q_1+q_2$ model; for simplicity let us consider that the degree distribution is Poissonian with average $c$, that we take as the SNR. Suppose that $q_1=3$ and fix the values of the $\mu_i$'s in such a way that $w>0$ in the expression of (\ref{eq_w_q1q2}). For $\om=0$ the analysis above shows the existence of reconstruction down to $c_{\rm sp}(\om=0)<c_{\rm KS}$, while $c_{\rm alg}(\om=0)=c_{\rm KS}$ as the bifurcating solution exists only for $\epsilon <0$. Suppose now that $\om$ is slightly increased to a small positive value; from (\ref{eq_q1q2_a1driven}) we conclude that a non-trivial perturbative solution exists for some range of $c>c_{\rm KS}$, but by continuity with the situation for $\om=0$ must undergo a bifurcation at some $c_{\rm alg}(\om)>c_{\rm KS}$. Also by continuity the spinodal of the high-accuracy branch $c_{\rm sp}(\om)$ must persist for small enough $\om >0$, hence the existence of a bifurcation diagram as in Fig.~\ref{fig:new}. When $\om$ is further increased the two spinodals $c_{\rm sp}(\om)$ and $c_{\rm alg}(\om)$ collides and for even larger values of $\om$ the bifurcation diagram becomes the one of the left panel of Fig.~\ref{fig:standard}. Indeed when $\om \to 1$ the model reduces to a symmetric SBM with $q_1=3$ communities. Preliminary numerical results obtained by a resolution of the cavity equations via the population dynamics algorithm suggest a rather narrow domain of parameters for which this phenomenon of coexistence of two non-trivial stable solutions is observable, for this reason we do not present numerical data to illustrate this model.

\subsection{On the definition of accuracy and the maximal overlap estimator}
\label{sec_estimator}

The covariance matrix $A_{\s \t}=\E[\delta_\s \delta_\t]$ appeared naturally in our expansions as a measure of the deviation between the fixed point distribution $P(\eta)$ solution of the cavity equations and the trivial fixed point $\delta(\eta-\oeta)$. Let us now briefly comment on its interpretation as the accuracy of an estimation procedure, and its connection with the maximal overlap estimator.

Consider the tree reconstruction problem explained in Sec.~\ref{sec_reconstruction}, and assume that the root had value $\t$ in the broadcast process; an observer, which has no direct knowledge of $\t$, computes its posterior probability distribution $\eta$ given the values of spins on far away vertices of the tree. If the observer proposes as an estimator of $\t$ a random spin value chosen with probability $\eta$, the probability of success of the reconstruction is, on average with respect to the broadcast, the estimation and the tree, $\E_\t[\eta_\t]$. The probability of success if one had discarded all the observations, i.e. if one draws the estimator with the prior probability $\oeta$, is $\oeta_\t$; the accuracy, defined as the difference of these two probabilities, is thus $\E_\t[\delta_\t]$. Averaging finally over the value $\t$ of the unknown spin, we obtain the accuracy
\beq
a=\sum_\t \oeta_\t \E_\t[\delta_\t] = \sum_\t A_{\t \t} = \text{Tr}\, A \ ,
\label{eq_estimator_accuracy}
\eeq
where we used the identity (\ref{eq_moment_Potts_1}) between conditional and unconditional distributions. For binary spins $\s=\pm 1$ there is an affine mapping between the MMSE and the accuracy.

Consider now the maximal overlap estimator defined in Sec.~\ref{sec_typology_Bayesian}, for which the observer estimates the value of the root as $\text{argmax}_\s \, \eta_\s$. The probability of correct estimation can be expressed in different forms involving the conditional or unconditional distributions of $\eta$, namely
\beq
P_{\rm corr}=
\sum_\t \oeta_\t \E_\t [\ind( \underset{\s}{\text{argmax}} \,\eta_\s = \t)] = 
\E\left[\sum_\t \eta_\t  \ind(\underset{\s}{\text{argmax}} \, \eta_\s = \t) \right] = \E[\max_\s \eta_\s ] \ ,
\eeq
the first step resulting from the general change of density between condition and unconditional distributions expressed in (\ref{eq_Bayes_Potts}). This quantity can easily be evaluated numerically from the resolution of the cavity equations by the population dynamics algorithm, it is however much more difficult to characterize it analytically than the accuracy (\ref{eq_estimator_accuracy}). If we could indeed determine systematically $A_{\s \t}$ as a perturbative expansion in the small parameter $\kappa$ in the neighborhood of the Kesten-Stigum transition, a similar determination is not possible for $P_{\rm corr}$. The scaling ansatz (\ref{eq_ansatz}) means that around the KS transition, the dominant behavior of $\delta$ under $P$ is described by $\delta / \sqrt{\kappa} \tod X$ (where $X$ has a symmetric distribution). We can thus conclude that $P_{\rm corr}$ should behave as $\max_\s \oeta_\s+ C \sqrt{\kappa}$, but with a prefactor $C$ that involves the whole distribution of the rescaled random vector $X$, and thus cannot be computed from a finite number of moments. An explicit determination of $P_{\rm corr}$ can be achieved in the large degree limit, thanks to the Gaussian simplifications explained in Sec.~\ref{sec_large_degree}. One finds indeed in this limit
\beq
P_{\rm corr}= \E\left[e^{\underset{\s}{\max} L_\s} \right] \ ,
\eeq
where the expectation is over a Gaussian vector $L$ defined by its moments in Eq.~(\ref{eq_largedegree_L}). For instance in the symmetric Ising case ($q=2$, $\om=0$), denoting $\lambda =\lim ( \tE[\ell] \theta^2) $ the SNR, a short computation yields
\beq
P_{\rm corr}= \frac{1}{2} + \int_0^{\sqrt{2 \lambda a}} \frac{\dd t}{\sqrt{2 \pi}} e^{-\frac{1}{2} t^2} \ ,
\eeq
where $a=a(\lambda)$ is the solution of 
\beq
a= e^{-\lambda a} \E\left[\frac{e^{2 \sqrt{2 \lambda a}Z} }{e^{\sqrt{2 \lambda a}Z}+ e^{-\sqrt{2 \lambda a}Z}  } \right] - \frac{1}{2} \ ,
\eeq
with $Z$ a standard Gaussian random variable (of zero mean and unit variance). Expanding now these expressions around the Kesten-Stigum transition, i.e. setting $\lambda = 1 + \epsilon$ with $\epsilon \to 0^+$, one obtains the leading behavior $a \sim \epsilon/2$, while $P_{\rm corr} \sim (1/2) + \sqrt{\epsilon/ 2 \pi}$.

\section{Moment expansions for $k$-wise interacting Ising variables}
\label{sec_Ising}

We turn now to the second specialization of the formalism of Sec.~\ref{sec_generic}: we will consider generic $k$-wise interactions, but we restrict now to binary variables ($q=2$), that for convenience we will represent as Ising spins, $\chi = \{-1,1\}$. As most of the reasonings are similar to the ones presented in Sec.~\ref{sec_Potts} we shall give less details on the computations and underline the main differences with the case of pairwise interacting Potts variables.

\subsection{Fourier transforms of Boolean functions}

The main ingredient defining the models under study here is a joint probability $p_{\rm j}(\s_1,\dots,\s_k)$ over $\{-1,+1\}^k$, invariant under all permutations of its arguments. This symmetry implies that $p_{\rm j}$ is a function of $(\s_1+\dots+\s_k)$ only, and can thus be specified by $k$ real numbers (taking into account the normalization condition). The occupation models of~\cite{ZdeborovaMezard08b}, whose definition was recalled in Sec.~\ref{sec_generic}, correspond to the special case of a $p_{\rm j}$ vanishing for some values of $(\s_1+\dots+\s_k)$, and constant otherwise. A convenient way to specify a generic permutation invariant $p_{\rm j}$ is via the following representation,
\beq
p_{\rm j}(\s_1,\dots,\s_k) = \frac{1}{2^k}\left[ 
1 + \gamma_1 \sum_{i=1}^k \s_i + \gamma_2 \sum_{i<j} \s_i \s_j + \dots
+ \gamma_k \, \s_1 \dots \s_k
\right] \ ,
\eeq
where the $\gamma_n$ are the Fourier coefficients of $p_{\rm j}$. They can be expressed as
\beq
\gamma_n = \E[\t_1 \dots \t_n] \ ,
\label{eq_occ_def_gamman}
\eeq
where the average is over a configuration $(\t_1,\dots,\t_k)$ drawn with probability $p_{\rm j}(\t_1,\dots,\t_k)$. Let us describe in these terms the conditional distribution $p_{\rm c}(\t_1,\dots,\t_{k-1}|\t)$ obtained from $p_{\rm j}$. Because of its invariance under the permutations of its $k-1$ first arguments, it is fully described by the averages of products of $n$ spins. These values are easily expressed in terms of the Fourier coefficients of $p_{\rm j}$, as
\beq
\E[\t_1 \dots \t_n | \t] = \frac{\gamma_n + \t \gamma_{n+1}}{1+\t \gamma_1} \qquad \text{for} \ \ \ n=1,\dots,k-1 \ .
\label{eq_occ_ave_spin_cond_without}
\eeq

In what follows we will assume that the stationary probability distribution $\oeta$ is unbiased, i.e. $\oeta_+=\oeta_-=\frac{1}{2}$. For the reversibility assumption of Sec.~\ref{sec_generic} to hold in this case one needs the marginal probability of a single variable drawn from $p_{\rm j}$ to be also unbiased. In terms of the Fourier coefficient representation this is equivalent to $\gamma_1=0$ (see Eq.~(\ref{eq_occ_def_gamman})). We will actually make a stronger hypothesis on $p_{\rm j}$, namely that it is invariant under a global spin reversal: $p_{\rm j}(\s_1,\dots,\s_k)=p_{\rm j}(-\s_1,\dots,-\s_k)$. This implies that not only $\gamma_1$ but all the Fourier coefficients $\gamma_p$ with $p$ odd vanish. With this assumption (\ref{eq_occ_ave_spin_cond_without}) can be simplified into:
\beq
\E[\t_1 \dots \t_{2p-1}|\t] =\t \gamma_{2p} \ , \qquad
\E[\t_1 \dots \t_{2p}|\t] = \gamma_{2p} \ .
\label{eq_occ_ave_spin_cond}
\eeq

\subsection{Cavity equations and free-entropy functional}

We have stated in Sec.~\ref{sec_generic}  recursive equations obeyed by the distributions $P_\t^{(n)}(\eta)$, $\hP_\t^{(n)}(\nu)$, and their unconditional versions $P^{(n)}(\eta)$, $\hP^{(n)}(\nu)$, see (\ref{eq_generic_conditional_hPtoP}-\ref{eq_generic_unconditional_PtohP}). As we are dealing now with Ising variables we can use a more succinct notation, a probability distribution over a Boolean variable being parametrized by a single real. We shall use the following notations,
\beq
\eta_\s = \frac{1+m \, \s}{2} \ , \qquad
\nu_\s = \frac{1+u \, \s}{2} \ ,
\eeq
with $m$ and $u$ in $[-1,1]$, and hence denote $P_\t^{(n)}(m)$, $P^{(n)}(m)$, $\hP_\t^{(n)}(u)$ and $\hP^{(n)}(u)$ for the above distributions expressed with this parametrization. The Belief Propagation equation (\ref{eq_generic_nutoeta}) can then be reformulated as
\beq
m=f(u^1,\dots,u^\ell) = \frac{\underset{i=1}{\overset{\ell}{\prod}} (1+u^i) -\underset{i=1}{\overset{\ell}{\prod}} (1-u^i) }{\underset{i=1}{\overset{\ell}{\prod}} (1+u^i) +\underset{i=1}{\overset{\ell}{\prod}} (1-u^i)} = \tanh\left(\sum_{i=1}^\ell \atanh(u^i) \right) \ ,
\label{eq_occ_BP_utom}
\eeq
with the normalization factor
\beq
z(u^1,\dots,u^\ell) = \frac{1}{2} \underset{i=1}{\overset{\ell}{\prod}} (1+u^i) +\frac{1}{2} \underset{i=1}{\overset{\ell}{\prod}} (1-u^i)\ .
\eeq
The other BP equation (\ref{eq_generic_etatonu}) reads in this notation:
\beq
u=\hf(m^1,\dots,m^{k-1}) = \frac{
\gamma_1 + \gamma_2 \underset{i=1}{\overset{k-1}\sum}m^i + \gamma_3 \underset{i<j}{\sum} m^i m^j + \dots + \gamma_k m^1 \dots m^{k-1}     }{
1 + \gamma_1 \underset{i=1}{\overset{k-1}\sum}m^i + \gamma_2 \underset{i<j}{\sum} m^i m^j + \dots + \gamma_{k-1} m^1 \dots m^{k-1} } \ .
\eeq
With the assumption of invariance of $p_{\rm j}$ under global spin reversal the odd Fourier coefficients vanish and one can simplify this equation into
\bea
u=\hf(m^1,\dots,m^{k-1}) &=& 
\frac{1}{\hz(m^1,\dots,m^{k-1})} 
\left( \gamma_2 \underset{i=1}{\overset{k-1}\sum} m^i + \gamma_4 \underset{i_1<i_2<i_3}{\sum} m^{i_1} m^{i_2} m^{i_3} + \dots  \right) \ , \nonumber \\
\hz(m^1,\dots,m^{k-1}) &=& 
1 + \gamma_2 \underset{i<j}{\sum} m^i m^j + \gamma_4 \underset{i_1<i_2<i_3<i_4}{\sum} m^{i_1} m^{i_2} m^{i_3} m^{i_4} +\dots \ .
\label{eq_occ_BP_mtou}
\eea

Let us now discuss the symmetry properties of the distributions $P_\t^{(n)}(m)$ and $P^{(n)}(m)$, and the relationships between them. The consequences of the Bayes theorem stated in (\ref{eq_Bayes}) become, for an unbiased stationary distribution $\oeta$,
\beq
P^{(n)}(m) = \frac{1}{2} P_+^{(n)}(m) + \frac{1}{2} P_-^{(n)}(m) \ , \qquad
P_\t^{(n)}(m) = (1+ \t m) P^{(n)}(m) \ , \qquad
\int \dd P^{(n)}(m) \, m = 0 \ .
\eeq
Actually the assumption of invariance under global spin reversal has further consequences: not only $P^{(n)}$ has zero average, but it is also symmetric. Hence one has
\beq
P^{(n)}(m) = P^{(n)}(-m) \ , \qquad P_+^{(n)}(m) = P_-^{(n)}(-m) \ .
\eeq
Combining these two set of properties yields particularly simple identities between moments of these distributions. The change of densities between $P_\t^{(n)}$ and $P^{(n)}$ means indeed that $\E_+^{(n)}[f(m)] = \E^{(n)}[(1+m)f(m)]$ for any function $f(m)$. Applying this identity with $f(m)=m^{2p}$ and  $f(m)=m^{2p-1}$ yields 
\beq
\E_+^{(n)}[m^{2p}] = \E^{(n)}[m^{2p}] + \E^{(n)}[m^{2p+1}] = \E^{(n)}[m^{2p}] \ , \qquad
\E_+^{(n)}[m^{2p-1}] = \E^{(n)}[m^{2p-1}] + \E^{(n)}[m^{2p}] = \E^{(n)}[m^{2p}] \ ,
\eeq
where we exploited the symmetry of $P^{(n)}$ that makes its odd moments vanish. These identities are well-known in the context of Low Density Parity Check Codes, see for instance lemma 3 in~\cite{Montanari05}. Spelling out these identities for the lowest order moments we obtain
\beq
\E_+^{(n)}[m] = \E_+^{(n)}[m^2] = \E^{(n)}[m^2] \ , \qquad
\E_+^{(n)}[m^3] = \E_+^{(n)}[m^4] = \E^{(n)}[m^4] \ .
\label{eq_occ_moment_ids}
\eeq
The distributions $\hP_\t^{(n)}$ and $\hP^{(n)}$ have exactly the same symmetry properties as $P_\t^{(n)}$ and $P^{(n)}$, hence the random variables $u$ enjoy the same identities as the $m$'s.

The identity $P_+^{(n)}(m) = P_-^{(n)}(-m)$ that follows from the invariance under global spin-flip allows to close the equations (\ref{eq_generic_conditional_hPtoP},\ref{eq_generic_conditional_PtohP}) on the two distributions $P_+^{(n)}(m)$ and $\hP_+^{(n)}(u)$, that are found to evolve with $n$ according to
\bea
P_+^{(n+1)}(m) &=& \sum_{\ell=0}^\infty \tp_\ell \int \dd \hP_+^{(n)} (u^1) \dots
\dd \hP_+^{(n)} (u^\ell) \, \delta(m - f(u^1,\dots,u^\ell)) \ , 
\label{eq_occ_conditional_hPtoP}
\\
\hP_+^{(n)}(u) &=& 
\sum_{\t_1,\dots,\t_{k-1}} p_{\rm c}(\t_1,\dots,\t_{k-1}|+)
\int \dd P_+^{(n)}(m^1) \dots \dd P_+^{(n)}(m^{k-1}) \, 
\delta(u-\hf(\t_1 m^1,\dots,\t_{k-1} m^{k-1})) \ ,
\label{eq_occ_conditional_PtohP}
\eea
where the function $f$ and $\hf$ are given in (\ref{eq_occ_BP_utom}) and (\ref{eq_occ_BP_mtou}) respectively. This form is very convenient for a numerical resolution by the population dynamics algorithm (see Appendix~\ref{app_numerical_resolution} for more details on this point).

Equivalently the unconditional version of the cavity equations read
\bea
P^{(n+1)}(m) &=& \sum_{\ell=0}^\infty \tp_\ell \int \dd \hP^{(n)} (u^1) \dots
\dd \hP^{(n)} (u^\ell) \, \delta(m - f(u^1,\dots,u^\ell)) \, z(u^1,\dots,u^\ell) \ , 
\label{eq_occ_unconditional_hPtoP}
\\
\hP^{(n)}(u) &=& 
\int \dd P^{(n)}(m^1) \dots \dd P^{(n)}(m^{k-1}) \, 
\delta(u-\hf( m^1,\dots, m^{k-1})) \, \hz( m^1,\dots, m^{k-1}) \ .
\label{eq_occ_unconditional_PtohP}
\eea

The free-entropy functional can be expressed in these two versions of the cavity formalism from (\ref{eq:phi},\ref{eq_phiint_conditional}) as
\bea
\phi(P,\hP) =
&-&\E[\ell] \int \dd P(m) \dd \hP(u) \, \ze(m,u) \ln \ze(m,u)  \nonumber \\
&+& \frac{\E[\ell]}{k} \int \dd P(m^1) \dots \dd P(m^k) \, \zc(m^1,\dots,m^k) \ln \zc(m^1,\dots,m^k) \nonumber \\
&+& \sum_{\ell=1}^\infty p_\ell \int \dd \hP(u^1) \dots \dd \hP(u^\ell)
\, \zv(u^1,\dots,u^\ell) \ln \zv(u^1,\dots,u^\ell) \ , \\
\phi(P_+,\hP_+) =
&-&\E[\ell] \int \dd P_+(m) \dd \hP_+(u) \, \ln \ze(m,u)  \nonumber \\
&+& \frac{\E[\ell]}{k} \sum_{\s_1,\dots,\s_k} \pj(\s_1,\dots,\s_k) \int \dd P_+(m^1) \dots \dd P_+(m^k) \,  \ln \zc(\s_1 m^1,\dots, \s_k m^k) \nonumber \\
&+& \sum_{\ell=1}^\infty p_\ell \int \dd \hP_+(u^1) \dots \dd \hP_+(u^\ell)
\,  \ln \zv(u^1,\dots,u^\ell) \ ,
\label{eq_phi_occ}
\eea
where
\bea
\ze(m,u) &=& 1+ m u \ , \\
\zc(m^1,\dots,m^k) &=& 
1 + \gamma_2 \underset{i<j}{\sum} m^i m^j + \gamma_4 \underset{i_1<i_2<i_3<i_4}{\sum} m^{i_1} m^{i_2} m^{i_3} m^{i_4} +\dots  \ , \\
\zv(u^1,\dots,u^\ell)&=& \frac{1}{2} \underset{i=1}{\overset{\ell}{\prod}} (1+u^i) +\frac{1}{2} \underset{i=1}{\overset{\ell}{\prod}} (1-u^i) \ .
\eea

\subsection{Stability analysis of the trivial fixed-point via moment expansions}

The functions $f(\{u^i\})$ and $\hf(\{m^i\})$ defined in the equations (\ref{eq_occ_BP_utom},\ref{eq_occ_BP_mtou}) vanish when all their arguments are equal to $0$, which traduces the stationarity of the unbiased distribution $\oeta$. As a consequence the distributions $P^{(n)}(m) = \delta(m)$, $\hP^{(n)}(u) = \delta(u)$ form a fixed point of the cavity equations. Following the same strategy as in Sec.~\ref{sec_expansion_Potts} we shall now investigate its stability, by first locating the Kesten-Stigum transition where the trivial fixed point go from stable to unstable, and then looking for a bifurcating non-trivial fixed point in a neighborhood of the transition.

\subsubsection{Linear analysis}

Linearizing for small arguments the functions $f(\{u^i\})$ and $\hf(\{m^i\})$  defined in (\ref{eq_occ_BP_utom},\ref{eq_occ_BP_mtou})  and inserting this expansion in the conditional distribution recursions (\ref{eq_occ_conditional_hPtoP},\ref{eq_occ_conditional_PtohP}) yields very easily
\beq
\E_+^{(n+1)}[m] = \tE[\ell]  \E_+^{(n)}[u] \ , \qquad \E_+^{(n)}[u] = (k-1) \gamma_2 \E[\t_1|+] \E_+^{(n)}[m] \ ,
\eeq
with $\E[\t_1|+]=\gamma_2$ as proven before in (\ref{eq_occ_ave_spin_cond}). Putting these two equations together, and expressing the conditional first moments in terms of the unconditional second moments according to (\ref{eq_occ_moment_ids}) gives
\beq
\E^{(n+1)}[m^2] = \tE[\ell] (k-1) \gamma_2^2 \E^{(n)}[m^2] \ .
\eeq
This implies that the Kesten-Stigum transition occurs when 
\beq
\tE[\ell] (k-1) \gamma_2^2 = 1 \ ;
\label{eq_Ising_kwise_KS}
\eeq
in the following we shall assume that the second Fourier coefficient does not vanish, $\gamma_2 \neq 0$, for this bifurcation to occur at a finite value of $\tE[\ell]$ (this excludes notably the case of a XORSAT constraint for $p_{\rm j}$ whenever $k \ge 3$).

\subsubsection{Second order expansion: the continuity of the Kesten-Stigum transition}
\label{sec:continuity}

We look now for a fixed point distribution $P(m)$ in the neighborhood of the Kesten-Stigum transition; we need to make an ansatz on the behavior of its moments in order to reduce the functional bifurcation problem to a finite-dimensional one, as already explained for the case of Potts variables in Sec.~\ref{sec_expansion_Potts}. Thanks to the spin reversal symmetry we known that for all odd moments $\E[m^{2p+1}]=\E[u^{2p+1}]=0$; we assume the existence of a small parameter $\kappa$ such that the even moments scale as $\E[m^{2p}]=O(\kappa^p)$, $\E[u^{2p}]=O(\kappa^p)$. Pushing the expansion of $f(\{u^i\})$ and $\hf(\{m^i\})$ to the order needed to obtain the first correction to the variances of $u$ and $m$ yields after a short computation:
\bea
\E[m^2] &=& \tE[\ell] \E[u^2] - \tE[\ell(\ell-1)]  (\E[u^2])^2 + O(\kappa^3) \ , \\
\E[u^2] &=& (k-1) \gamma_2^2 \E[m^2] - (k-1)(k-2) \gamma_2^3  (\E[m^2])^2 + O(\kappa^3) \ .
\eea
Eliminating the variance of $u$ gives us a single equation on $\E[m^2]$; we shall denote $a=\E[m^2]/2$ in such a way that this quantity corresponds to the one of Sec.~\ref{sec_expansion_Potts}, which obeys
\beq
a = \tE[\ell] (k-1) \gamma_2^2 \, a - 2 \, \tE[\ell] (k-1)(k-2) \gamma_2^3  \, a^2 - 2 \, \tE[\ell(\ell-1)] ((k-1) \gamma_2^2 a)^2 + O(\kappa^3) \ .
\eeq
Denoting $\tE[\ell] (k-1) \gamma_2^2 = 1 + \epsilon$, with $\epsilon$ parametrizing the distance to the Kesten-Stigum transition, dividing the equation by $a$, and taking $\epsilon=0$ in the correction term gives
\beq
0 = \epsilon -  2 \left[ 
(k-2) \gamma_2 + \frac{\tE[\ell(\ell-1)]}{\tE[\ell]^2}
\right] a \ .
\label{eq_occ_bifurcation}
\eeq
We shall show now that the expression in the square bracket is non-negative for all the models encompassed by the study of this section, hence that the bifurcating solution (which must certainly obey the positivity condition $a=\E[m^2] >0$) always exists for $\epsilon >0$, i.e. in the regime of parameters where the trivial fixed point is unstable. To prove our claim we first notice that the Fourier coefficient $\gamma_2$ lies necessarily in the interval $\left[-\frac{1}{k-1},1 \right]$. Indeed, from the interpretation of these coefficients as spin averages given in (\ref{eq_occ_def_gamman}) we obtain
\beq
\E[(\t_1+\dots+\t_k)^2] = \E[k \t_1^2 + k(k-1)\t_1 \t_2 ] 
= k (1+ (k-1) \gamma_2) \ge 0 \Rightarrow \gamma_2 \ge -\frac{1}{k-1} \ ,
\eeq
where we have merely exploited the permutation invariance and the fact that Ising spins square to 1. Moreover the upper-bound $\gamma_2 = \E[\t_1 \t_2] \le 1$ is obvious. Then we rewrite the square bracket of (\ref{eq_occ_bifurcation}) as
\beq
\frac{\tE[\ell^2]-\tE[\ell]^2}{\tE[\ell]^2} + 1 -\frac{1}{\tE[\ell]} +(k-2) \gamma_2 = \frac{\tE[\ell^2]-\tE[\ell]^2}{\tE[\ell]^2} +( 1 -(k-1)\gamma_2^2  +(k-2) \gamma_2 ) \ ,
\label{eq_occ_continuity}
\eeq
where we used the Kesten-Stigum condition $\tE[\ell] (k-1) \gamma_2^2 = 1$. The first term above is proportional to the variance of the offspring distribution, hence non-negative; as $1-(k-1)x^2 + (k-2) x \ge 0$ for $x \in [-1/(k-1),1]$, the authorized interval of variation of $\gamma_2$, the second term in (\ref{eq_occ_continuity}) is also non-negative.

\subsubsection{Third order expansion}

We have also pushed the expansion in moments to the next order; compared to the Potts case the computations are much easier, thanks on the one hand to the binary nature of the variable (hence the moments to be determined are scalar quantities that do not bear any spin index), and on the other hand to the spin-flip symmetry (which cancels the moments of odd order). Let us introduce for convenience some more compact notations, that are reminiscent of the ones used in Sec.~\ref{sec_Potts}:
\beq
a=\frac{1}{2} \E[m^2] \ , \qquad b=\E[m^4] \ , \qquad \ha=\frac{1}{2} \E[u^2] \ , \qquad \hb=\E[u^4] \ .
\eeq
These quantities obey the following set of equations:
\bea
a &=& \tE[\ell] \, \ha - 2 \, \tE[\ell(\ell-1)]  \, \ha^2 
+ \tE[\ell(\ell-1)] \, \ha \, \hb + \frac{20}{3} \, \tE[\ell(\ell-1)(\ell-2)] \, \ha^3 
 \ , \label{eq_occ_expansion_1} \\
b &=& \tE[\ell] \, \hb + 12 \, \tE[\ell(\ell-1)]  \, \ha^2 \ , \nonumber \\
\ha &=& (k-1) \gamma_2^2 \, a - 2 (k-1)(k-2) \gamma_2^3  \, a^2 
+ (k-1)(k-2) \gamma_2^4 \, a \, b 
+
4 (k-1)(k-2) (k-3) \left(\frac{\gamma_4^2}{6} - \gamma_2^2 \gamma_4 + \frac{5 \gamma_2^4 }{2} \right) 
\, a^3  \ , \label{eq_occ_expansion_3} \nonumber \\
\hb &=& (k-1) \gamma_2^4 \, b + 12 \, (k-1)(k-2) \gamma_2^4  \, a^2  \label{eq_occ_expansion_4} \ ,
\nonumber
\eea
where in $a$ and $\ha$ (resp. in $b$ and $\hb$) we neglected terms of order $\kappa^4$ (resp. $\kappa^3$).

The expansion of the free-entropy to the corresponding order yields
\bea
\phi (a,b,\ha,\hb) &=& 
 (k-1) \gamma_2^2 \, a^2 - \frac{4}{3} (k-1)(k-2) \gamma_2^3 \, a^3 + \frac{1}{24} (k-1) \gamma_2^4 \, b^2 + (k-1)(k-2) \gamma_2^4 \, a^2 \, b \\ && + 2 (k-1)(k-2) (k-3) \left(\frac{\gamma_4^2}{6} - \gamma_2^2 \gamma_4 + \frac{5 \gamma_2^4 }{2} \right) \, a^4 \nonumber \\ &&
+ \tE[\ell] \, \ha^2 - \frac{4}{3} \, \tE[\ell(\ell-1)] \, \ha^3 + \frac{1}{24} \, \tE[\ell] \, \hb^2
+ \tE[\ell(\ell-1)] \, \ha^2 \, \hb + \frac{10}{3} \, \tE[\ell(\ell-1)(\ell-2)] \, \ha^4
\nonumber \\ && -2 \, a \, \ha -\frac{1}{12} \, b \, \hb \ ; \nonumber
\eea
one can check that the derivatives of this expression vanish when the cavity equations (\ref{eq_occ_expansion_1}) are fulfilled, a property which arises from the variational character of the free-entropy. Another short computation also reveals that the free-entropy, computed on the non-trivial perturbative fixed-point which here always exists for $\epsilon >0$, behaves as $\epsilon^3$ (with a positive prefactor) when $\epsilon \to 0^+$.

\subsection{Numerical results for two examples}

In order to confirm our analytical expansions, and to complement them with a global bifurcation analysis, we have solved numerically the cavity equations for two cases of occupation models, as we shall now detail.

\subsubsection{Bicoloring}
\label{sec_bicoloring}

We considered first the bicoloring problem, for which the joint probability $p_{\rm j}(\s_1,\dots,\s_k)$ is $0$ if $\s_1=\dots=\s_k = \pm 1$, $1/(2^k-2)$ otherwise. This constraint forbids monochromatic hyperedges, and give an equal weight to all configurations with at least one positive and at least one negative variable around it. One easily finds the corresponding value of the Fourier coefficients,
\beq
\gamma_n = - \frac{1}{2^{k-1}-1}
\eeq
for all even $n$ between $2$ and $k$ (the odd coefficients vanish due to the up-down symmetry). We used for the degree distributions $p_\ell = \tp_\ell$ Poisson laws of average $\alpha k$, that corresponds to a random hypergraph with $M = \alpha N$ constraints. The criterion (\ref{eq_Ising_kwise_KS}) shows then that the Kesten-Stigum transition happens at
\beq
\alpha_{\rm KS}= \frac{(2^{k-1}-1)^2}{k(k-1)} \ .
\label{eq_bicoloring_alphaKS}
\eeq

\begin{figure}
\includegraphics[width=5.5cm]{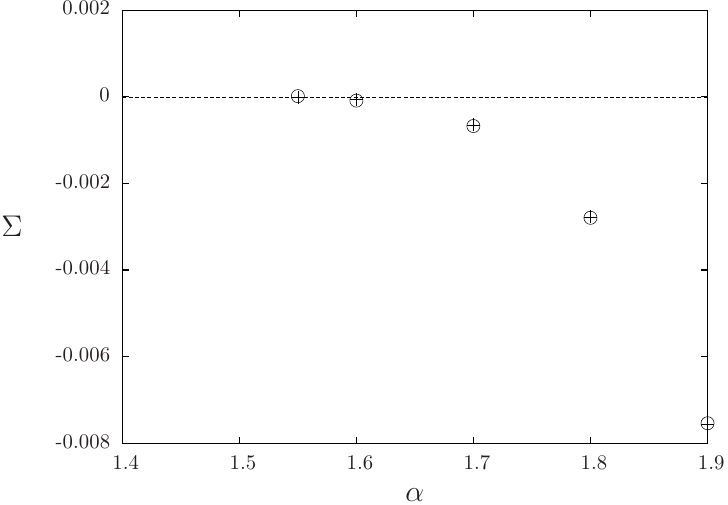}
\includegraphics[width=5.5cm]{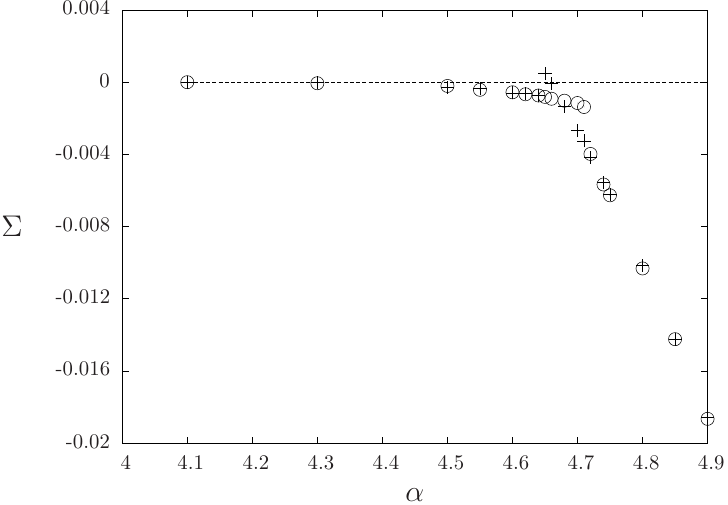}
\includegraphics[width=5.5cm]{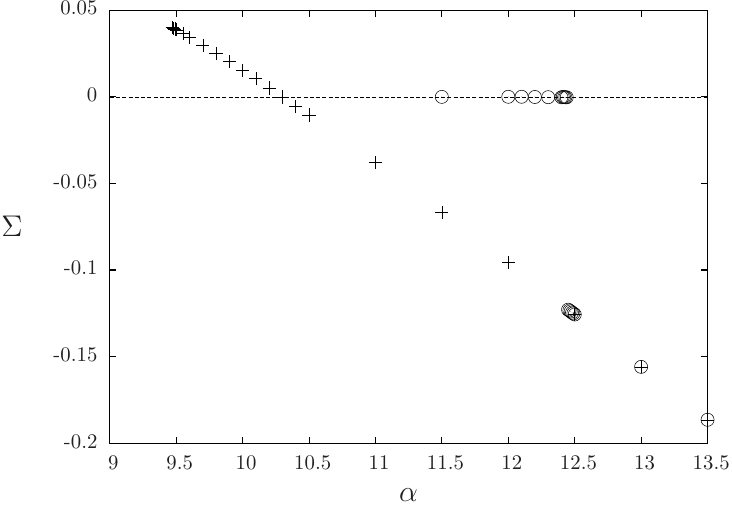}
\caption{The complexity $\Sigma = - \phi$ for the fixed point solutions of the cavity equations for the hypergraph bicoloring problem, for $k=3,4,5$ from left to right; the symbols are the same as in Fig.~\ref{fig_bicoloring_q_main}, crosses correspond to an informative initial condition, circles to an uninformative one.}
\label{fig_bicoloring_Sigma}
\end{figure}

The figure \ref{fig_bicoloring_q_main} discussed in Sec.~\ref{sec_mainresults_occ} presents the variance $a=\E[m^2]/2$ of the fixed-point solutions of the cavity equations reached from the reconstruction (informative) and robust reconstruction (uninformative) initial condition. As discussed there our analytical prediction on the existence of a bifurcating solution above the Kesten-Stigum transition is confirmed by this numerical results, and for $k=3,4,5$ we find explicit realizations of the bifurcation diagrams presented in the left panel of Fig.~\ref{fig:standard} and in both panels of Fig.~\ref{fig:new}. On Fig.~\ref{fig_bicoloring_Sigma} we plot the complexity (i.e. minus the free-entropy) of these fixed-points; the location of the discontinuity in the derivative of the solution that minimizes the complexity defines the threshold $\alpha_{\rm IT}$ reported in Table~\ref{tab_bicoloring_main}. We chose here to plot the opposite of the free-entropy as it is a more familiar quantity in the context of random, instead of planted, constraint satisfaction problems. The interpretation of this coexistence of solutions in this context has been discussed, also on the example of the bicoloring, in~\cite{gabrie2017phase}.

\subsubsection{2-in-4 satisfiability}
\label{sec:2in4}

We have also considered the so-called 2-in-4 satisfiability model in the nomenclature of~\cite{ZdeborovaMezard08b}, defined by $k=4$ and $p_{\rm j}(\s_1,\s_2,\s_3,\s_4) = \ind(\s_1 + \s_2 + \s_3 + \s_4 =0)/6$: the constraint imposes that among the four variables exactly two are positive and two negative. The corresponding Fourier coefficients are 
\beq
\gamma_1=\gamma_3=0 \ , \qquad \gamma_2 = -\frac{1}{3} , \qquad 
\gamma_4 = 1 \ ;
\eeq
note that this case saturates the lower-bound on $\gamma_2$.

In our numerical investigation we have used truncated Poisson distributions for $p_\ell$ and $\tp_\ell$, namely
\beq
p_\ell = \frac{1}{e^c - 1 - c} \frac{c^\ell}{ \ell!} \, \ind(\ell \ge 2) \ ,
\qquad
\tp_\ell = \frac{1}{e^c - 1} \frac{c^\ell}{ \ell!} \, \ind(\ell \ge 1) \ .
\eeq
The averages of these distribution are
\beq
\E[\ell] = c \frac{1-e^{-c} }{1-(c+1)e^{-c}}\ , \qquad
\tE[\ell] = \frac{c}{1-e^{-c}} \ .
\eeq
In terms of this parameter $c$ the Kesten-Stigum transition is thus found to happen at $c_{\rm KS} \approx 2.82144$.

Note that this is a locked model in the sense of~\cite{ZdeborovaMezard08b}, as the minimal degree of a variable is 2 and no pair of configurations allowed by $p_{\rm j}$ differ from one single spin flip: the typical solutions of such a Constraint Satisfaction Problem on a locally tree-like factor graph are separated by an Hamming distance that diverges with the graph size. Nevertheless this property does not have any effect on the existence of a continuously bifurcating solution above the Kesten-Stigum transition. In some sense the locked property is a ``high-frequency condition'' on $p_{\rm j}$, whereas we have shown that the criterion (\ref{eq_occ_continuity}) only depends on the low frequency Fourier coefficient $\gamma_2$.

Indeed our numerical resolution of the cavity equations showed that a continuously growing solution exists right above the Kesten-Stigum transition, and disappears at $c_{\rm alg} \approx 2.84$. Note that the interval $[c_{\rm KS},c_{\rm alg}]$ is very small, which explains why it remained unnoticed in~\cite{ZdeborovaKrzakala09}. This model thus falls in the scenario sketched on the right panel of Fig.~\ref{fig:new}, with the other thresholds $c_{\rm sp} \approx 1.256$ and $c_{\rm IT} \approx 1.853$ correctly determined in~\cite{ZdeborovaKrzakala09}. Actually for locked models the high-accuracy fixed point is perfectly informative, as it corresponds to $P_\t(m)=\delta(m-\t)$; the transition at $c_{\rm sp}$ corresponds to a change of stability of this perfectly informative fixed point, that can be tested by taking $\varepsilon \to 1$ after $n \to \infty$ in (\ref{eq_initialcondition}), a procedure termed small noise reconstruction in~\cite{ZdeborovaMezard08b}.

\section{Numerical experiments on single samples}
\label{sec_numerical}

This section is devoted to a study of the behaviour of the Belief
Propagation (BP) algorithm run on given instances of two inference
problems defined and studied analytically in the previous sections, 
namely the planted hypergraph bicoloring and the asymmetric SBM with two 
groups. As a matter of fact, most of the analytical results presented in the 
rest of the paper directly apply to (infinite) tree reconstruction problems, 
their interpretation in terms of inference on (large but) finite size graphs
relying on delicate conjectures of the cavity method, as briefly discussed in
Sec.~\ref{sec_link_graphs}. The experiments reported in this section will
allow us to test this connection, and to confront quantitatively the 
predictions of the tree reconstruction problem, studied numerically via the 
population dynamics algorithm, and the results of BP on single large instances.
Because of the local convergence of the graph problems towards trees it is 
quite simple to see that a finite number of iterations of BP can be described
analytically, in the thermodynamic limit, by a finite number of the tree 
distributional iterations (\ref{eq_generic_conditional_hPtoP},\ref{eq_generic_conditional_PtohP}). There are, however, two aspects that make the graph-tree 
connection much less trivial and justify these numerical tests: (i) on finite graphs BP can be run until convergence to a fixed point, i.e. for a number of iterations much larger than the girth of the graph, this regime cannot a priori be described in terms of tree-like local properties and could be sensitive to the long cycles of the graph; (ii) in the graph problems there is initially no observation of the labels on the vertices, the infinitesimal information of the tree robust reconstruction problem must thus arise from the amplification of noise in the initial condition of BP.

\subsection{Generation of planted problems, BP equations and observables}

The models that we study numerically, namely random
hypergraph bicoloring and asymmetric SBM with 2 groups,
have been already defined in detail in
Sec.~\ref{sec_definitions_graphs}.
Here we just give additional informations about the generation of
the instances with a planted solution, and about the
BP equations and the way we solve them, for the convenience of the reader
who would like to repeat our numerical tests.
Both models have Ising (i.e.\ binary) variables
$\s_i\in\{-1,1\}$, and thus the marginals can be written
in terms of a single scalar variable.

\subsubsection{Random hypergraph bicoloring}

The random ensemble of hypergraph bicoloring problems has three 
parameters: the number of variables $N$, the number
of constraints $M=\alpha N$, and the degree 
of the factor nodes $k$ (i.e.\ the number of variables per
constraint).
The $N$ variables are divided in 2 groups of equal size
and the planted configuration is defined as $\s_i^*=1$ in
the first group and $\s_i^*=-1$ in the second one.
For each of the $M$ constraints we extract $k$ variables
uniformly at random, conditional on the $k$ variables not
belonging all to the same group (we implemented this condition by a
rejection method).
The random graph thus obtained can be described by the
set $E$ of edges $(ia)$ connecting variables nodes and
factor nodes ($1\le i\le N$, $1\le a\le M$ and $|E|=KM$).

One can write the posterior distribution of $\us$ given the observation of 
the graph and treat this probability measure with the BP algorithm, that
can be put after some simplifications under the form of messages 
$\eta_{i\to a}$ and $\hat\eta_{a\to i}$ passed between variables and interactions
obeying the following equations:
\begin{eqnarray}
\eta_{i\to a}^{(t)} &=& \frac{\prod_{b\in\dima} \hat\eta_{b\to i}^{(t-1)}}
{\prod_{b\in\dima} \hat\eta_{b\to i}^{(t-1)} +
\prod_{b\in\dima} (1-\hat\eta_{b\to i}^{(t-1)})} \ , \label{eq:RHBeq1}\\
\hat\eta_{a\to i}^{(t)} &=& \gamma\,\hat\eta_{a\to i}^{(t-1)} + (1-\gamma)
\frac{1-\prod_{j\in\dami} \eta_{j\to a}^{(t)}}
{2-\prod_{j\in\dami} \eta_{j\to a}^{(t)}
-\prod_{j\in\dami} (1-\eta_{j\to a}^{(t)})} \ ,
\label{eq:RHBeq2}
\end{eqnarray}
where $\di=\{a:(ia)\in E\}$ and $\da=\{i:(ia)\in E\}$ are the local neighborhoods of variable and factor nodes respectively, and $\gamma$ is a damping factor set to $0.5$ in most of our numerical simulations.
The BP messages $\eta_{i\to a}$ and $\hat\eta_{a\to i}$ represent the probability that variable $i$ belongs to a given group (say the first group) in a modified graph where some of the edges have been removed: in particular $\eta_{i\to a}$ considers the graph where constraint $a$ has been removed, while $\hat\eta_{a\to i}$ considers the graph where constraints in $\partial i \setminus a$ have been removed.

BP messages are initialized in the following way:
with probability $q_0$ we set $\hat\eta_{a\to i}^{(0)}=\eta_{i\to a}^{(0)}=\ind[\s_i^*=1]$ and with probability $1-q_0$ we set $\hat\eta_{a\to i}^{(0)}=0.5$ and $\eta_{i\to a}^{(0)}\in[0.45,0.55]$ uniformly at random. The parameter $q_0$ (the $q_0$ and $q$ of this section should not be confused with the number of states of the Potts model of the rest of the paper) thus controls the amount of direct information on the planted configuration we use in this initial condition.
Of course only $q_0=0$ should be considered if we want to study BP as
an inference algorithm that does not use any information on the hidden
labels. It is, however, useful to allow arbitrary values of $q_0$ as a tool to investigate the connections between the tree and graph problems. The choice of distributing the messages in the interval $[0.45,0.55]$ for the non-informed vertices is made to avoid the trivial fixed point with all messages $\eta=\hat\eta=0.5$.

BP messages are not updated all in parallel, as the time indices in Eqs.~(\ref{eq:RHBeq1},\ref{eq:RHBeq2}) may suggest.
In each step of BP we visit all the variables once in a random order: for each variable $i$ we compute all the outgoing messages $\eta_{i\to a}$ according to Eq.~(\ref{eq:RHBeq1}) and immediately we update all the BP messages leaving neighbouring factor nodes according to Eq.~(\ref{eq:RHBeq2}).
By these tricks (and the use of damping) we avoid any undesirable oscillation and improve convergence to fixed points.
Convergence is declared achieved if, in a given step of BP, all messages change by less than a pre-fixed threshold, i.e.\ $|\eta_{i\to a}^{(t+1)}-\eta_{i\to a}^{(t)}|<10^{-8}$.

At a fixed point $\{\eta_{i\to a}^\star, \hat\eta_{a\to i}^\star\}$ of the BP equations, the local magnetizations are given by
\beq
m^\star_i = \frac{\prod_{a\in\di} \hat\eta_{a\to i}^\star - \prod_{a\in\di} (1-\hat\eta_{a\to i}^\star)}{\prod_{a\in\di} \hat\eta_{a\to i}^\star + \prod_{a\in\di} (1-\hat\eta_{a\to i}^\star)} \ .
\eeq
Because of the global spin-flip symmetry of the model the mean
magnetization $m =\sum_i m^\star_i / N$ is zero, while detection of
the planted solution is signalled by the following order parameters:
the staggered magnetization $m_s$ (i.e.\ the absolute value of the
mean overlap with the planted configuration), the magnetization
variance $m_2$ (recall that $\om=0$ here), that equals the mean overlap between two real replicas, and the maximum overlap $q$ with the planted configuration, defined as
\beq
m_s = \frac1N \left|\sum_i m^\star_i s_i^*\right| \,, \qquad m_2 = \frac1N
\sum_i (m^\star_i)^2\,, \qquad q = \frac1N \left|\sum_i \sign(m^\star_i)
  s_i^*\right|\,.
\label{eq:orderParam}
\eeq
The Bethe (replica symmetric) entropy is given by the following expression
\bea
S &=& \frac{1}{N}\sum_i \log\left[\prod_{a\in\di} \hat\eta_{a\to i}^\star+\prod_{a\in\di} (1-\hat\eta_{a\to i}^\star)\right]
+ \frac{1}{N} \sum_a \log\left[ 1 - \prod_{i\in\da} \eta_{i\to a}^\star - \prod_{i\in\da} (1-\eta_{i\to a}^\star)\right]\nonumber\\
&&- \frac{1}{N} \sum_{(ia)} \log\left[\hat\eta_{a\to i}^\star \eta_{i\to a}^\star + (1-\hat\eta_{a\to i}^\star) (1-\eta_{i\to a}^\star) \right] \ ,
\label{eq_Bethe_S_bicol}
\eea
that reduces to $S_\text{para} = \big[\log(2)+\alpha \log(1-2^{1-k})\big]$ on the paramagnetic (trivial) BP fixed point, where all messages are non-informative $\hat\eta_{a\to i}^\star=\eta_{i\to a}^\star=1/2$.

\subsubsection{Asymmetric two groups SBM}

An instance of the asymmetric SBM with 2 groups is generated using the
following 4 parameters: the number of variables $N$, the mean degree $d$,
the asymmetry $\om$ and the parameter $\theta$ related to the SNR. The
$N$ variables are divided in 2 groups of sizes $N_1=\frac{1+\om}{2}N$
and $N_2=\frac{1-\om}{2}N$, such that the planted configuration is
$\s_i^*=1$ in the first group and $\s_i^*=-1$ in the second
group. Then the edge set $E$ of the interaction graph between the
variable nodes is generated by selecting uniformly at random
$M_{11}=\frac{N_1^2}{2}\frac{c_{++}}{N}$ edges among variables in the
first group, $M_{22}=\frac{N_2^2}{2}\frac{c_{--}}{N}$ edges among
variables in the second group and $M_{12}= N_1 N_2 \frac{c_{+-}}{N}$
edges joining variables in different groups, where
$c_{\s\t}=d(1+\theta\frac{(\s-\om)(\t-\om)}{1-\om^2})$. It is easy to
check that
$\frac{M_{11}}{N_1}+\frac{M_{12}}{2N_1}=\frac{M_{22}}{N_2}+\frac{M_{12}}{2N_2}=\frac{d}{2}$,
i.e.\ the mean degree in each group is the same.  It is worth noticing
that these rules define a ``microcanonical'' ensemble of random graphs
having a \textit{fixed} number of edges. In the ``canonical'' ensemble
of random graphs, each possible edge is chosen independently with a
given probability and thus the number of edges is a random variable.
The difference between the two ensemble vanishes in the large $N$
limit, and produces fluctuations of order $O(1/\sqrt{N})$ in intensive 
observables
measured in finite graphs. The BP equations for the posterior measure of the SBM were written in the general case in~\cite{DecelleKrzakala11,Decelle_SBM_long}, we reproduce here this derivation for the asymmetric two group model.

We shall rewrite the posterior probability as the Gibbs measure of an Ising model
\beq
\mathbb{P}[\us] \propto \exp\bigg[\sum_i H_i \s_i + \sum_{i<j} J_{ij} \s_i \s_j \bigg]\ ,
\eeq
where here and in the following $\propto$ denotes proportionality up to a constant independent of the spin variables.
Indeed the prior is given by
\beq
\prod_i \frac{1+\om\,\s_i}{2} \propto e^{H \sum_i \s_i} \quad
\text{with} \quad \tanh(H) = \om\,,
\eeq
while the likelihood is proportional to
\beq
\prod_{(ij)\in E} \frac{1}{N} c_{\s_i\s_j} \prod_{(ij)\notin E}
\left(1-\frac{1}{N} c_{\s_i\s_j}\right) \simeq \prod_{(ij)\in E}
\frac{1}{N} c_{\s_i\s_j}\;\exp\bigg(-\frac{1}{N} \sum_{i<j}
c_{\s_i\s_j}\bigg)\,,
\eeq
where 
$c$ is the affinity matrix with which the edges have been generated.
We have approximated the product over the non-edges (i.e.\
pairs of vertices not connected by an edge) with a sum over all pairs
of variables, the difference being a correction of order $O(1/N)$ in
the large $N$ limit. The following algebraic relation that holds for
$\s,\t\in\{-1,1\}$
\bea
c_{\s\t} &=& d \left(1+\theta\frac{(\s-\om)(\t-\om)}{1-\om^2}\right) \propto
e^{J\s\t-K(\s+\t)} \quad\text{with}\\
J &=& \frac14
\log\left[\frac{(1+\theta)^2-\om^2(1-\theta)^2}{(1-\theta)^2(1-\om^2)}\right] \ ,\\
K &=& \frac14
\log\left[\frac{1-\om^2+\theta(1+\om)^2}{1-\om^2+\theta(1-\om)^2}\right] \ ,
\eea
allows to rewrite the first term of the likelihood in exponential
form
\beq
\exp\left[J \sum_{(ij)\in E} \s_i \s_j - K \sum_i d_i \s_i \right]\;.
\eeq
The second term of the likelihood (the one given by all the edges, including those absent from the graph)
provides a term proportional to
\beq
\exp\left[-\frac{d\theta}{2(1-\om^2)N}\left(\sum_i \s_i - N \om\right)^2\right] \ .
\eeq
So the posterior distribution that we have to study is given by the
following expression
\beq
\mathbb{P}[\us] \propto \exp\bigg[ \sum_i \Big(H - d_i K +
\frac{d\theta\om}{1-\om^2}\Big) \s_i + J \sum_{(ij)\in E} \s_i \s_j -
\frac{d\theta}{(1-\om^2)N} \sum_{i<j} \s_i \s_j \bigg] \,,
\label{eq:IsingSBM}
\eeq
that corresponds to an Ising model with local fields and couplings
given by
\bea
H_i &=& H - d_i K + \frac{d\theta\om}{1-\om^2} \ ,\\
J_{ij} &=& \left\{
\begin{array}{rl}
J-\frac{d\theta}{(1-\om^2)N} \simeq J & \quad\text{if}\;\;(ij)\in E\,,\\
-\frac{d\theta}{(1-\om^2)N} & \quad \text{if}\;\; (ij)\notin E\,.
\end{array}
\right.
\eea
The corresponding BP equations are very well known
\beq
u_{i\to j} = \atanh\left[ \tanh(J_{ij}) \tanh\bigg( H_i + \sum_{k\neq j} u_{k\to i}\bigg)\right] \ .
\eeq
Unfortunately these BP equations involve $N(N-1)$ messages $u_{i\to
  j}$ with $i\neq j$ (we assume $u_{i \to i}=0$) and we need to
simplify the if we want BP to run in a time linear in the system size.
The main observation to achieve such a simplification is that BP
messages running along the non-edges, i.e.\ sent between vertices not
connected in $E$, are $O(1/N)$ and thus very small. The following
equations hold up to terms of order $O(1/N)$ if $(ij)\notin E$:
\beq 
u_{i \to j} = J_{ij} \tanh\bigg(H_i + \sum_{k \neq j} u_{k\to i}\bigg)
= J_{ij} m_i\, ,
\eeq
where the first equality comes from $J_{ij}$ being $O(1/N)$, while the
second one from ignoring $u_{j\to i}\sim O(1/N)$ in the definition
of local magnetization
\beq
m_i \equiv \tanh\bigg(H_i + \sum_k u_{k\to i}\bigg) \, .
\label{eq:BPlocalMag}
\eeq
Eliminating BP messages on non-edges via the substitution $u_{i\to j}
= -d\theta m_i/((1-\om^2)N)$ we end up working only with the $O(N)$ BP
messages running on the edges in $E$. These are updated according to the
following iterative equations
\beq
u_{i\to j}^{(t+1)} = \gamma u_{i\to j}^{(t)} + (1-\gamma) \atanh\left[
  \tanh(J) \tanh\bigg(\widetilde{H}_i^{(t)} + \sum_{k\in \partial i
    \setminus j} u_{k\to i}^{(t)}\bigg)\right]\, ,
\label{eq:BPforSBM}
\eeq
where $\partial i = \{j:(ij)\in E\}$ and we have redefined the local
fields including also the effect of the non-edges
\beq
\widetilde{H}_i^{(t)} = H - d_i K +  \frac{d\theta}{1-\om^2}\bigg(\om -
\frac{1}{N} \sum_\ell m_\ell^{(t)}\bigg)\, .
\eeq
The local magnetizations are defined via
\beq
m_i^{(t)} = \tanh\left(\widetilde{H}_i^{(t)} + \sum_{k\in \partial i}
  u_{k\to i}^{(t)}\right)\,.
\eeq

We solve equations (\ref{eq:BPforSBM}) starting from a set of BP
messages strongly biased towards a configuration $\ut^0$, that is
$u_{i\to j}^{(0)} = L \t_i^0$, where $L$ is a large number (we fix
$L=100$ in our simulations, but the results are independent on
$L$). 
Again we call $q_0$ the similarity between the starting
configuration and the planted one, that is the components of $\ut^0$
are independent random variables distributed according to
\beq
\mathbb{P}[\t^0_i = s] = q_0\,\delta_{s,\s^*_i} +(1-q_0)
\frac{1+\om\,s}{2} \, .
\label{eq:tau0distr}
\eeq
BP messages are updated in a random order and convergence is
determined by the condition
\beq
\left|m_i^{(t+1)} - m_i^{(t)}\right| < 10^{-8} \quad
\forall i\,.
\eeq
At the BP fixed point, the local magnetizations are given by
\beq
m^\star_i = \tanh\bigg(\widetilde{H}^\star_i + \sum_{j \in \partial i}
u^\star_{j\to i}\bigg)\,.
\eeq
The parameters $m_s$, $m_2$ and $q$ defined in
Eq.~(\ref{eq:orderParam}) signal again the 
detection of the planted configuration (for $\om \neq 0$ the absolute
value in the definition of $m_s$ is not strictly required, but keeping
it is not an error). It is worth noticing that in this case the
non-informative fixed point is characterized by the BP messages $u_{i\to
  j}^\star=K$ and by the order parameters $m_s=m_2=\om^2$
and $m_2=\om$, due to the asymmetry in the model.

With a little bit of algebra, the Bethe RS free-entropy can be
simplified to the following form,
\bea
\log(Z) &=& N(1-d)\log(2) - \sum_i \frac{1-d_i}{2}
\log\Big(1-(m^\star_i)^2\Big) + 
\frac{d\theta}{2(1-\om^2)N} \bigg(\sum_i m^\star_i\bigg)^2+\nonumber\\
&&\sum_{(ij)\in E} \log \sum_{\s_i,\s_j}
\exp\bigg[\Big(\widetilde{H}^\star_i+\sum_{k\in\partial i\setminus j}
u^\star_{k\to i}\Big)\s_i +
\Big(\widetilde{H}^\star_j+\sum_{k\in\partial j\setminus i}
u^\star_{k\to j}\Big)\s_j+J\s_i\s_j\bigg]\, .
\eea

\subsection{Results for random hypergraph bicoloring}

We present results for $k=3,4,5$ and system size $N=10^6$ (some
samples of size $N=10^5$ are shown just for the scaling of the
convergence time). Every time we present three plots in a row, the
left one refers to $k=3$, the central one to $k=4$ and the right one
to $k=5$.

\begin{figure}[htbp]
\begin{center}
\hspace{-0.15\textwidth}
\includegraphics[width=0.44\textwidth]{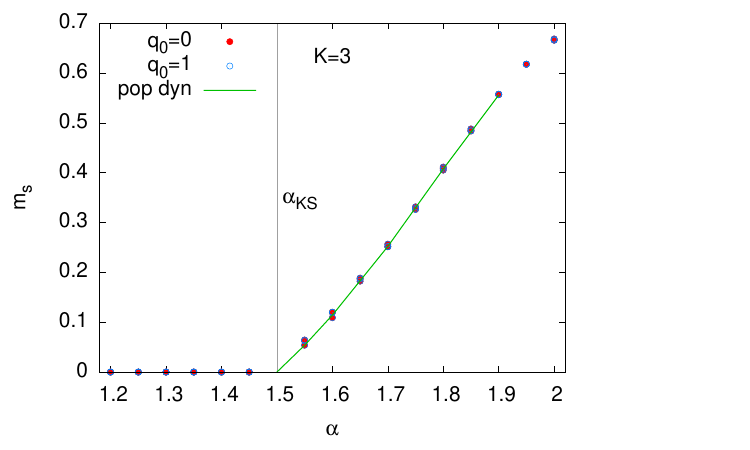}
\hspace{-0.1\textwidth}
\includegraphics[width=0.44\textwidth]{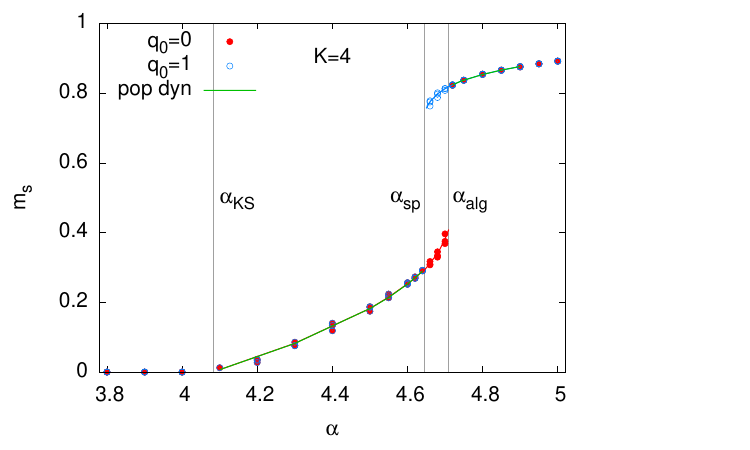}
\hspace{-0.1\textwidth}
\includegraphics[width=0.44\textwidth]{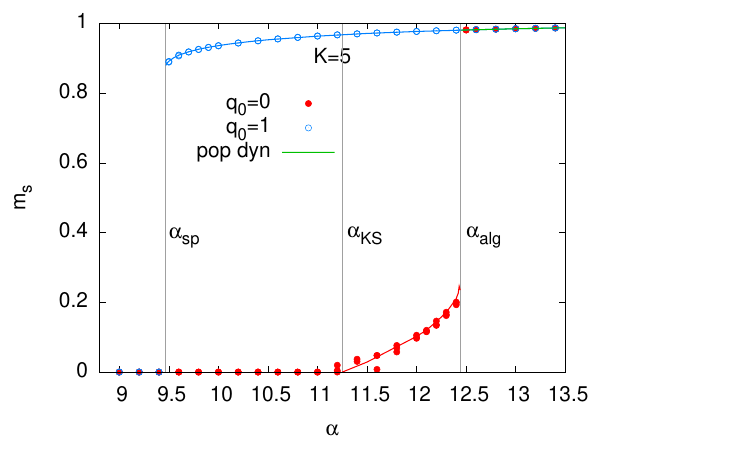}
\hspace{-0.2\textwidth}
\caption{The order parameter $m_s$ detecting the planted configuration
  for $k=3$ (left), $k=4$ (center) and $k=5$ (right), obtained by
  running BP on three samples of size $N=10^6$ for each 
  value $\alpha$. The lines are the predictions obtained by solving the cavity equations for
  the reconstruction problem on a tree, and plotted also in Fig.~\ref{fig_bicoloring_q_main}, with the correspondence $m_s=2 a$ explained in the text.}
\label{fig:bicol_ms}
\end{center}
\end{figure}

In Figure \ref{fig:bicol_ms} we show the order parameter $m_s$ that
detects the planted configuration. Red filled points correspond to
the BP fixed point reached starting with the $q_0=0$ initial condition
(i.e. with no direct information on the planted configuration), while blue open
points correspond to the BP fixed point reached from the $q_0=1$
initial condition (complete information on the planted configuration).  
For comparison we also draw with lines the
results obtained solving via population dynamics the cavity equations:
red (resp.\ blue) line corresponds to results obtained with an initial
condition (\ref{eq_initialcondition}) having $\varepsilon=10^{-3}$ (resp.\ $\varepsilon=1$), while
green line is the result that is obtained with both initial
condition (i.e.\ it is independent on the initial condition). These cavity equations results were already plotted in Fig.~\ref{fig_bicoloring_q_main} in slightly different units, namely $a=m_s/2$. Let us explain the reasoning behind this conversion. A technical statement of the tree-graph connection hypothesis of the cavity method is that the empirical distribution of the magnetizations $m_i^\star$ computed at a BP fixed point of a large finite graph, conditional on $\s_i^* = \t$, should be well approximated by the distribution $P_\t(m)$ (on the infinite tree), in formula
\beq
\frac{\sum_i g(m_i^\star) \delta_{\s_i^*,\t} }{\sum_i \delta_{\s_i^*,\t}} \approx \E_\t[g(m)] \ ,
\label{eq_conversion_graph_tree}
\eeq
for any function $g$. Applying this identity with $g(m)=m$, and recalling that for models with a global spin-flip symmetry $\E_+[m]=-\E_-[m]=\E[m^2]=2 a$ yields the identification $m_s=2 a$.

\begin{figure}[htbp]
\begin{center}
\hspace{-0.07\textwidth}
\includegraphics[width=0.45\textwidth]{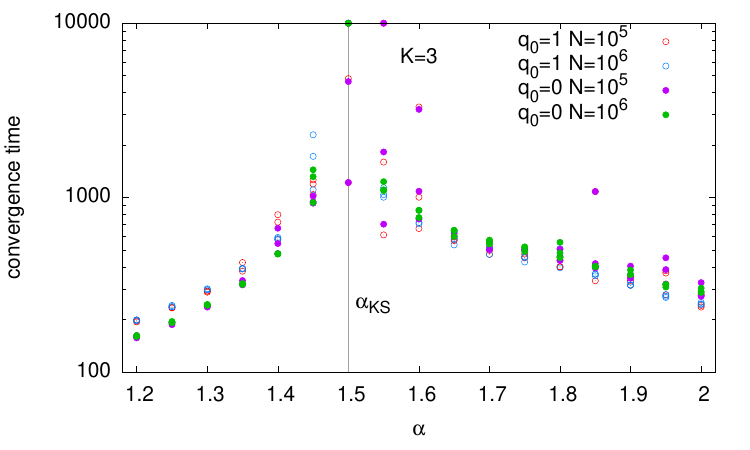}
\includegraphics[width=0.45\textwidth]{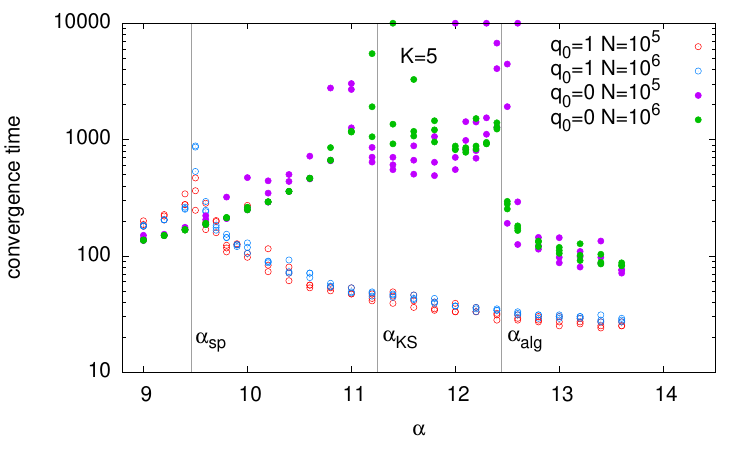}
\caption{Times to meet the BP convergence criterion (i.e.\ any
  marginal must change by less than $10^{-8}$) for $k=3$ (left) and
  $k=5$ (right), 2 system sizes ($N=10^5,10^6$) and 2 different
  initial conditions ($q_0=0,1$). A data point with a time $10^4$
  means BP did not reach convergence.}
\label{fig:bicol_times}
\end{center}
\end{figure}

For each $\alpha$ value we run BP on three different samples and we
report in the plots only the data corresponding to cases where the
convergence criterion was met within $10^4$ BP steps 
(Fig.~\ref{fig:bicol_times} shows that this happens most of the time). 
Zooming in the
plots one should be able to see three data points very nearby; if one
fails to see three different points it is because the data from different
samples perfectly coincide or because BP on some samples did not reach
convergence (the last case happen especially at $\alpha_{\rm KS}$ and
very close to it).

We notice that for all the cases where BP reaches a fixed point, the
latter is very well described by the cavity equations derived in the
thermodynamic limit for the tree reconstruction
problem. When different initial conditions of BP lead to different fixed points one observes, as expected, that the one reached with $q_0=0$ is associated to the robust reconstruction version of the tree problem ($\varepsilon \to 0$), while $q_0=1$ reproduces the reconstruction one ($\varepsilon=1$).
Sample-to-sample fluctuations can be appreciated only
slightly on the right of $\aKS$ and close to the spinodal points at
$\aSP$ and $\aALG$.

We observe that for $\alpha>\aKS$ BP with $q_0=0$, i.e. the version of the algorithm that does not use any direct information on the planted configuration, can actually detect it
in all the samples we have studied, although for
$\alpha<\aALG$ the detection is sub-optimal (hybrid-hard phase). For
$\alpha<\aALG$ BP can achieve the optimal detection only if
initialized with a $q_0$ value large enough (we shall discuss this
point later, when searching for the unstable fixed point, separating
the two stable fixed points already found).

In Figure \ref{fig:bicol_times} we show the convergence times of BP for $k=3$
(left) and $k=5$ (right). For each system size ($N=10^5,10^6$) and
initial condition ($q_0=0,1$) we study 3 samples.

For $k=3$ we observe that convergence times strongly increase around
the KS threshold, where also strong sample-to-sample fluctuations
arise and the size dependence can be appreciated (consider that
at $\aKS$ BP did not reach a fixed point for any of the $N=10^6$
samples). Away from $\aKS$ convergence times are size-independent, and
only weakly dependent on the initial condition for $\alpha<\aKS$.

For $k=5$ convergence times increase not only at $\aKS$, but also at
the spinodal points $\aSP$ and $\aALG$. The behavior of BP strongly
depends on the initial condition (recall that in the region
$\aSP < \alpha < \aALG$ BP reaches two different fixed points
depending on the value of $q_0$, see right panel in Figure
\ref{fig:bicol_ms}). The dependence on the system size between
$N=10^5$ and $N=10^6$ is not evident and sample-to-sample fluctuations
for $N=10^5$ are definitely larger.

\begin{figure}[htbp]
\begin{center}
\hspace{-0.15\textwidth}
\includegraphics[width=0.44\textwidth]{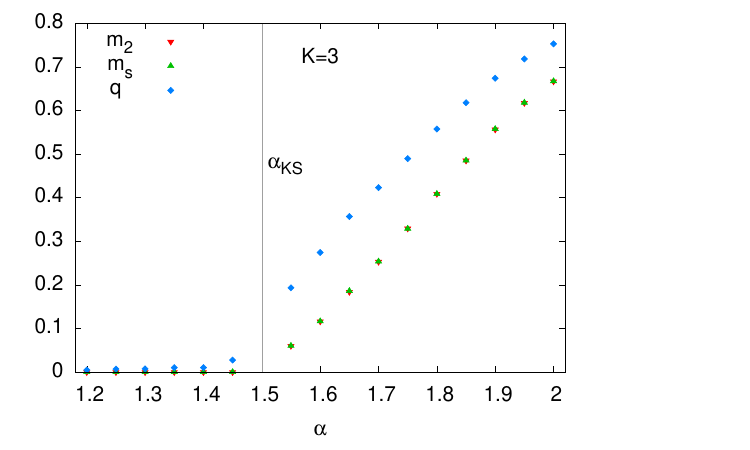}
\hspace{-0.1\textwidth}
\includegraphics[width=0.44\textwidth]{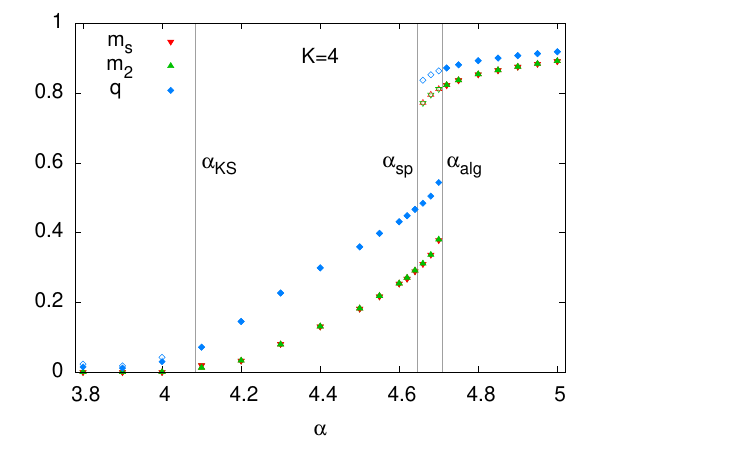}
\hspace{-0.1\textwidth}
\includegraphics[width=0.44\textwidth]{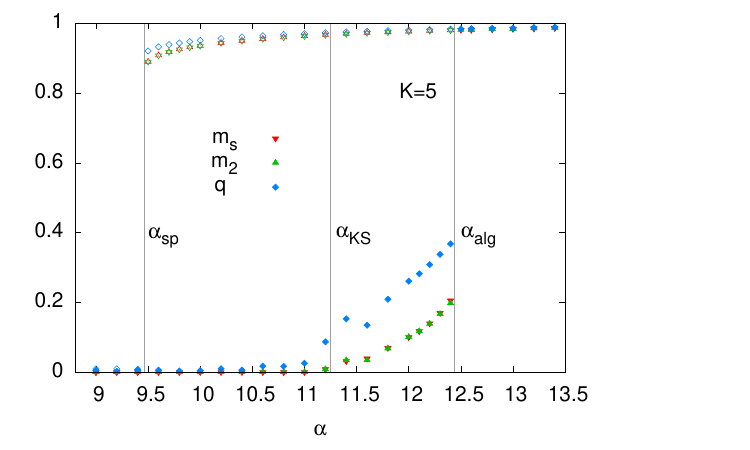}
\hspace{-0.2\textwidth}
\caption{Comparison between different order parameters detecting the
  planted configuration for $k=3$ (left), $k=4$ (center) and $k=5$ (right). Data have been averaged over the samples
  where BP reached a fixed point. As in Figure \ref{fig:bicol_ms},
  filled (resp.\ empty) points correspond to $q_0=0$ (resp.\
  $q_0=1$).}
\label{fig:bicol_cmp}
\end{center}
\end{figure}

In Figure \ref{fig:bicol_cmp} we show several order parameters that
can detect the planted configuration: $m_s$, $m_2$ and $q$. In order
to make the plot cleaner we have averaged over the samples where BP
reached a fixed point.

We notice that the Nishimori equality $m_s=m_2$ is very well satisfied in all
the BP fixed points we reached.  This is expected from the fact that
the planted configuration is a typical configuration of the posterior distribution, the mean
overlap with the planted configuration $m_s$ should be equal to the
mean overlap between two replicas $m_2$. One can also derive this property from the correspondence
(\ref{eq_conversion_graph_tree}) with the tree computation, and the consequence of the Bayes theorem stated as moment identities in Sec.~\ref{sec_Ising}.

As expected, the maximum overlap $q$ is the largest order parameter
(larger than $m_s$). According to Sec.~\ref{sec_estimator} $q$ should have 
a square root singularity  at $\aKS$; this is not visible 
from the data of because of strong fluctuations in the neighborhood of
$\aKS$.

\begin{figure}[htbp]
\begin{center}
\hspace{-0.15\textwidth}
\includegraphics[width=0.44\textwidth]{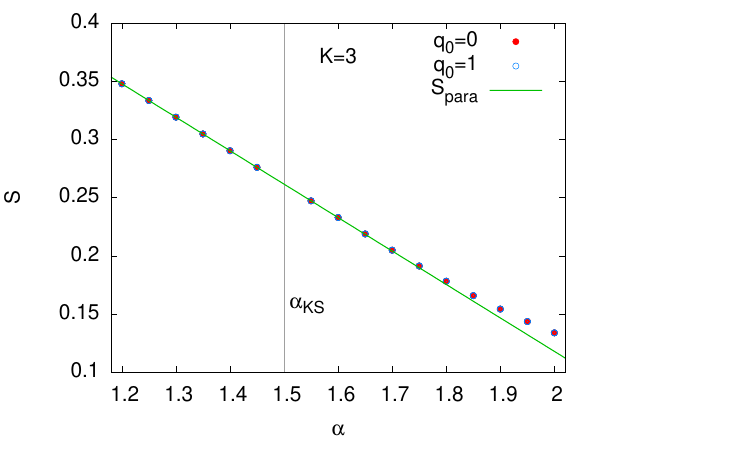}
\hspace{-0.1\textwidth}
\includegraphics[width=0.44\textwidth]{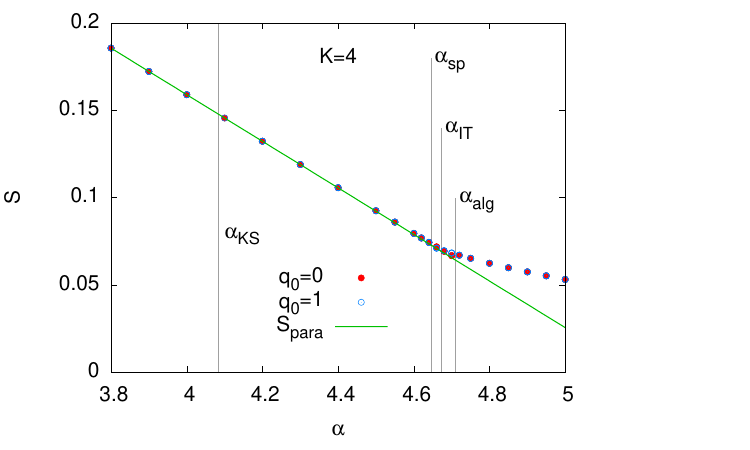}
\hspace{-0.1\textwidth}
\includegraphics[width=0.44\textwidth]{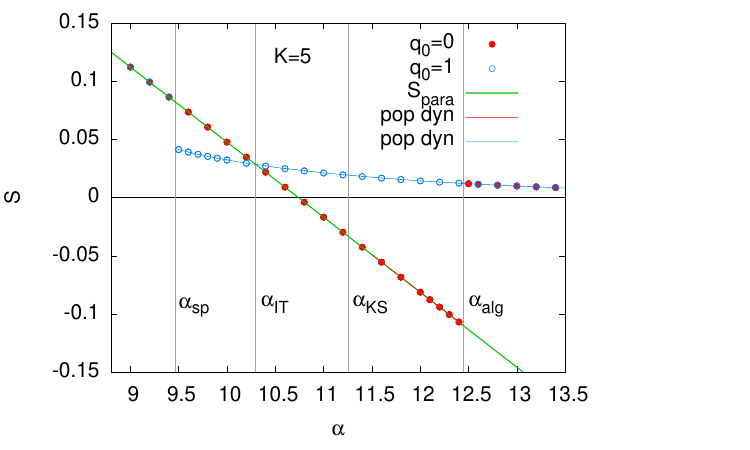}
\hspace{-0.2\textwidth}

\hspace{-0.15\textwidth}
\includegraphics[width=0.44\textwidth]{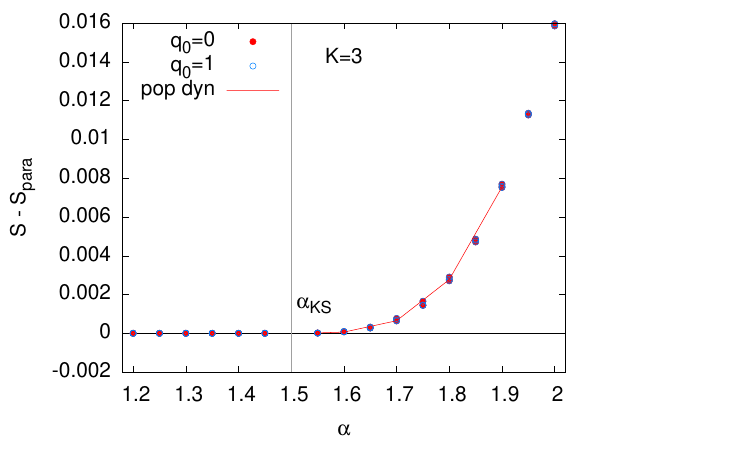}
\hspace{-0.1\textwidth}
\includegraphics[width=0.44\textwidth]{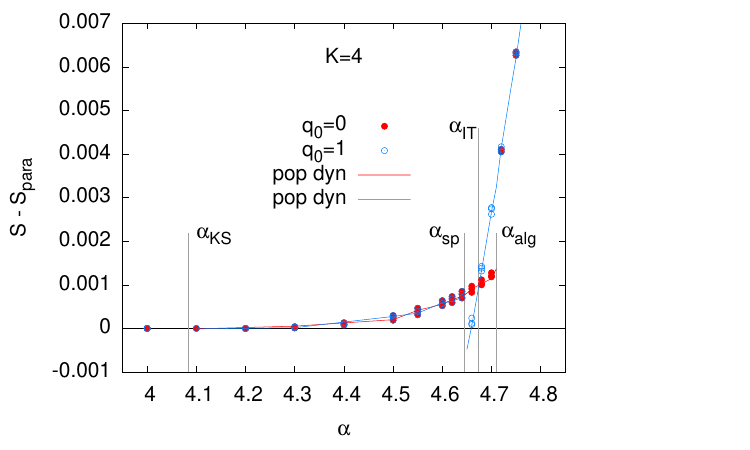}
\hspace{-0.1\textwidth}
\includegraphics[width=0.44\textwidth]{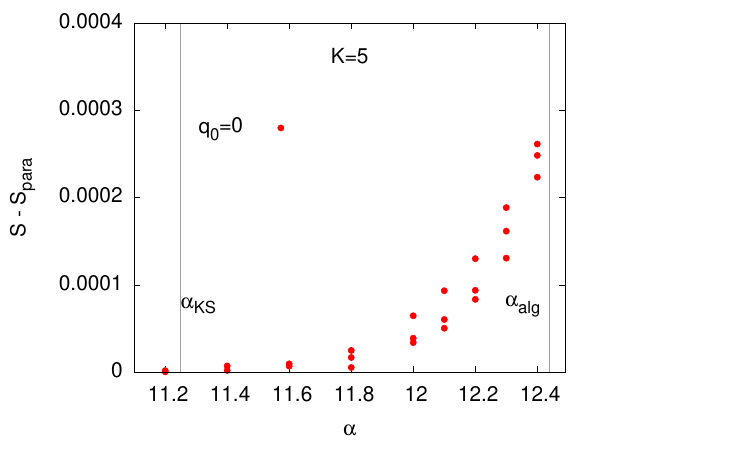}
\hspace{-0.2\textwidth}
\caption{
Bethe RS entropy computed at the BP fixed points for $k=3$ (left), $k=4$ (center) and $k=5$ (right). The lower panels show the 
difference with respect to the paramagnetic entropy in order to show better the 
difference which is tiny in some cases. For $k= 5$ the difference is too 
small to be estimated reliably via population dynamics and so we report 
only data obtained via BP.
}
\label{fig:bicol_entro}
\end{center}
\end{figure}

We investigated further the properties of the BP fixed points, by computing their Bethe entropy given in 
Eq.~(\ref{eq_Bethe_S_bicol}), the results being displayed in Figure \ref{fig:bicol_entro}.
In the top panels we also show the paramagnetic entropy (the one of the trivial fixed point) 
with a green straight line.  The comparison with the results of the cavity equations is made with the
correspondence $S=S_\text{para}+\phi$, with $\phi$ the free-entropy of Eq.~(\ref{eq_phi_occ}), to 
compensate for a different choice of normalization. The dominating fixed
point, that is the fixed point providing the right entropy of the
posterior probability distribution, is the one with with the largest
entropy at each value of $\alpha$.  The case $k=5$ clearly shows that,
while on some branches the entropy can be negative, the dominating one is always
positive (as it should be by definition of the entropy of a probability
measure over a discrete set).  

The lower panels in Figure \ref{fig:bicol_entro} serve to highlight
the way the entropy departs from the paramagnetic curve at $\aKS$. In
particular for $k=4$ it is possible to appreciate the crossing of the
entropies obtained with the two different initial conditions ($q_0=0$
and $q_0=1$) taking place at $\aIT$, that is hardly visible in the
corresponding upper panel.  Note that for $k=5$ the
difference $S-S_{\rm para}$ is extremely small, hence we did not manage to estimate it
from the population dynamics algorithm.
Nonetheless the entropy computed on the BP fixed point
clearly shows a stable increase with respect to the paramagnetic
value, with quite strong sample-to-sample fluctuations (given by the
spread of the three points in the figure).

We have shown evidence that for $k\ge 4$, in the range
$\aSP \le \alpha \le \aALG$ there exist at least two different fixed
points of BP (along with their symmetric partners under the global spin-flip symmetry). 
At $\aIT$ the entropies of these two BP fixed point
cross and this corresponds, in the thermodynamic limit, to a first
order transition between two different thermodynamic states. 
Running BP without information on the planted
configuration ($q_0=0$), the drastic change in the posterior
distribution taking place at $\aIT$ is not visible: the algorithm
remains confined to the low-informative branch until $\aALG$. 
By analogy with the bifurcation theory of fixed-point systems of equations (even if $\alpha$ is not here
a parameter modifying smoothly a set of equations of constant dimension) we expect a branch of unstable fixed
points of BP to exist in the range of parameters $\aSP \le \alpha \le \aALG$, and to connect these two branches as in the scalar toy model sketched in Fig.~\ref{fig:new}.
However, getting evidence of the existence of this unstable fixed
point is very difficult, because it is repulsive
by its very nature.

\begin{figure}[htbp]
\begin{center}
\includegraphics[width=0.49\textwidth]{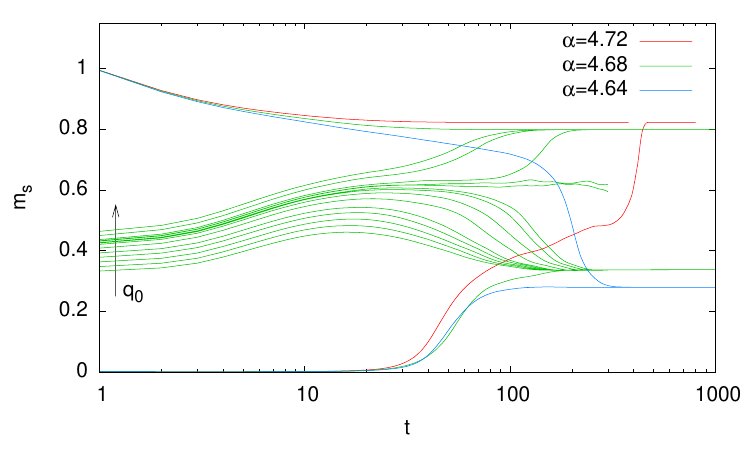}
\includegraphics[width=0.49\textwidth]{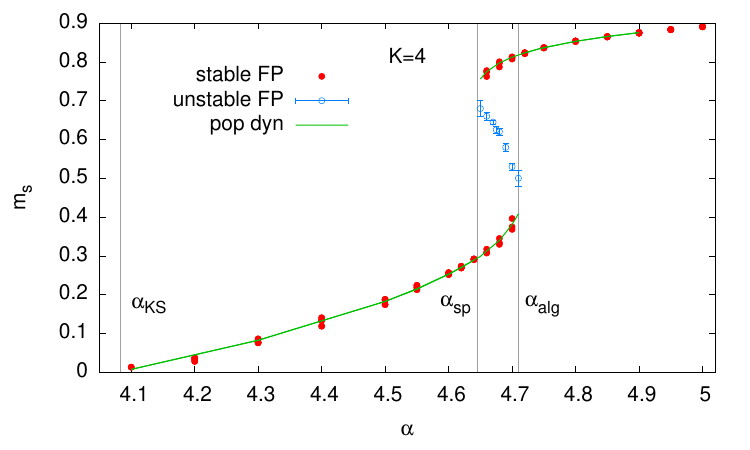}
\caption{Search for the unstable fixed point of BP in the range
  $[\aSP,\aALG]$. In the left panel, lowermost (resp.\ uppermost)
  curves have been obtained with $q_0=0$ (resp.\ $q_0=1§$); the rest
  of the curves for $\alpha=4.68$ have been obtained varying $q_0$ by
  $\Delta q_0=0.01$ in the range $[0.2,0.29]$ and by
  $\Delta q_0=0.002$ in the range $[0.26,0.27]$.}
\label{fig:bicol_unstable}
\end{center}
\end{figure}

We report in Figure~\ref{fig:bicol_unstable} the evidence we have gathered in favour of this hypothesis.
On its left panel we show the evolution of the order parameter $m_s$ as a function of the number of
iterations of the BP updates, for different values of the initialization parameter $q_0$.
During the evolution all variables are updated;
in order words $q_0$ is \emph{not} the fraction of variables
\emph{pinned} to the value of the planted configuration. So any fixed
point we find is a fixed point of the standard BP algorithm. The
initial condition is just used to aim the algorithm to different
fixed points, but then the BP algorithm is unconstrained.
The evolution of BP at $\alpha=4.68$ with different values of $q_0$
reported in the left panel of Figure~\ref{fig:bicol_unstable} clearly
shows that for $q_0$ large enough BP converges to the same high-information
fixed point reached for $q_0=1$, while for $q_0$ small enough only the low-information
fixed point can be reached. Note that it is very hard to
precisely define a \emph{separatrix} value $q_0^*$ that separates the
two regimes, because BP is a stochastic algorithm, so its behaviour is
not deterministically fixed by the initial condition, but also by the
random numbers used during the evolution to choose the order of update
of the BP messages. This means that, for the same $q_0\approx q_0^*$,
it may happen that two evolutions converge to different fixed
points. Nevertheless the left panel of Figure~\ref{fig:bicol_unstable}
show evidence that some evolutions of BP (those with $q_0=0.266$ and
$q_0=0.268$ in the present figure) remain for a long time on a
stationary regime which is neither of the two stable fixed points
already found: we define this regime as the unstable fixed point of
BP. Although the precise value of $q_0$ is not significant for the
argument we just made, the value of $m_s$ on the unstable fixed point
can be measured with a reasonable uncertainty.
We repeated the above procedure for several $\alpha$ values in the range
$[\aSP,\aALG]$ and we report in the right panel of
Figure~\ref{fig:bicol_unstable} the summary of the results for $k=4$,
with open blue symbols showing the unstable branch in addition to the results already presented
in Fig.~\ref{fig:bicol_ms}.

\subsection{Results for the asymmetric stochastic block model with two groups}

As discussed in Sec.~\ref{sec_Ising_asym} the asymmetric SBM with two groups 
undergoes different kinds
of phase transitions depending on the level of the asymmetry: for
$|\om|<\omc=1/\sqrt{3}$ the transition is continuous, while for
$|\om| >\omc$ the transition is discontinuous. We focused our analysis on
2 values of $\om$, namely $\om=0.4$ belonging to the former case and
$\om=0.8$ belonging to the latter case. We shall present results for a
rather small average degree $d=4$, for which the Kesten-Stigum transition 
occurs at $\tKS=0.5$, but
we have checked that the same conclusions apply to the case $d=8$.
We run BP only on problems of a very large size ($N=10^7$) in order to
reduce as much as possible the finite size effects.

\begin{figure}[htbp]
\begin{center}
\hspace{-0.07\textwidth}
\includegraphics[width=0.52\textwidth]{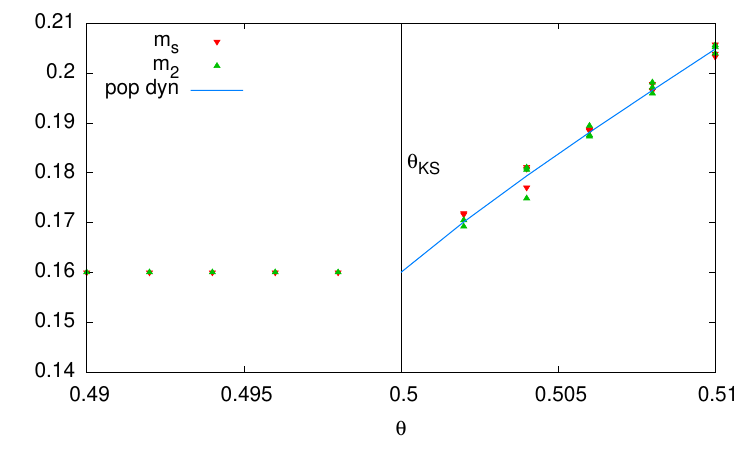}
\includegraphics[width=0.52\textwidth]{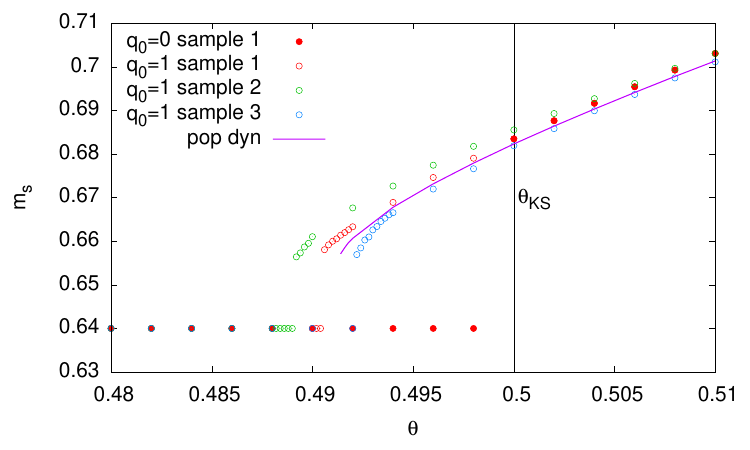}
\caption{Comparison between BP and population dynamics in detecting
  the planted configuration in the asymmetric SBM with $d=4$. For each
  $\theta$ value we have simulated 3 samples of size $N=10^7$. The
  asymmetry is $\om=0.4<\omc$ in the left panel and $\om=0.8>\omc$ in
  the right panel. In both cases the planted configuration can be detected
  purely from the graph ($q_0=0$) 
for $\theta>\tKS$. Despite visible sample-to-sample fluctuations the
  prediction by population dynamics is always very faithful.}
\label{fig:asymmSBM}
\end{center}
\end{figure}

We show in Figure~\ref{fig:asymmSBM} a comparison of the 
properties of the fixed-point reached by BP on single samples with
the results of the population dynamics study of the tree problem, 
concentrating on values of $\theta$ close to the Kesten-Stigum transition, 
where fluctuations become more
significant.
In the left panel we show data for $\om=0.4$ where the
transition is continuous: we report both the values of $m_s$ and $m_2$
at the fixed point in order to show how well the Nishimori condition
is satisfied on a given sample. For each sample (we run BP on 3
different samples for each $\theta$ value) both the runs with $q_0=0$
and $q_0=1$ converge to exactly the same fixed point or did not
converge within the maximum number of iterations $t_\text{max}=10^4$
(the latter happened for all the 3 samples at $\theta=0.5$ and for one
sample at $\theta=0.502$). We conclude that the main effects of being
close to the critical point are the lack of convergence and some
visible sample-to-sample fluctuations (remember we run BP on samples
of size $N=10^7$). Nevertheless the prediction coming from population
dynamics is accurate even in the vicinity of $\tKS$.

In the right panel of Figure~\ref{fig:asymmSBM} we show data for the
overlap with the planted configuration $m_s$ measured in 3 samples of
size $N=10^7$ and asymmetry $\om=0.8>\omc$.  Starting from an
non-informative initial condition ($q_0=0$) the planted configuration can
be detected only for $\theta\ge\tKS$, while with an informative
initial condition ($q_0=1$) the planted configuration can be detected for
$\theta\ge\td$. The Nishimori condition $m_2=m_s$ is verified with such
a good accuracy that we do not plot the data for $m_2$, the points
being almost perfectly superimposed.
The prediction of the population dynamics is faithful, although sample-to-sample
fluctuations are still clearly visible, even for such large sizes. The
quantity that mostly changes from sample to sample is the location of
the spinodal point $\td$, where the informative fixed point first
appears: given $\td$, the rest of the curve is pretty well conserved
and sample-independent.

One may wonder how we can `follow' a sample varying $\theta$ (see
right panel in Figure~\ref{fig:asymmSBM}) given that at different
$\theta$ values the graph has a different number of links within and
between the communities. The construction of the graph for each value
of $\theta$ is a random process and we identify the sample with the
random seed used by the algorithm. The stochastic algorithm that
builds the graph is such that when it is run with the same seed (and
so with the same sequence of pseudo-random numbers) and two not too
different $\theta$ values most of the links are in common between the
two graphs. This is the reason why we can `follow' a sample in
$\theta$.

\begin{figure}[htbp]
\begin{center}
\hspace{-0.15\textwidth}
\includegraphics[width=0.44\textwidth]{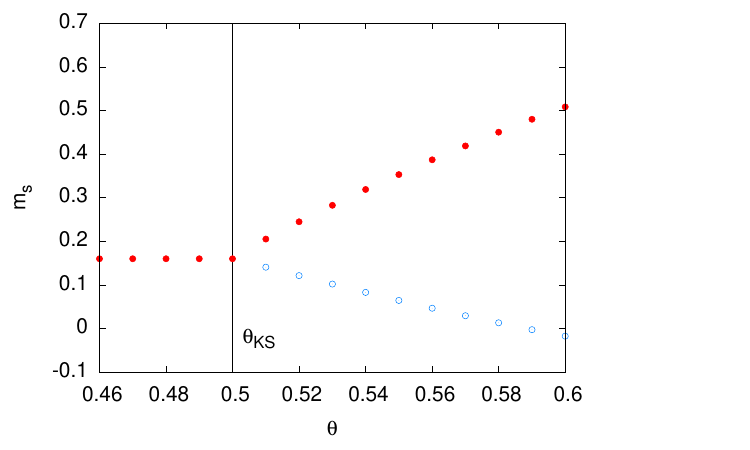}
\hspace{-0.1\textwidth}
\includegraphics[width=0.44\textwidth]{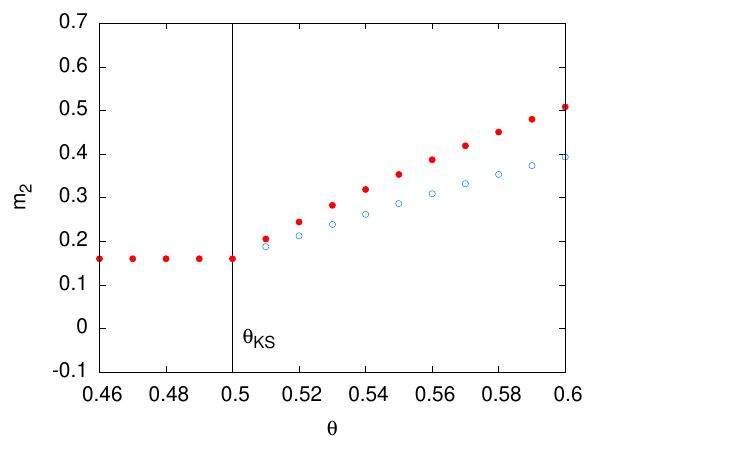}
\hspace{-0.1\textwidth}
\includegraphics[width=0.44\textwidth]{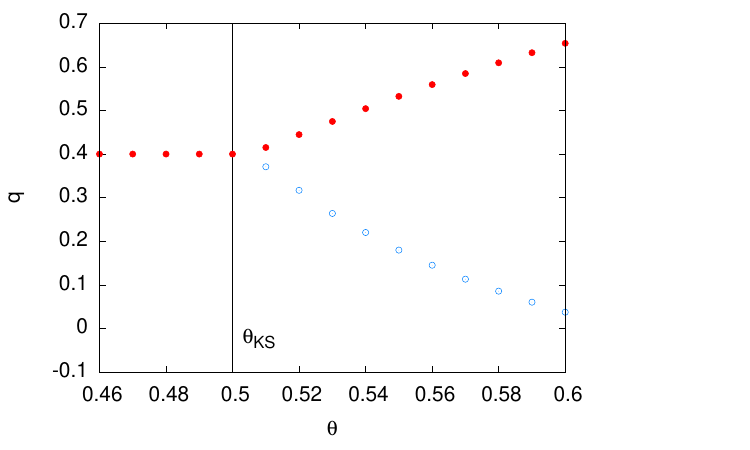}
\hspace{-0.2\textwidth}

\hspace{-0.15\textwidth}
\includegraphics[width=0.44\textwidth]{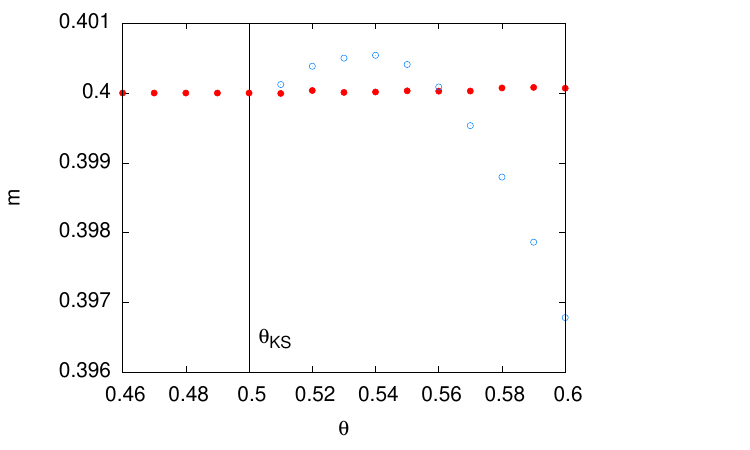}
\hspace{-0.1\textwidth}
\includegraphics[width=0.44\textwidth]{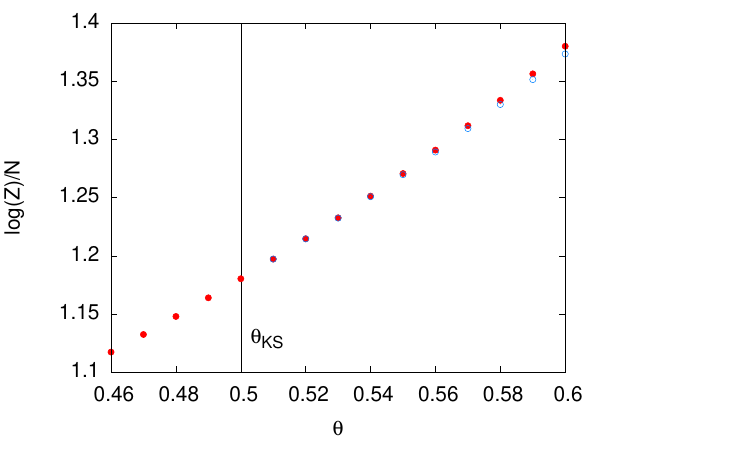}
\hspace{-0.1\textwidth}
\vspace{2mm}
\includegraphics[width=0.44\textwidth]{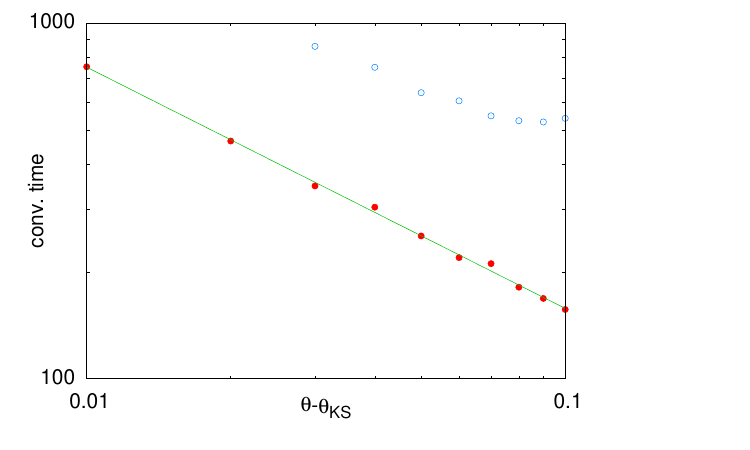}
\hspace{-0.2\textwidth}
\caption{Different fixed points reached by BP run with a
  non-informative initial condition on samples of the asymmetric SBM
  ($d=4$, $\om=0.4$, $N=10^7$). Red full points refer to the
  dominating fixed point having the largest $\log(Z)$, while blue
  empty points refer to the sub-dominating fixed point.}
\label{fig:mBar04_wide}
\end{center}
\end{figure}

The very good agreement between the BP results and the population
dynamics shown in Fig.~\ref{fig:asymmSBM} actually hides a subtle
point that we shall now explain in the case $\om=0.4$, for which the
transition at $\tKS$ is continuous. We have indeed observed that, with the
non-informative initial condition ($q_0=0$), different runs of BP on the same
sample converge to two different fixed points when $\theta > \tKS$, 
with comparable probability (as soon as $q_0>0$ this issue disappears, but
as an inference algorithm BP should not use any information on the planted 
configuration).
The properties of these two fixed points are displayed in 
Fig.~\ref{fig:mBar04_wide} (for $\theta\le\tKS$ the trivial fixed point is 
unique and has $m_s=m_2=\om^2$ and $m=q=\om$). We report with 
red full points the data corresponding to the
dominating fixed point, i.e.\ the one having the largest value of
$\log(Z)$, and with blue empty points the data corresponding to the
sub-dominant fixed point. Comparing the first two panels in
Figure~\ref{fig:mBar04_wide} we see that the dominating fixed point satisfies
the Nishimori condition $m_s=m_2$, while the sub-dominant one does
not. Also the condition $m=\om$ is satisfied by the dominant fixed point
(within finite size fluctuations), while the sub-dominant fixed point typically
violates such a condition. 
For all these 
reasons we consider the dominant fixed-point as the correct one, that is 
why we reported in Fig.~\ref{fig:asymmSBM} only the data corresponding to
this one.

The existence of 2 fixed points beyond the KS threshold can be justified by the
following reasoning. Consider first the symmetric case $\om=0$, which 
enjoys an exact global spin-flip symmetry (corresponding to the exchange of 
the two communities in the SBM interpretation). As a consequence of this
symmetry there must be (at least) two non-trivial fixed point that arise at the
Kesten-Stigum transition from the bifurcation of the trivial one, these two 
fixed points having opposite values of the local magnetizations. Consider now a slight 
increase of $\om$; by continuity one expects the two fixed points to persist, even if
they are not anymore symmetric one of the other, and in particular they
do not have the same value for their free-entropy $\log(Z)$.
What is somehow unexpected is the observation that BP reaches both fixed points
with similar probabilities. The only difference is in the time BP
takes to reach these 2 fixed points: this is shown in the sixth panel in
Figure~\ref{fig:mBar04_wide} where it is evident that reaching the
dominating fixed point requires a smaller number of iterations (the
interpolating power law fit has an exponent $\approx -2/3$).
It seems that the small difference in $\log(Z)$ between the 2 fixed points does
not influence enough the BP evolution at the very beginning, when the
choice of the basin of attraction is made; the only effect is to speed
up the evolution in case the basin of attraction of the dominating fixed point
is eventually chosen.

From the data shown in Figure~\ref{fig:mBar04_wide} it is clear the
dominating fixed point is the one we would like BP to reach, since it is the
one better correlated with the planted configuration satisfying the right
conditions. However, when running BP from a non-informative initial
condition we may likely end up in the sub-dominating fixed point.
The simplest solution is to run BP several times, until both fixed points are
found and the one with largest $Z$ is then chosen.
Actually this procedure may be slow, and so we have found a more
efficient way to reach both fixed points in just one run of BP.

At $\tKS$ the bifurcation of the trivial fixed point produces 2 fixed points which
are on opposite sides, that is the unstable fixed point stays more or less on
the midpoint of the line joining the 2 stable fixed points. This property
remains approximately true also for $\theta>\tKS$ and we can exploit
it in order to `jump' from one fixed point to the other one. Once BP reaches a
fixed point with messages $\{u_{i\to j}^\star\}$ we can produce an initial
condition for a second BP run with the transformation
\beq
u_{i\to j}^{(0)} = 2 K - u_{i\to j}^\star
\label{eq:messTrans}
\eeq
(remember the non-informative fixed point has all messages equal to $K$). We have
checked this initial condition always brings BP to the other fixed point. And
so we can find both fixed points with just 2 runs (the second being very fast,
thanks to the initial condition which is pretty close to the fixed point).

The BP initial condition for the first run,
$u_{i\to j}^{(0)} = L \tau_i^0$ with $L=100$ and $\tau_i^0$
distributed according to Eq.~(\ref{eq:tau0distr}), is quite `drastic'
(i.e.\ has very strong messages) in order to be far from the
non-informative fixed point.  For $\theta>\tKS$ this is actually not necessary,
given that the non-informative fixed point is locally unstable.  So, in order to
check that all what we have described above does not depend on such a
drastic first initial condition, we have repeated the numerical
experiments in the regime $\theta>\tKS$ by using a different initial
condition, which is much closer to the non-informative fixed point
\beq
u_{i\to j}^{(0)} = K (1 + \varepsilon \tau_i^0)
\eeq
with $\varepsilon=10^{-3}$ and $\tau_i^0$ again distributed according
to Eq.~(\ref{eq:tau0distr}). The results are identical to those with
the drastic first initial condition.

For the sake of completeness 
let us also mention that
for $\theta>\tKS$ and $\om$
large enough (e.g.\ at $\theta=0.6$ in the range $\om \gtrsim 0.46$)
we observed a divergence of the BP convergence time to the subdominant fixed point.
Increasing further $\om$ some runs of BP converge to the dominant fixed point,
others keep on wandering in a region of the message space that is far from
the dominant fixed point, but that does not contain anymore a strict fixed point. This
phenomenon could be interpreted as a spontaneous replica symmetry 
breaking of the sub-dominant fixed point.
To fix the algorithm in such a way that it always converges to the dominant 
fixed point we adopted the following rule:
stop BP after $t_\text{max}$ iterations and apply
the tranformation in (\ref{eq:messTrans}) to the current BP messages.
Even if the BP messages did not corresponds to a true fixed point their
transformation is actually close enough to the other fixed point to make the
second BP run converge to the dominant fixed point with high probability
(this trick worked for all the hundreds of samples we
tried).

\begin{figure}[htbp]
\begin{center}
\hspace{-0.07\textwidth}
\includegraphics[width=0.47\textwidth]{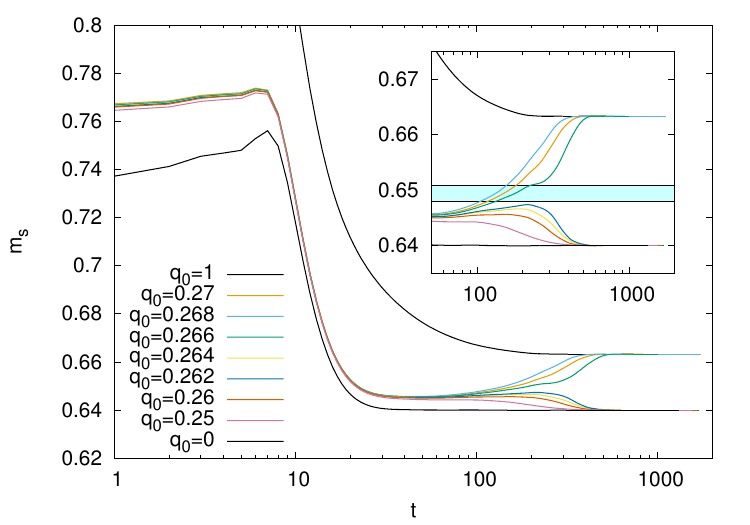}
\includegraphics[width=0.55\textwidth]{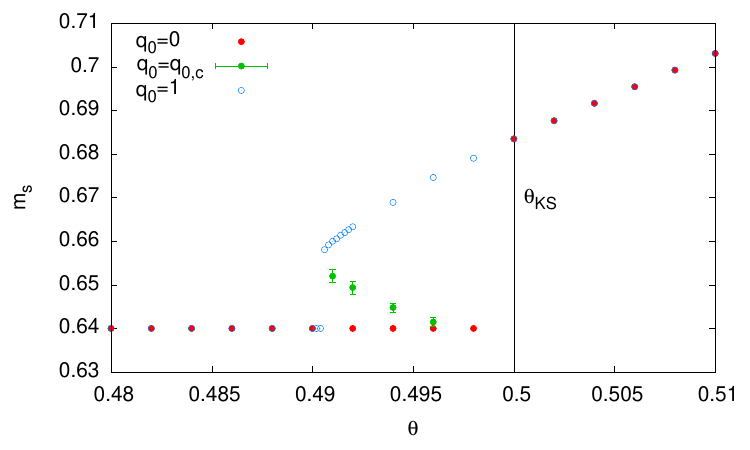}
\caption{The left panel illustrates the way we estimate the value of
  $m_s$ for the unstable fixed point of BP (here $\theta=0.492$), that
  plays the role of separatrix for the BP dynamics. The right panel
  reports these values obtained in the coexistence region
  $\td<\theta<\tKS$ (here $\om = 0.8$).}
\label{fig:asymmSBMunstable}
\end{center}
\end{figure}

Finally we present in Fig.~\ref{fig:asymmSBMunstable} a study of the 
unstable branch of fixed points of BP for the discontinuous case, $\om>\omc$,
in the range $\td<\theta<\tKS$. Similarly to what we did for the hypergraph
bicoloring (cf. Fig.~\ref{fig:bicol_unstable}), 
well chosen parameters $q_0$ for the initial
condition allow to 
identify the value of $m_s$ that plays the
role of separatrix for the BP evolution (this value with its
uncertainty is marked by a cyan strip in the inset).
In the right panel of Figure~\ref{fig:asymmSBMunstable} we report the
data of the first sample already shown in the right panel of
Figure~\ref{fig:asymmSBM} together with the values of $m_s$ measured
at the unstable fixed point in the coexistence region $\td<\theta<\tKS$. 

\section{Conclusions}
\label{sec_conclu}

We have discussed in this paper the typology of phase transitions in
inference problems, emphasizing in particular the possible existence
of hybrid-hard phases; our main technical contribution is a rather
generic expansion of the functional cavity equations for sparse models
around their trivial fixed point. Let us briefly sketch some possible
directions for future work. 

Some of our results could probably be
proven rigorously,
in particular the techniques of~\cite{Sly11} combined with the higher order 
expansions we obtained should yield the non-tightness of the Kesten-Stigum 
bound for the
reconstruction of the $q=4$ anti-ferromagnetic Potts model on a tree
with low enough degree (more directly than using the bound on the 
information theoretic threshold of~\cite{AbSa15}). 

We performed our expansions assuming either pairwise interactions between arbitrary discrete spins, or $k$-wise interactions between binary valued variables; one could relax this assumption and consider $k$-wise interactions between $q$-valued variables, as was done in~\cite{CoEfJaKaKa17} (but truncating the expansion to a low order). 

We believe the $q_1 + q_2$ SBM we introduced would deserve a further investigation; we argued that its phase diagram must contain some hybrid-hard phases, but probably in a narrow range of parameters. As a first step the large degree limit of this model could be studied, the corresponding dense model also exhibits this hybrid-hard phenomenon while being much easier to study, with evolution equations bearing on covariance matrices of Gaussian random variables instead of probability distributions.

\section*{Acknowledgment}

We would like to thank Andrea Montanari for discussions about the
asymmetric SBM with 2 communities, Amin Coja-Oghlan for useful
discussions, in particular about~\cite{CoEfJaKaKa17}, and Florent
Krzakala for discussions about bifurcation phase diagrams.

LZ acknowledges funding from the European Research Council (ERC) under
the European Unions Horizon 2020 research and innovation programme (grant agreement No 714608 - SMiLe). GS is part of the PAIL grant of the French Agence Nationale de la Recherche, ANR-17-CE23-0023-01.

\appendix
\section{Numerical resolution of the cavity equations}
\label{app_numerical_resolution}

\subsection{Population dynamics algorithm}

We shall give in this appendix some details on the numerical procedure we used to solve the cavity equations and produce the curves presented in the main part of the text. It has become customary to solve recursive distributional equations like (\ref{eq_Potts_recursion_Pt}) by a population dynamics method~\cite{ACTA73,MezardParisi01}. The main idea of this method is to a approximate a probability distribution, for instance $P^{(n)}_\t$, by the empirical distribution of a large number $\N \gg 1$ of representants of $P^{(n)}_\t$, namely
\beq
P^{(n)}_\t(\eta) = \frac{1}{\N} \sum_{i=1}^{\N} \delta(\eta - \eta^{(n,\t,i)}) \ ,
\eeq
where the $\eta^{(n,\t,i)}$ are i.i.d. samples from $P^{(n)}_\t$. The iterative equation (\ref{eq_Potts_recursion_Pt}) is then translated into a rule to generate the representative elements (i.e. the population) at iteration $n+1$ from the one at iteration $n$, as follows. For each of the $\t$ and independently for $i=1,\dots,\N$:
\begin{itemize}
\item draw $\ell$ from $\tp_\ell$.
\item draw $\t_1,\dots,\t_\ell$ with probability $M_{\t \t_1} \dots M_{\t \t_\ell}$.
\item draw $i_1,\dots,i_\ell$ independently, uniformly in $\{1,\dots,\N\}$.
\item set $\eta^{(n+1,\t,i)} = f(\eta^{(n,\t_1,i_1)},\dots,\eta^{(n,\t_\ell,i_\ell)})$.
\end{itemize}
Average values over the conditional distributions $P_\t^{(n)}$ or over the unconditional distribution $P^{(n)}$ are then evaluated as empirical averages over the population: for an arbitrary function $f$ one computes
\beq
\E_\t^{(n)}[f(\eta)] = \frac{1}{\N} \sum_{i=1}^{\N} f(\eta^{(n,\t,i)}) \ , \qquad
\E^{(n)}[f(\eta)] =\sum_\t \oeta_\t \frac{1}{\N} \sum_{i=1}^{\N} f(\eta^{(n,\t,i)}) \ .
\eeq
This approach is very natural, simple to implement, and becomes exact in the limit $\N\to\infty$. Unfortunately it suffers in many cases from an instability problem at finite $\N$ that requires additional care. In order to explain the origin of this difficulty we recall that the unconditional distribution must obey, for all iterations $n$, the condition $\E^{(n)}[\eta]=\oeta$. A consequence of this identity and of the Bayes theorem as stated in (\ref{eq_Bayes}) is that, for all $\t$, the conditional distributions should obey
\beq
\E^{(n)}_\t \left[ \frac{\oeta_\t}{\eta_\t} \right] = 1 \ .
\eeq
The elements of the population at iteration $n+1$ should thus be such that
\beq
\frac{1}{\N} \sum_{i=1}^{\N} \frac{\oeta_\t}{\eta^{(n+1,\t,i)}_\t} = 1 \ .
\label{eq_numerical_symmetry}
\eeq
If the condition $\E^{(n)}[\eta]=\oeta$ is verified exactly then it will also be the case at iteration $n+1$; however, when $\N$ is finite there are necessarily some small fluctuations around this value, that turn out to be amplified under the iteration whenever $\tE[\ell]|\theta_2| > 1$ (as in the expression of the KS transition, but without the square on the eigenvalue). In that case the implementation presented above is bound to fail because of this instability.

Depending on the model there are different ways to treat this problem. If the problem has an explicit symmetry, like the permutation invariance for the symmetric $q$-state Potts model, or the up-down symmetry for the Ising models studied in Sec.~\ref{sec_Ising}, this can be exploited to tame the instability (see below for more details on the practical implementation).

The most difficult cases arise when there is no explicit symmetry to use, for instance for the asymmetric Ising model studied in Sec.~\ref{sec_Ising_asym} with $\om \neq 0$. In such a case one needs to compute, at each iteration, the left hand side of (\ref{eq_numerical_symmetry}) from the newly generated elements, and if this average is not equal to 1 apply some transformation rules on the population samples in order to bring it closer to this target value. There are still many ways to implement this idea, we will be more explicit below for the asymmetric Ising case.

\subsection{Symmetric problems}

As mentioned above the population dynamics algorithm can be simplified and stabilized when the problem to solve exhibit some additional symmetries. Consider for instance the symmetric Potts model with $q$ states, as defined in Sec.~\ref{sec_applications_symPotts}. The invariance under any permutation $\pi$ of the $q$ states imply that $P^{(n)}(\eta) = P^{(n)}(\eta \circ \pi )$, where $(\eta \circ \pi)_\s=\eta_{\pi(\s)}$ is the probability measure on $\chi$ obtained from $\eta$ by the reshuffling $\pi$ of the $q$ states. This observation allows to express all conditional distributions $P_\t$ in terms of a single one, for instance $\t=1$, and close the iteration equation on $P_1$ as
\beq
P^{(n+1)}_1(\eta) = \sum_{\ell=0}^\infty \tp_\ell \sum_{\pi_1,\dots,\pi_\ell} \rho(\pi_1) \dots \rho(\pi_\ell) \int \dd P_1^{(n)}(\eta^1) \dots  \dd P_1^{(n)}(\eta^\ell) \, \delta(\eta - f(\eta^1 \circ \pi_1 ,\dots,\eta^\ell \circ \pi_\ell)) \ ,
\label{eq_cavity_sym_numerical}
\eeq
where the $\pi$'s are permutations on $q$ colors, $\rho$ is a probability distribution on them, such that with probability $\frac{1}{q} + \theta \left(1- \frac{1}{q} \right)$ the permutation $\pi$ is uniform under the condition $\pi(1)=1$, otherwise it is uniform under the condition $\pi(1) \neq 1$. This equation is much simpler than the generic one, as it involves only one population instead of $q$, and also much more stable numerically.

Even simpler is the treatment of Ising spin models ($q=2$) which are symmetric under the spin reversal operation, as for instance those studied in Sec.~\ref{sec_Ising}. In that case the probability laws on $\chi$ are encoded by a single real $m$, the magnetization, and the conditional and unconditional distributions verify the identities $P^{(n)}(m) = P^{(n)}(-m)$, $P^{(n)}_-(m) = P^{(n)}_+(-m)$. This allows to close the cavity equation on a single population $P^{(n)}_+(m)$, see in particular equations (\ref{eq_occ_conditional_hPtoP},\ref{eq_occ_conditional_PtohP}).

\subsection{The asymmetric Ising model}

Let us finally turn to the case of the asymmetric Ising model, defined in Sec.~\ref{sec_Ising_asym}. The general BP equation of Eq.~(\ref{eq_BP}) for $\eta_\s$ can be written as a scalar recursion by parametrizing $\eta_\s = \frac{1+\s m}{2}$. The BP equation reads in this parametrization $m=f(m^1,\dots,m^\ell)$, with
\bea
&&m = \frac{z_+(m^1,\dots,m^\ell) - z_-(m^1,\dots,m^\ell)}{z(m^1,\dots,m^\ell)} \ , \qquad
z(m^1,\dots,m^\ell) = z_+(m^1,\dots,m^\ell) + z_-(m^1,\dots,m^\ell) \ ,  \\
&& z_+(m^1,\dots,m^\ell) = \frac{1+\om}{2} \prod_{i=1}^\ell \left(1 + \theta \frac{m^i - \om}{1+\om} \right) \ , \qquad
z_-(m^1,\dots,m^\ell) = \frac{1-\om}{2} \prod_{i=1}^\ell \left(1 - \theta \frac{m^i - \om}{1-\om} \right) \ .
\eea
One sees immediately that $z_\sigma(\om,\dots,\om)=\frac{1+ \sigma \om}{2}$, hence $z(\om,\dots,\om)=1$ and $\om=f(\om,\dots,\om)$ is a fixed point, as it should.

The ``naive'' implementation of the population dynamics algorithm explained above corresponds to a representation of $P_+^{(n)}(m)$ (resp. $P_+^{(n)}(m)$) by $\N$ reals $m^{(n,+,i)}$ (resp. $m^{(n,-,i)}$). In order to cure the instability and to enforce the condition (\ref{eq_numerical_symmetry}) at each iteration we have proceeded as follows: once the $2 \N$ magnetizations $m^{(\s,i)}$ have been generated according to the standard procedure (we remove the iteration index to lighten the notation) we compute 
\beq
\beta^+ = \frac{2}{1-\om} \left(\frac{1}{\N} \sum_{i=1}^{\N} \frac{1+\om}{1+m^{(+,i)}} - \frac{1+\om}{2} \right) \ , \qquad
\beta^- = \frac{2}{1+\om} \left(\frac{1}{\N} \sum_{i=1}^{\N} \frac{1-\om}{1-m^{(-,i)}} - \frac{1-\om}{2} \right) \ ,
\eeq
and replace the elements of the population by
\beq
{m'}^{(+,i)} = \frac{\beta^+ (1+ m^{(+,i)}) - (1- m^{(+,i)})}{\beta^+ (1+ m^{(+,i)}) + (1- m^{(+,i)}) }  \ , \qquad
{m'}^{(-,i)} = \frac{(1+ m^{(-,i)}) - \beta^- (1- m^{(-,i)})}{ (1+ m^{(-,i)}) + \beta^- (1- m^{(-,i)}) } \ .
\eeq
One can check that this procedure strictly enforces the identity (\ref{eq_numerical_symmetry}), the coefficients $\beta$ measuring the deviation of the naively generated population elements from the symmetry respecting probability distributions.

\section{Moment expansion for pairwise interacting Potts variables}
\label{sec_app_expansion_Potts}

The goal of this appendix is to justify some of the statements made in Sec.~\ref{sec_expansion_Potts}, in particular the ansatz (\ref{eq_ansatz}) for the scaling of the centered moments of the perturbative non-trivial fixed point, and the third order expansion given in (\ref{eq_Potts_cavity_A}-\ref{eq_Potts_cavity_C}).

Let us consider a fixed point solution $P(\eta)$ of the recursion equation (\ref{eq_Potts_recursion_P}), and look for a hierarchy of equations between its centered moments. As in the main text we denote $\delta_\s=\eta_\s - \oeta_\s$ the centered random variable with distribution $P$, and $\hdelta_\s = \sum_{\s'} \hM_{\s \s'} \delta_{\s'}$ a linearly transformed version of $\delta$. Denoting $\E[\bullet]$ the average with respect to $P$ we have by definition $\E[\delta_\s]=0$, and we assume $\E[\delta_\s \delta_{\s'}]$ to be non-zero but small, of an order denoted $\kappa$. From the self-consistent equation (\ref{eq_Potts_recursion_P}) on $P$ and the expression (\ref{eq_BP_delta}) of the function $f$ we can write the $p$-th centered moment of $P$ ($p \ge 2$ is understood below) as:
\beq
\E [\delta_{\s_1} \dots \delta_{\s_p}] = 
\E \left[ \left( \frac{z_{\s_1}}{z} - \oeta_{\s_1} \right) 
\cdots
             \left( \frac{z_{\s_p}}{z} - \oeta_{\s_p} \right)  z 
\right] \ ,
\eeq
where on the right hand side $z_\s=\oeta_\s \prod_{i=1}^\ell(1+\hdelta_\s^i)$, $z=\sum_\g z_\g$, $\ell$ is drawn from $\tp_\ell$ and for $i=1,\dots, \ell$ the $\hdelta^i$ are  linearly transformed (according to (\ref{eq_BP_def_hdelta})) versions of $\eta^i$'s i.i.d. samples drawn from $P(\eta)$. To simplify this expression we introduce the notation $\ve_\s = \prod_{i=1}^\ell(1+\hdelta_\s^i) - 1$, in such a way that $z_\s = \oeta_\s (1+\ve_\s)$ and $z = 1 + \sum_\g \oeta_\g \ve_\g$. The previous equation thus becomes
\bea
\E[\delta_{\s_1} \dots \delta_{\s_p}] &=& 
\oeta_{\s_1} \dots \oeta_{\s_p} \E[
(\ve_{\s_1} - \sum_{\g_1} \oeta_{\g_1} \ve_{\g_1}) 
\dots
(\ve_{\s_p} - \sum_{\g_p} \oeta_{\g_p} \ve_{\g_p}) 
(1 + \sum_\g \oeta_\g \ve_\g)^{1-p}
]  \label{eq_ave_proddelta} \\
&=& \oeta_{\s_1} \dots \oeta_{\s_p} \sum_{m=0}^\infty
\binom{1-p}{m} \sum_{\g_1,\dots,\g_{p+m}} (\delta_{\s_1,\g_1} - \oeta_{\g_1})
\dots
(\delta_{\s_p,\g_p} - \oeta_{\g_p}) 
\oeta_{\g_{p+1}} \dots \oeta_{\g_{p+m}}
\E[\ve_{\g_1} \dots \ve_{\g_{p+m}} ] \ , \nonumber
\eea
where the binomial coefficient with negative argument takes its conventional value $\binom{1-p}{m} = (-1)^m \binom{p-2+m}{m}$. In order to close these equations we should now express the average of products of $\ve$'s in terms of centered moments of $P$. As an intermediate step let us further define $\mu_\s=\ve_\s + 1 =\prod_{i=1}^\ell(1+\hdelta_\s^i)$; exploiting the independence of the random variables $\eta^i$ for distinct $i$ one can easily compute the average of products of $\mu$'s,
\beq
\E[\mu_{\s_1} \dots \mu_{\s_p}] = \sum_{\ell=0}^\infty \tp_\ell \left( 1 + \sum_{S} \E\left[\prod_{j \in S} \hdelta_{\s_j} \right]  \right)^\ell \ ,
\eeq
where $S$ is summed over subsets of $\{1,\dots,p\}$ of cardinality $|S| \ge 2$ (because of the property $\E[\hdelta_\s]=0$). Expanding the last power we rewrite this as
\beq
\E[\mu_{\s_1} \dots \mu_{\s_p}] = 1 +  \sum_{r=1}^\infty \tE\left[\binom{\ell}{r} \right] \sum_{S_1,\dots,S_r} \E\left[\prod_{j \in S_1} \hdelta_{\s_j} \right] \dots \E\left[\prod_{j \in S_r} \hdelta_{\s_j} \right] \ ,
\label{eq_aveprodmu}
\eeq
where the subsets $S$ satisfy the same properties as above, and the average denoted $\tE$ is over $\ell$ drawn with probability $\tp_\ell$, the requirement $\ell \ge r$ being kept understood. We can now come back to the computation of the average of products of $\ve$'s, noting that
\beq
\E[\ve_{\s_1} \dots \ve_{\s_p}] = \E[(\mu_{\s_1}-1) \dots (\mu_{\s_p}-1)] =
\sum_T (-1)^{p-|T|} \E\left[\prod_{j \in T} \mu_{\s_j} \right] \ ,
\label{eq_aveprodve}
\eeq
where $T$ runs over all subsets of $\{1,\dots,p\}$. Before proceeding let us recall the following version of the inclusion-exclusion principle: if $E$ is an arbitrary finite set, $T$ runs over all the subsets of $E$, and $F$ is a subset of $E$, then
\beq
\sum_{T \subset E} (-1)^{|E|-|T|} \ind(F \subset T) = (1-1)^{|E|-|F|} = \ind (F=E) \ ,
\eeq
which can be easily proven by enumerating the number of subsets $T$ that contains $F$ and that are contained in $E$. Consider now an arbitrary function $g(S)$ of the subsets of $E$, and an integer $r$; as a consequence of the above identity we have
\bea
\sum_{T \subset E} (-1)^{|E|-|T|} \sum_{S_1 \subset T , \dots , S_r \subset T } g(S_1) \dots g(S_r) &=& \sum_{S_1 \subset E , \dots ,S_r \subset E } g(S_1) \dots g(S_r) \sum_{T \subset E} (-1)^{|E|-|T|} \ind(S_1 \subset T) \dots \ind(S_r \subset T) \nonumber \\
&=& \sum_{S_1 \subset E , \dots ,S_r \subset E } g(S_1) \dots g(S_r) \ind(S_1 \cup \dots \cup S_r = E ) \ .
\eea 
Inserting (\ref{eq_aveprodmu}) in (\ref{eq_aveprodve}) and using this last form of the inclusion-exclusion principle yields finally
\beq
\E[\ve_{\s_1} \dots \ve_{\s_p} ] = \tE[\ell] \E[\hdelta_{\s_1} \dots \hdelta_{\s_p} ] + \sum_{r=2}^\infty \tE\left[\binom{\ell}{r} \right] 
\sum_{S_1,\dots,S_r } \E\left[\prod_{j \in S_1} \hdelta_{\s_j} \right] \dots \, \E\left[\prod_{j \in S_r} \hdelta_{\s_j} \right] \ ,
\label{eq_aveprodve2}
\eeq
where we treated the term $r=1$ separately as it is simpler than the general terms for $r \ge 2$. In the latter the summation is over $S_1,\dots,S_r$ subsets of $\{1,\dots,p\}$ of cardinality $|S_i| \ge 2$ whose union must cover $\{1,\dots,p\}$ (but they do not need to be disjoint). Each term in this sum can be associated to a diagram, i.e. an hypergraph on $p$ vertices with $r$ hyperedges that connect at least $2$ vertices, in such a way that no vertex remains isolated. Furthermore, in (\ref{eq_ave_proddelta}) some of the $\ve_\s$ appears multiplied by $\oeta_\s$ and summed over $\s$; in that case the only diagrams that contribute are those in which the corresponding vertex has degree at least 2, because of the property $\sum_\s \oeta_\s \hdelta_\s =0$.

The two expressions (\ref{eq_ave_proddelta}) and (\ref{eq_aveprodve2}) form an infinite hierarchy of equations that, formally, determine all the centered moments of the possible distributions $P(\eta)$ fixed point solutions of (\ref{eq_Potts_recursion_P}); as $\eta$ is bounded these moments are enough to characterize $P$ itself. The results presented in Sec.~\ref{sec_expansion_Potts} have been obtained by truncating this hierarchy under the hypothesis that
\beq
\E[\delta_{\s_1} \dots \delta_{\s_p}] = O( \kappa^{\lceil p/2 \rceil}) \ ,
\qquad
\E[\ve_{\s_1} \dots \ve_{\s_p}] = O( \kappa^{\lceil p/2 \rceil}) \ ,
\eeq
where $\kappa$ is a small parameter controlling the distance between the studied fixed point $P(\eta)$ and the trivial one $\delta(\eta - \oeta)$. The self-consistency of this ansatz can be checked on (\ref{eq_ave_proddelta},\ref{eq_aveprodve2}); using the hypothesis on $\E[\ve_{\s_1} \dots \ve_{\s_p}]$ in the right hand side of (\ref{eq_ave_proddelta}) leads to a compatible scaling for $\E[\delta_{\s_1} \dots \delta_{\s_p}]$, the dominant contribution coming from the term $m=0$ if $p$ is even, and from $m=0$ and $m=1$ if $p$ is odd. Similarly one can insert the hypothesis on $\E[\delta_{\s_1} \dots \delta_{\s_p}]$ in the right hand side of (\ref{eq_aveprodve2}); the first term corresponding to $r=1$ is obviously compatible with the ansatz. The order in $\kappa$ of the generic term with $r \ge 2$ is
\beq
\left\lceil \frac{|S_1|}{2} \right\rceil + \dots + \left\lceil \frac{|S_r|}{2} \right\rceil \ge \left\lceil  \frac{|S_1| + \dots + |S_r|}{2} \right\rceil \ge \left\lceil \frac{p}{2}\right\rceil \ ,
\eeq
as the sets $S_1,\dots,S_r$ have to cover $\{1,\dots,p\}$, hence these terms are not strictly dominant with respect to the case $r=1$, which concludes our verification of the consistency of the ansatz. From (\ref{eq_ave_proddelta},\ref{eq_aveprodve2}) it is relatively easy to obtain the third order expansion stated in the main text in Eqs.~(\ref{eq_Potts_cavity_A}-\ref{eq_Potts_cavity_C}), exploiting the ansatz to truncate the hierarchy of equations at the desired order in $\kappa$, and the diagrammatic representation of the terms in (\ref{eq_aveprodve2}) to organize their bookkeeping.

\section{Expansion of the free-entropy for pairwise interacting Potts variables}
\label{sec_app_expansion_Potts_phi}

We provide in this Appendix some details on the derivation of the expansion (\ref{eq_Potts_phi}) for the free-entropy of pairwise interacting models. Let us start from the expression (\ref{eq_Potts_phi_uncond}) that we rewrite in a more compact notation,
\beq
\phi(P) = \E[\zv \ln \zv] -  \frac{1}{2} \E[\ell] \E[\ze \ln \ze] \ ,
\eeq
where the expressions of $\zv$ and $\ze$ are given in (\ref{eq_BP}) and (\ref{eq_Potts_ze}) respectively.

To deal with the second term of $\phi$ we write $\ze=1+\sum_\s \delta_\s^1 \hdelta_\s^2$, where $\delta_\s^1=\eta_\s^1 - \oeta_\s$ and $\hdelta_\s^2 = \sum_{\s'} \hM_{\s \s'} (\eta_{\s'}^2-\oeta_{\s'})$ with $\eta^1$ and $\eta^2$ two independent samples from $P(\eta)$. Using the power series expansion of $(1+x) \ln(1+x)$ around $x=0$ gives
\beq
\E[\ze \ln \ze] = \sum_{p=2}^\infty \frac{(-1)^p}{p(p-1)} \sum_{\s_1,\dots,\s_p} \E[\delta_{\s_1} \dots \delta_{\s_p} ] \E[\hdelta_{\s_1} \dots \hdelta_{\s_p} ] \ .
\eeq
Neglecting the terms with $p \ge 5$ yields the first line in (\ref{eq_Potts_phi}).

In the first term we write $\zv = 1 + \sum_\s \oeta_\s \ve_\s$, where the notation $\ve_\s = \prod_{i=1}^\ell(1+\hdelta_\s^i) - 1$ is the same as in Appendix \ref{sec_app_expansion_Potts}, the only difference being that $\ell$ is now drawn from the distribution $p_\ell$ instead of $\tp_\ell$. This yields
\beq
\E[\zv \ln \zv] =\sum_{p=2}^\infty \frac{(-1)^p}{p(p-1)} \sum_{\s_1,\dots,\s_p} \oeta_{\s_1} \dots \oeta_{\s_p} \E[\ve_{\s_1} \dots \ve_{\s_p} ] \ ,
\eeq
where the expression of $\E[\ve_{\s_1} \dots \ve_{\s_p} ]$ can be read off from (\ref{eq_aveprodve2}) with the modification $\tp_\ell \to p_\ell$. Enumerating the various terms that contribute up to order $\kappa^4$ using the diagrammatic representation explained in Appendix \ref{sec_app_expansion_Potts} yields the remaining terms of (\ref{eq_Potts_phi}). We factored out in these terms a common factor $\E[\ell]$, using the identity
\beq
\E[\ell (\ell-1) \dots (\ell - r +1)] = \E[\ell] \tE[\ell (\ell-1) \dots (\ell - r +2)]
\eeq
that follows from the size bias between $p_\ell$ and $\tp_\ell$ expressed in (\ref{eq_tpell}).

Finally let us write down an expression that is equivalent to the expansion (\ref{eq_Potts_phi}) but where the centered moments are expressed in the basis of eigenvectors of $M$:
\bea
\frac{1}{\E[\ell]} \phi(P) &=& 
\frac{1}{4} \sum_{jk} (A'_{jk})^2 \theta_j \theta_k (\tE[\ell]\theta_j \theta_k - 1)
- \frac{1}{12} \sum_{jkl} (B'_{jkl})^2 \theta_j \theta_k \theta_l (\tE[\ell]\theta_j \theta_k \theta_l - 1) \label{eq_Potts_phip} \\
&&+ \frac{1}{24} \sum_{jklm} (C'_{jklm})^2 \theta_j \theta_k \theta_l \theta_m (\tE[\ell]\theta_j \theta_k \theta_l \theta_m - 1) \nonumber \\
&& + \frac{1}{12}\tE[\ell (\ell-1)] \left[
\sum_{\substack{j_1 j_2 j_3 \\ k_1 k_2 k_3}} 
A'_{j_1 k_1} A'_{j_2 k_2} A'_{j_3 k_3} \theta_{j_1} \theta_{k_1} \theta_{j_2} \theta_{k_2} \theta_{j_3} \theta_{k_3} f_{j_1 j_2 j_3} f_{k_1 k_2 k_3}
- 2 \sum_{j_1 j_2 j_3} A'_{j_1 j_2} A'_{j_2 j_3} A'_{j_3 j_1} \theta_{j_1}^2 \theta_{j_2}^2 \theta_{j_3}^2 
\right] \nonumber \\
&& -\frac{1}{2} \tE[\ell (\ell-1)] \sum_{\substack{j k l \\ j_1 j_2 }} B'_{jkl} A'_{j_1 k} A'_{j_2 l} \theta_j \theta_{j_1} \theta_{j_2} \theta_k^2 \theta_l^2 f_{j j_1 j_2} + \frac{1}{4} \tE[\ell (\ell-1)] \sum_{jklm} C'_{jklm} A'_{jk} A'_{lm} \theta_j^2 \theta_k^2 \theta_l^2 \theta_m^2
\nonumber \\
&& + \frac{1}{48} \tE[\ell (\ell-1)(\ell-2)]\sum_{\substack{j_1 j_2 j_3 j_4\\ k_1 k_2 k_3 k_4}} A'_{j_1 k_1} A'_{j_2 k_2} A'_{j_3 k_3} A'_{j_4 k_4} \theta_{j_1} \theta_{k_1} \theta_{j_2} \theta_{k_2} \theta_{j_3} \theta_{k_3} \theta_{j_4} \theta_{k_4} h_{k_1 k_2 k_3 k_4}^{j_1 j_2 j_3 j_4} \ ,
\nonumber
\eea
with the definitions of the tensors that encode the structure of the eigenvectors of $M$:
\bea
f_{j_1 j_2 j_3} &=& \sum_\s \oeta_\s r_\s^{(j_1)} r_\s^{(j_2)} r_\s^{(j_3)} \ , \\
g_{j_1 j_2 j_3 j_4} &=& \sum_\s \oeta_\s r_\s^{(j_1)} r_\s^{(j_2)} r_\s^{(j_3)} r_\s^{(j_4)} \ , \\
h_{k_1 k_2 k_3 k_4}^{j_1 j_2 j_3 j_4} &=& g_{j_1 j_2 j_3 j_4} g_{k_1 k_2 k_3 k_4} - 6 g_{j_1 j_2 j_3 j_4} \delta_{k_1 k_2} \delta_{k_3 k_4} - 12 f_{j_1 j_2 j_3} f_{k_1 k_2 j_4} \delta_{k_3 k_4} + 3 \delta_{j_1 j_2} \delta_{k_1 k_2} \delta_{j_3 j_4} \delta_{k_3 k_4} + 12 \delta_{k_4 j_1} \delta_{k_1 j_2} \delta_{k_2 j_3} \delta_{k_3 j_4} \ . \nonumber
\eea

\bibliography{biblio}

\end{document}